\documentclass[12pt]{article}

\usepackage{graphicx}
\usepackage{epstopdf}
\usepackage[body={17.5cm, 21cm},right=2cm]{geometry}
\usepackage{amssymb}
\usepackage{amsmath}
\usepackage{cite}
\usepackage{hyperref}

\newcommand\eea{\end{eqnarray}}
\newcommand\bea{\begin{eqnarray}}

\def\beq{\begin{equation}}
\def\eeq{\end{equation}}

\def\leq{\raise 0.4ex\hbox{$<$}\kern -0.8em\lower 0.62ex\hbox{$-$}}
\def\geq{\raise 0.4ex\hbox{$>$}\kern -0.7em\lower 0.62ex\hbox{$-$}}
\def\lsim{\raise 0.4ex\hbox{$<$}\kern -0.8em\lower 0.62ex\hbox{$\sim$}}
\def\gsim{\raise 0.4ex\hbox{$>$}\kern -0.7em\lower 0.62ex\hbox{$\sim$}}
\newcommand{\be}{\begin{equation}}
\newcommand{\ee}{\end{equation}}
\newcommand{\ba}{\begin{align}}
\newcommand{\ea}{\end{align}}
\newcommand{\bg}{\begin{gather}}
\newcommand{\eg}{\end{gather}}
\newcommand{\bseq}{\begin{subequations}}
\newcommand{\eseq}{\end{subequations}}

\begin{document}

\vspace{5mm}
\vspace{0.5cm}
\begin{center}

\def\thefootnote{\fnsymbol{footnote}}

{\Large \bf Soft-Pion Theorems for Large Scale Structure}
\\[0.5cm]

{\small
Bart Horn, Lam Hui and Xiao Xiao \\
\textit{Institute for Strings, Cosmology and Astroparticle Physics (ISCAP) and Physics Department\\ Columbia University, New York, NY 10027, USA}}\\

\end{center}

\vspace{.8cm}

\hrule \vspace{0.3cm}
{\small  \noindent \textbf{Abstract} \\[0.3cm]
\noindent
Consistency relations -- which relate an N-point function to a squeezed (N+1)-point 
function -- are useful in large scale structure (LSS) because 
of their non-perturbative nature: they hold even if the N-point
function is deep in the nonlinear regime, and even if they involve astrophysically 
messy galaxy observables. 
The non-perturbative nature of the consistency relations is guaranteed by the fact
that they are symmetry statements, in which the velocity plays the role of the soft pion.
In this paper, we address two issues: 
(1) how to derive the relations systematically using the
residual coordinate freedom in the Newtonian gauge, and relate
them to known results in $\zeta$-gauge (often used in studies of inflation);
(2) under what conditions
the consistency relations are violated.
In the non-relativistic limit, 
our derivation reproduces the Newtonian consistency relation
discovered by Kehagias \& Riotto and Peloso \& Pietroni. 
More generally, there is an infinite set of consistency relations, as
is known in $\zeta$-gauge. There is a one-to-one
correspondence between symmetries in the two gauges;
in particular, the Newtonian consistency relation
follows from the dilation and special conformal symmetries in $\zeta$-gauge.
We probe the robustness of the consistency relations by
studying models of galaxy dynamics and biasing. 
We give a systematic
list of conditions under which the consistency relations are violated;
violations occur if the galaxy bias is non-local in an infrared
divergent way. 
We emphasize the relevance of the adiabatic mode condition,
as distinct from symmetry considerations.
As a by-product of our investigation, we discuss a simple fluid Lagrangian for LSS.

\vspace{0.5cm}  \hrule
\def\thefootnote{\arabic{footnote}}
\setcounter{footnote}{0}

\vspace{.8cm}

\section{Introduction}

In studies of large scale structure (LSS), we are familiar with
the consequence of {\it linearly realized} symmetries. For instance,
consider a time-independent spatial translation:
\begin{eqnarray}
\vec x \rightarrow \vec x + \vec \epsilon\, ,
\end{eqnarray}
where $\vec \epsilon$ is constant in time and space.
For a small $\vec \epsilon$, we can think of its effect
on the density field $\delta$ as:
\begin{eqnarray}
\delta \rightarrow \delta + \Delta \delta \quad {\rm where} \quad
\Delta \delta \equiv - \vec \epsilon \cdot \vec \nabla \delta \, .
\end{eqnarray}
Notice how the change in the field $\Delta\delta$ depends
{\it linearly} on the field i.e.\ the symmetry is linearly
realized. Its consequence for an N-point correlation function
is well known:
\begin{eqnarray}
\label{DeltaCorr}
\Delta \langle \delta(\vec x_1) ... \delta(\vec x_N) \rangle
= - \vec \epsilon \cdot \sum_{a=1}^N \vec \nabla_a \langle 
\delta (\vec x_1) ... \delta (\vec x_N) \rangle  = 0\, ,
\end{eqnarray}
where $\vec \nabla_a$ stands for $\partial / \partial \vec x_a$.
This is nothing other than an infinitesimal form of the statement
that the correlation function is translationally invariant:\ $\langle \delta(\vec x_1) ... \delta(\vec x_N) \rangle
= \langle \delta(\vec x_1 - \vec \epsilon) ... \delta(\vec x_N - \vec
\epsilon) \rangle$.  Assuming the initial conditions
were translationally invariant, the correlation function at late times
has the same property since the dynamics itself does.

What is somewhat less familiar in LSS is the consequence of 
{\it nonlinearly realized} symmetries. By this, we mean that the symmetry
involves transforming the field of interest in a nonlinear way.
In this paper, a field that transforms this way will turn out to be
the velocity field $\vec v$, i.e.
\begin{eqnarray}
\Delta \vec v = \Delta_{\rm lin.} \vec v + \Delta_{\rm nl.} \vec v \, ,
\end{eqnarray}
where $\Delta_{\rm lin.} \vec v$ depends linearly on the fluctuating
variable $\vec v$, while $\Delta_{\rm nl.} \vec v$ does not depend on
the fluctuating field, or any fluctuating field for that matter --
in this sense, it might be more precise to call $\Delta_{\rm nl.} \vec
v$ sublinear. An example is a {\it time-dependent} spatial
translation:
\begin{eqnarray}
\label{timedepTranslate}
\vec x \rightarrow \vec x + \vec n(\eta)\, ,
\end{eqnarray}
where $\vec n$ depends on (conformal) time $\eta$ but
not space. It is evident that
under such a transformation, in addition to the usual
linear transformation $\Delta_{\rm lin.} \vec v = - \vec n \cdot \vec
\nabla \vec v$, the velocity experiences a nonlinear
shift:
\begin{eqnarray}
\label{DeltaNLv}
\Delta_{\rm nl.} \vec v = \vec n' \, ,
\end{eqnarray}
where ${}'$ denotes a time derivative.
It was pointed out recently by Kehagias \& Riotto and
Peloso \& Pietroni \cite{Kehagias:2013yd,Peloso:2013zw} (KRPP) that 
just such a time-dependent spatial translation
is in fact a symmetry of the familiar system of
equations for pressureless fluid coupled to gravity in the Newtonian limit:\ the continuity, Euler and Poisson equations.
As pointed out by the same authors, the consequence
of such a nonlinearly-realized symmetry is {\it not}
the simple invariance of a general correlation function
(such as in Eq.\ \ref{DeltaCorr}), but rather a relation
between an (N+1)-point function and an N-point function:
\begin{eqnarray}
\label{KRPPrelationIntro}
\lim_{\vec q \rightarrow 0} \, 
{\langle \vec v (\vec q) {\cal O}_{\vec k_1} {\cal O}_{\vec k_2} ...
{\cal O}_{\vec k_N} \rangle^{c'} \over P_v (q)} \sim 
\langle {\cal O}_{\vec k_1} {\cal O}_{\vec k_2} ... {\cal O}_{\vec
  k_N} \rangle^{c'} \, ,
\end{eqnarray}
where $P_v$ is the velocity power spectrum,
the superscript $c'$ denotes 
a {\it connected} correlation function with
the overall delta function removed, and ${\cal O}$ denotes
some observable, which could be different for each $\vec k$.

This sort of {\it consistency relation}, which relates a squeezed
(N+1)-point function to an N-point function, is well known in
the context of single field inflation. The first example was pointed out by
Maldacena \cite{Maldacena:2002vr} (see also \cite{Creminelli:2004yq}):
\begin{eqnarray}
\lim_{\vec q \rightarrow 0} 
{\langle \zeta_{\vec q} \zeta_{\vec k_1}
... \zeta_{\vec k_N} \rangle^{c'} \over P_\zeta (q)} = 
- \left( 3(N-1) + \sum_{a=1}^N \vec k_a \cdot
{\partial \over \partial \vec k_a} \right)
\langle \zeta_{\vec k_1} ... \zeta_{\vec k_N} \rangle^{c'} \, ,
\end{eqnarray}
where $\zeta$ is the curvature perturbation. It arises from a spatial
dilation symmetry, which is non-linearly realized on $\zeta$.
Recently, more non-linearly realized symmetries were uncovered,
including the special conformal symmetry 
\cite{Creminelli:2012ed,Hinterbichler:2012nm}
and in fact a whole
infinite tower of symmetries \cite{Hinterbichler:2013dpa} (H2K hereafter).
Recent work has emphasized the non-perturbative nature of these
consistency relations as Ward/Slavnov-Taylor identities
\cite{Assassi:2012zq,Assassi:2012et,Kehagias:2012pd,Goldberger:2013rsa,Hinterbichler:2013dpa,Pimentel:2013gza,Berezhiani:2013ewa}. They can be viewed as 
the cosmological analog of the classic soft-pion theorems,
which relate a scattering amplitude with N particles to another
one with the same set of particles plus a soft Nambu-Goldstone boson. 
In the case of the strong interactions, the symmetry in question is chiral
symmetry, which is spontaneously broken. The corresponding
Nambu-Goldstone boson is the pion, which shifts nonlinearly under
chiral symmetry. In the examples above, $\zeta$ and the velocity play the 
role of the pion -- both shift nonlinearly under the respective
symmetries.

The non-perturbative nature of the consistency relations, such as
Eq.\ (\ref{KRPPrelationIntro}), makes them very interesting for LSS studies.
The hard modes, with momenta labeled by $\vec k$, can be highly
nonlinear and can even be astrophysically messy observables, such as
galaxy density as opposed to mass density. The robustness of these
consistency relations is not completely surprising:\ we are
familiar with the statement that correlation functions
should be translationally invariant (Eq.\ \ref{DeltaCorr}), 
even if they involve fluctuations
that are highly nonlinear and/or astrophysically complex. 
The robustness stems from the fact that these statements,
whether the analogs of Eq.\ (\ref{DeltaCorr}) or
(\ref{KRPPrelationIntro}), are derived from
symmetries, 
linearly or nonlinearly realized.
Nonetheless, the consistency relations from nonlinearly realized
symmetries are less familiar in the context of LSS, and deserve
further investigation.

First, we wish to probe the consistency
relations by studying them in the context of concrete models of
galaxy dynamics and biasing. An often instructive way to understand
supposedly robust relations is to ask when they actually break down.
We will delineate the assumptions underlying the
consistency relations and study examples where they are violated.
We will see that the consistency relations stand on three
legs:\ the existence of nonlinearly realized symmetries, the single field
initial condition, and the adiabatic mode condition.
We emphasize especially the role of the last in
the breakdown of consistency relations.
Using the language of galaxy-biasing, we can phrase
a general violation as the sign of a particular kind of non-local
galaxy bias.
As a by-product of our analysis, we also write down 
a Goldstone boson effective Lagrangian for a pressureless 
fluid coupled to gravity, similar to effective actions for quintessence
\cite{Boubekeur:2008kn, Creminelli:2008wc}.

Second, we wish to establish the connection between the 
Newtonian consistency relation found by KRPP
(Eq.\ \ref{KRPPrelationIntro}),
and the relativistic consistency relations in the context of
inflation. How are they related? Given the infinite number
of consistency relations known for inflation, could there be more such
relations for the context of large scale structure?
The inflationary consistency relations are usually phrased
in unitary or $\zeta$ gauge, while the Newtonian gauge is
more natural for LSS. We will show, not surprisingly, that there is a
one-to-one correspondence between symmetries in the two gauges.
The time-dependent spatial translation of KRPP
(Eq.\ \ref{timedepTranslate}) turns out to correspond to the
special conformal symmetry in $\zeta$ gauge.
The full list of symmetries in the Newtonian gauge can
be inferred from the known list in $\zeta$ gauge.
The KRPP consistency relation is the only one with a non-trivial
Newtonian/sub-Hubble limit, as we will show.

While this paper was in preparation, several papers appeared on the archive
which treated related issues. The issue of multiple species and
velocity bias is taken up in \cite{Peloso:2013spa}, the role of
non-local galaxy bias is discussed in \cite{Kehagias:2013rpa}, and
the importance of the equivalence principle is emphasized in
\cite{Creminelli:2013mca,Kehagias:2013rpa,Creminelli:2013nua}. 
Redshift-space and non-perturbative issues are discussed in
\cite{Creminelli:2013poa}. 
A different kind of approximate consistency relation, one
not based on symmetries in the traditional sense, is discussed in
\cite{Valageas:2013zda,Kehagias:2013paa}. 
Our overlap with \cite{Creminelli:2013mca} is the greatest, in that we both
view the consistency relations through the prism of diffeomorphism
invariance. We also overlap with \cite{Valageas:2013cma}
in the attempt to spell out the assumptions underlying the consistency relations.
We are especially indebted to Filippo Vernizzi for initial collaboration
and many illuminating discussions.

The paper is organized as follows. 
Section \S \ref{newtonian} treats the Newtonian system while \S \ref{diffGR} treats the full general
relativistic problem. In \S \ref{KRPPreview}, we review the symmetry
discovered by KRPP and the derivation of the corresponding
consistency relation, paying special attention to the underlying
assumptions. In \S \ref{robustness}, we discuss the
robustness and limitations of the consistency relations, and delineate
the underlying assumptions concerning the galaxy dynamics and biasing.
We mention in \S \ref{fluidLagLSS} a simple Lagrangian that reproduces
the Newtonian LSS equations for dark matter, 
though this Lagrangian is not needed for the analyses in this paper.
In \S \ref{diffGR}, we place the KRPP consistency relation in a larger context:\
all consistency relations result from residual coordinate transformations 
(diffeomorphisms) allowed within
a given gauge. We establish the connection between such symmetries
in the $\zeta$ gauge and the Newtonian gauge. The KRPP symmetry
is the sub-Hubble limit of one (or more precisely, two) of these symmetries.
We conclude in \S \ref{discuss}. Readers not interested in the
detailed arguments can find here a summary of our main results.
A few technical results are collected
in the Appendices.

A few words are in order on our notation and terminology. We use the symbol $\pi$
to represent the Nambu-Goldstone boson of a 
non-linearly realized symmetry (the pion), in accordance with 
standard practice. For our LSS application, $\pi$ is the velocity potential.
The same symbol is also used to denote the numerical
value $3.14...$. Which is meant should be obvious from the context. Essentially,
the numerical $\pi$ always precedes the Newtonian constant $G$ in the
combination $4\pi G$. In cases where both could potentially appear, we
use $M_P^{2} \equiv 1/(8\pi G)$ to avoid confusion.
Also, we use the term {\it nonlinear} to refer to quantities that are
not linear in the LSS observables (fields such as density or velocity). 
Sometimes, this has the usual meaning that such quantities go like the
fields raised to higher powers:\ quadratic and so on.
But, sometimes, this means the quantities of interest do not
depend on the field variables at all, such as the nonlinear part of
certain symmetry transformations e.g.\ Eq.\ (\ref{DeltaNLv}).
We rely on the context to differentiate between the two.


\section{Consistency Relation from Time-dependent Translation -- a
  Newtonian Symmetry}
\label{newtonian}

In this section, we focus on the Newtonian symmetry uncovered by KRPP.
\S \ref{KRPPreview} is a review of the symmetry and its implied consistency
relation. In \S \ref{robustness} we discuss the robustness and
limitations of the consistency relation, and go over what assumptions can or cannot
be relaxed, especially concerning the nonlinear, astrophysically messy,
galaxy observables. We discuss what kind of 
galaxy dynamics, and what sort of galaxy selection, could lead to
violations of the consistency relation. 
As a by-product of our investigation, we describe a simple Lagrangian
for LSS in \S \ref{fluidLagLSS}.

\subsection{Time-dependent Translation Symmetry and the Background Wave Argument -- a Review}
\label{KRPPreview}

We begin with a review of the Newtonian symmetry discovered by KRPP.
We go over the background wave derivation of the consistency relation
in some detail, emphasizing the underlying assumptions, and making the
derivation easily generalizable to the general relativistic case.
Two fundamental concepts are:\ (1) the existence of a non-linearly realized symmetry (one that shifts at least some
of the LSS observables by an amount that is independent of the
observables), and (2) an adiabatic mode
condition, which is an additional condition that dictates the time-dependence of the symmetry.

\vspace{0.2cm}

\noindent {\bf Time-dependent Translation Symmetry.} 
The set of Newtonian equations of motion for LSS is:
\footnote{Throughout this paper, we will be cavalier about
the placement of indices for objects with Latin indices, e.g.
$v^i, x^i$ are the same as $v_i, x_i$.}
\begin{eqnarray}
\label{NewtonianLSS}
\delta' + \partial_i [(1 + \delta) v^i] = 0 \quad , \quad 
v^i {}' + v^j \partial_j v^i + {\cal H} v^i = - \partial_i \Phi \quad , \quad 
\nabla^2 \Phi = 4\pi G \bar\rho a^2 \delta \, ,
\end{eqnarray}
where $\delta$ is the mass overdensity, 
${}' \equiv \partial/\partial\eta$ denotes the derivative with respect to conformal time $\eta$, 
$\partial_i$ denotes the derivative with respect to the comoving coordinate $x^i$,
$v^i$ is the peculiar velocity $dx^i/d\eta$, $\Phi$ is the gravitational potential,
$G$ is Newton's constant, $\bar\rho$ is the mean mass density at the time of interest,
$a$ is the scale factor, and ${\cal H} \equiv a'/a$ is the comoving Hubble parameter.
The first equation expresses continuity or mass conservation. The second equation is the Euler equation
or momentum conservation for a pressureless fluid. The third equation is the Poisson equation.
Let us start with this basic set. We will later consider generalizations to include pressure,
relativistic corrections, and even complex galaxy formation processes.

KRPP pointed out that this system of equations admits the following symmetry:
\begin{eqnarray}
\label{KRPPsymm}
\eta \rightarrow \tilde\eta = \eta \,\,\, , \,\,\,
x^i \rightarrow \tilde x^i = x^i + n^i \,\,\, , \,\,\,
v^i \rightarrow \tilde v^i = v^i + n^i {}'  \,\,\, , \,\,\,
\Phi \rightarrow \tilde \Phi = \Phi - ({\cal H} n^i {}' + n^i {}'') x^i \,\,\, , \,\,\,
\delta \rightarrow \tilde \delta = \delta\, ,
\end{eqnarray}
where $n^i$ is a function of time but not space. It can be shown that under
this set of transformations, Eq.\ (\ref{NewtonianLSS}) takes on exactly the same
form, with all the variables replaced by ones with a $\, \tilde{}\, $ on top.
To see that this is true, it is important to keep in mind:
\begin{eqnarray}
{\partial \over \partial \tilde\eta} = - n^i {}' {\partial \over \partial x^i} + {\partial \over \partial \eta}
\, ,
\end{eqnarray}
where on the left, $\tilde x^i$ is held fixed, and on the right,
$\partial/\partial x^i$ is at a fixed $\eta$, and
$\partial/\partial \eta$ is at a fixed $x^i$.
On the other hand $\partial/\partial \tilde x^i = \partial/\partial x^i$. 
The symmetry transformation described by Eq.\ (\ref{KRPPsymm}) is a 
time-dependent spatial translation. (Henceforth, we will occasionally
refer to this somewhat sloppily as simply translation.)
Under this translation,
the velocity gets shifted in the expected manner.
The gravitational
potential needs to be shifted correspondingly to preserve the form
of the Euler equation. The density $\delta$, on the other hand, does not
change at all, in the sense that $\tilde\delta(\tilde {x}) =
\delta({x})$. 

Eq.\ (\ref{KRPPsymm}) is a symmetry regardless of the time-dependence
of $n^i$. 
For the purpose of deriving the consistency relations, however, we need to impose
an additional condition. Suppose we start with $v^i = 0$
in Eq. (\ref{KRPPsymm});\ we would like the velocity generated by the
transformation, i.e.\ $\tilde v^i = n^i {}'$,
to be the long wavelength limit of an actual physical mode that satisfies the
equations of motion. We say long wavelength because $n^i {}'$ has no spatial
dependence and so is strictly speaking a $q=0$ mode ($q$ being the
wavenumber/momentum in Fourier space). What we want to impose is
this:\ if we take a physical velocity mode
at a finite $q$, and make its $q$ smaller and smaller, we would like
$n^i {}'$ to match its time-dependence.
Following terminology used in general relativity, we refer to it as the
{\it adiabatic mode condition} \cite{Weinberg:2003sw}. 
This condition ensures the symmetry transformation
generates a velocity mode that evolves in a physical way. 
It is easy to see that at long wavelength, where the equations can be linearized,
Eq. (\ref{NewtonianLSS}) can be combined into
a single equation:
\begin{eqnarray}
\label{divV}
\partial_i v^i {}' + {\cal H} \partial_i v^i = 4\pi G \bar\rho a^2 \partial_i \int d\eta \, v^i \, ,
\end{eqnarray}
or written in a more familiar form:
\begin{eqnarray}
\label{linearizeDelta}
\delta'' + {\cal H} \delta' = 4\pi G \bar\rho a^2 \delta \quad {\rm with} \quad \delta' = - \partial_i v^i
\, .
\end{eqnarray}
One might be tempted to say a velocity going like $n^i {}'$ satisfies
Eq.\ (\ref{divV}) trivially for an $n^i {}'$ of arbitrary
time-dependence,
since $n^i {}'$ has no spatial dependence. 
What we want, however, is for $n^i {}'$ to have the same time-dependence as that of
a velocity mode at a low, but finite momentum. 
In other words, we impose the
{\it adiabatic mode condition}:
\begin{eqnarray}
\label{nadiabatic}
n^i {}'' + {\cal H} n^i {}' = 4\pi G \bar\rho a^2 n^i
\end{eqnarray}
i.e.\ $n^i (\eta)$ has the same time-dependence as
the linear growth factor $D(\eta)$, assuming growing mode initial
conditions.
Effectively, we demand that our symmetry-generated velocity-shift
(or more precisely, the nonlinear part thereof)
satisfy Eq.\ (\ref{divV}) with the
spatial gradient removed.

\vspace{0.2cm}

\noindent {\bf The Background Wave Argument.}
Next, we give the background wave derivation
of the consistency relation. More sophisticated and rigorous
derivations exist \cite{Assassi:2012zq,Hinterbichler:2013dpa,Goldberger:2013rsa,Pimentel:2013gza,Berezhiani:2013ewa}, but the background wave argument has the virtue of
being fairly intuitive. Our goal here is to go over the
underlying assumptions, and formulate the argument in such a way
to ease later generalizations. The form of our expressions follow
closely those in H2K \cite{Hinterbichler:2013dpa}.

Before we carry out the argument,
it is convenient (especially for later discussions) to introduce the velocity potential $\pi$,
assuming potential flow on large scales,\footnote{
\label{potentialflow}
Assuming potential flow is not strictly necessary, as the argument can
be made using the velocity $v^i$ itself in place of $\partial_i \pi$. 
The reason we make this assumption is that the velocity enters
into our derivation mainly as a large scale or low momentum mode.
Assuming the growing mode initial condition, the large scale
velocity does take the form of a potential flow.
Indeed, vorticity remains zero until orbit crossing.
We do {\it not} assume potential flow on small scales.
}
in which case, the symmetry transformation
of Eq.\ (\ref{KRPPsymm}) tells us:
\begin{eqnarray}
\pi \rightarrow \tilde \pi = \pi + n^i {}' x^i \quad {\rm with} \quad
v^i = \partial_i \pi \, .
\end{eqnarray}
The velocity potential $\pi$ has the hallmark of a pion or 
Nambu-Goldstone boson:\ it experiences
a {\it nonlinear} shift under the symmetry transformation.\footnote{Note that $\Phi$ also experiences a nonlinear shift
(Eq.\ \ref{KRPPsymm}), and thus can also be used as the pion
in this derivation. The two give the same result, see \S \ref{robustnessGR}.
}
Note that $\pi$ also implicitly has a {\it linear} shift: 
\begin{eqnarray}
\tilde \pi (\tilde x) = \pi (x) + n^i {}' x^i \sim \pi (\tilde x) -
n^i \partial_i \pi + n^i {}' x^i \, 
\end{eqnarray}
where we have Taylor expanded $\pi(x)$ to first order in $\tilde x^i - x^i$, assuming
a small $n^i$. 
Here, linear and nonlinear refer to whether or not the transformation is linear in $\pi$ (or 
any other LSS fields).
The total shift in $\pi$, i.e.\ $\tilde \pi (\tilde x) - \pi (\tilde x)$, thus has both linear and nonlinear
pieces:
\begin{eqnarray}
\label{pishift}
\Delta_{\rm lin.} \pi = - n^i \partial_i \pi \quad , \quad
\Delta_{\rm nl.} \pi = n^i {}' x^i \, .
\end{eqnarray}
Similarly, the overdensity changes by the amount
$\tilde\delta(\tilde x) - \delta(\tilde x)$:
\begin{eqnarray}
\label{deltashiftN}
\Delta_{\rm lin.} \delta = - n^i \partial_i \delta \quad , \quad \Delta_{\rm nl.} \delta = 0 \, ,
\end{eqnarray}
i.e.\ $\delta$ experiences no nonlinear shift.

Consider an N-point function involving a product of N LSS observables.
Let us denote each as ${\cal O}$, labeled by momentum, so that the N-point function is
$\langle {\cal O}_{\vec k_1} {\cal O}_{\vec k_2} ... {\cal O}_{\vec k_N} \rangle$. 
Here, the
${\cal O}$'s at different momenta need not be the same observable. For instance,
one can be $\delta$, the other can be the gravitational potential, et cetera. 
They need not even be evaluated at the same time.
We are interested in this N-point function in the presence of 
some long wavelength (soft) $\pi$. Let us imagine splitting all
fluctuations into hard and soft modes, with ${\vec k_1}, ..., {\vec
  k_N}$ falling into the hard category. The N-point function obtained
by integrating over the hard modes, but leaving the soft modes of
$\pi$ unintegrated, can be Taylor expanded as:
\begin{eqnarray}
\langle  {\cal O}_{\vec k_1}  ... {\cal O}_{\vec k_N}
\rangle_{\pi_{\rm soft}}
\approx \langle  {\cal O}_{\vec k_1} ... {\cal O}_{\vec k_N} \rangle_{0}
+ \int \frac{d^3 \vec{p}}{(2\pi)^3}{\delta \langle  {\cal O}_{\vec
    k_1}  ... {\cal O}_{\vec k_N} \rangle_{\pi_{\rm soft}}
\over \delta \pi_{\vec p}^*}
\Big|_{0} \pi_{\vec p}^*\, ,
\end{eqnarray}
where we have taken the functional derivative with respect to,
and summed
over the Fourier modes of, $\pi$ with {\it soft} momenta $\vec p$.
Multiplying both sides by $\pi_{\vec q}$ (where $\vec q$ is also soft)
and ensemble averaging over the soft modes, one finds
\begin{eqnarray}
\label{ratioO}
\frac{\langle \pi_{\vec{q}} {\cal O}_{\vec k_1}  ... {\cal O}_{\vec k_N} \rangle}{P_{\pi}(q)}
=  \frac{\delta \langle  {\cal O}_{\vec k_1}  ... {\cal O}_{\vec k_N}
  \rangle_{\pi_{\rm soft}}}{\delta \pi_{\vec q}^*} \Big|_{0} \, .
\end{eqnarray}
We have used the definition of the power spectrum:
$\langle \pi_{\vec{q}} \pi_{\vec p}^* \rangle = (2\pi)^3 \delta_D (\vec{q} - \vec p) P_\pi (q)$, with
$\delta_D$ being the Dirac delta function.  
We can on the other hand compute the derivative on the right hand side
this way:
\begin{eqnarray}
\label{derivO}
\int \frac{d^3 \vec{p}}{(2\pi)^3}{\delta \langle  {\cal O}_{\vec k_1}
  ... {\cal O}_{\vec k_N} \rangle_{\pi_{\rm soft}}
\over \delta \pi_{\vec p}^*} \Big|_{0} \Delta_{\rm nl.} \pi_{\vec p}^*
= \Delta_{\rm lin.} \langle  {\cal O}_{\vec k_1}  ... {\cal O}_{\vec k_N} \rangle
\, .
\end{eqnarray}
This statement says that the change to 
the N-point function induced by the symmetry transformation (the right hand side)
is equivalent to the change to the N-point function by adding a long-wavelength
background $\pi$ induced by the same symmetry (the left hand side). 
We will unpack it a bit more in \S \ref{robustness}.
A careful reader might note that there is no reason why one
should include on the right hand side
only the {\it linear} part of the transformation
of the N-point function. That is true:\ by including only
the linear transformation, we are effectively dealing with
the {\it connected} N-point function. For a proof, see H2K.
\footnote{The restriction to the {\it connected} N-point function
will not be so relevant for the Newtonian (KRPP) consistency relation or the
general relativistic dilation consistency relation,
i.e.\ they take the same form whether the consistency relation is phrased
in terms of {\it connected} or {\it general}
correlation functions. The restriction is relevant for those
consistency relations that involve more than one derivative
on the right hand side. For them, it is important to keep in mind that the
N momenta on the right hand side sum to zero
(see \cite{Goldberger:2013rsa});
deriving {\it general} consistency relations from
{\it connected} ones necessarily introduce extra terms.
}
Combining Eqs.\ (\ref{ratioO}) and (\ref{derivO}), and adding
the superscript $c$ for connected N-point function, we have:
\begin{eqnarray}
\label{backWave}
\boxed{
\int {d^3 q\over (2\pi)^3} \, {\langle \pi_{\vec q} {\cal O}_{\vec k_1}  ... {\cal O}_{\vec k_N} \rangle^c
\over P_\pi (q)} \Delta_{\rm nl.} \pi_{\vec q}^*
= \Delta_{\rm lin.} \langle  {\cal O}_{\vec k_1}  ... {\cal O}_{\vec k_N} \rangle^c
\, .}
\end{eqnarray}
It is important to note that
the connected correlation functions on both
sides contain delta functions. 
This way of writing the consistency relation follows the Ward identity treatment
of H2K, and is applicable to any symmetries with $\pi$ as the Nambu-Goldstone boson.
The background wave argument has the advantage of being intuitive,
but is a bit heuristic. Readers interested in subtleties
can consult e.g.\ H2K. Our final result here matches theirs.

Let us apply Eq.\ (\ref{backWave}) to the translation symmetry.
To be specific, let us take our observable ${\cal O}$ to be the mass overdensity $\delta$.
We use the following convention for the Fourier transform of some function $f$:
\begin{eqnarray}
f(\vec q) = \int d^3 x f(\vec x) e^{i \vec q \cdot \vec x} \, , \, f(\vec x) = \int \frac{d^3 q}{(2 \pi)^3} f(\vec q) e^{-i \vec q \cdot \vec x} \, .
\end{eqnarray}
Eqs. (\ref{pishift}) and (\ref{deltashiftN}) thus implies:
\begin{eqnarray}
\label{pideltashiftFourier}
\Delta_{\rm nl.} \pi_{\vec q}^* =  i \, n^j {}' {\partial\over \partial q^j} [(2\pi)^3 \delta_D (\vec q)]
\quad , \quad
\Delta_{\rm lin.} \delta_{\vec k} = i n^j k^j \delta_{\vec k} \, .
\end{eqnarray}
Substituting into Eq. (\ref{backWave}), we find
\begin{eqnarray}
 \lim_{\vec q \rightarrow 0} \, n^j {}' (\eta)
{\partial \over \partial q^j}
\left[ {\langle \pi_{\vec q} \delta_{\vec k_1}  ... \delta_{\vec k_N} \rangle^c
\over P_\pi (q)} \right] = -
\sum_{a=1}^N n^j (\eta_a) k_a^j \, \langle \delta_{\vec k_1} ... \delta_{\vec k_N} \rangle^c \, ,
\end{eqnarray}
where $\eta$ is the time implicitly assumed
for $\pi_{\vec q}$, and $\eta_a$ is the time for $\delta_{\vec k_a}$. 
Note that $k_a^j$ refers to the $j$-component of the vector $\vec k_a$.
Since $n^j(\eta)$ has the time-dependence of the linear growth factor
$D(\eta)$ (from the adiabatic mode condition), but can otherwise point
in an arbitrary direction, we conclude:
\begin{eqnarray}
\label{KRPPrelation0}
 \lim_{\vec q \rightarrow 0} \, 
{\partial \over \partial q^j}
\left[ {\langle \pi_{\vec q} \delta_{\vec k_1}  ... \delta_{\vec k_N} \rangle^c
\over P_\pi (q)} \right] = -
\sum_{a=1}^N {D(\eta_a) \over D'(\eta)} k_a^j \, \langle \delta_{\vec k_1} ... \delta_{\vec k_N} \rangle^c \, .
\end{eqnarray}
We have yet to remove the delta functions from both sides.
To do so, we use the following, pure shift, symmetry:
\begin{eqnarray}
\pi \rightarrow \tilde\pi = \pi + {\,\rm const.} \, ,
\end{eqnarray}
This symmetry does not involve transforming space-time at all,
and so none of the observables receive a {\it linear} shift.
The argument leading to Eq.\ (\ref{backWave}) thus tells us
\begin{eqnarray}
\label{KRPPrelation00}
 \lim_{\vec q \rightarrow 0} \, {\langle \pi_{\vec q} \delta_{\vec k_1}  ... \delta_{\vec k_N} \rangle^{c'}
\over P_\pi (q)} = 0 \, .
\end{eqnarray}
We use the superscript $c'$ to denote the connected correlation function with the overall delta function removed:
\begin{eqnarray}
\langle {\cal O}_{\vec k_1}  ... {\cal O}_{\vec k_N} \rangle^c =
(2\pi)^3 \delta_D (\vec k_1 + ... + \vec k_N)
\langle {\cal O}_{\vec k_1}  ... {\cal O}_{\vec k_N} \rangle^{c'} 
\end{eqnarray}
Combining Eqs.\ (\ref{KRPPrelation0}) and (\ref{KRPPrelation00}), we
have the Newtonian {\it translation consistency relation}:
\begin{eqnarray}
\label{KRPPrelation1}
\boxed{\lim_{\vec q \rightarrow 0} \, 
{\partial \over \partial q^j}
\left[ {\langle \pi_{\vec q} \delta_{\vec k_1}  ... \delta_{\vec k_N} \rangle^{c'}
\over P_\pi (q)} \right] = -
\sum_{a=1}^N {D(\eta_a) \over D'(\eta)} k_a^j \, \langle
\delta_{\vec k_1} ... \delta_{\vec k_N} \rangle^{c'} \, ,}
\end{eqnarray}
where $\eta$ is the time for the soft mode
$\pi_{\vec q}$, and each $\eta_a$ is the time for the corresponding
hard mode $\delta_{\vec k_a}$. 
This turns out to be a common feature for all consistency relations
as we will see:\ Eqs.\ (\ref{KRPPrelation00}) and (\ref{KRPPrelation0})
are two consistency relations that differ by one derivative with
respect to $q$;\ the former allows us to remove the delta function from the latter in a
straightforward way --
the result is Eq.\ (\ref{KRPPrelation1}).
The form adopted by KRPP is to integrate the above over $q$, and using
Eq.\ (\ref{KRPPrelation00}), to obtain:\footnote{
We are grateful to Lasha Berezhiani and Justin Khoury for helping us
understand this point.
}
\begin{eqnarray}
\label{KRPPrelation2}
\lim_{\vec q \rightarrow 0} \, 
\left[ {\langle \pi_{\vec q} \delta_{\vec k_1}  ... \delta_{\vec k_N} \rangle^{c'}
\over P_\pi (q)} \right] = -
\sum_{a=1}^N {D(\eta_a) \over D'(\eta)} \vec q \cdot \vec k_a \, \langle \delta_{\vec k_1} ... \delta_{\vec k_N} \rangle^{c'} \, .
\end{eqnarray}

The soft mode $\pi_{\vec q}$, by the linearized continuity equation,
is related to $\delta_{\vec q}$ by\footnote{In the context of a consistency relation which
is purported to be non-perturbative, one might wonder
if using the linear relation between $\delta$ and $\pi$ 
(for the soft mode only) is justified. It can be shown that including
nonlinear corrections to this relation leads to terms 
subdominant in the squeezed limit of the correlation function.
See \cite{Creminelli:2013poa}.
}
\begin{eqnarray}
\delta_{\vec q} (\eta) = q^2 {D(\eta)\over D'(\eta)} \pi_{\vec q}
(\eta) \, .
\end{eqnarray}
One can therefore rewrite Eq.\ (\ref{KRPPrelation2}) as\footnote{
The use of the equation of motion within the {\it connected} correlation
function does not lead to contact terms.
See e.g. H2K and \cite{Goldberger:2013rsa} on the role of contact terms in Ward identity arguments.
}
\begin{eqnarray}
\label{KRPPrelation3}
\lim_{\vec q \rightarrow 0} \, 
\left[ {\langle \delta_{\vec q} \delta_{\vec k_1}  ... \delta_{\vec k_N} \rangle^{c'}
\over P_\delta (q)} \right] = -
\sum_{a=1}^N {D(\eta_a) \over D(\eta)} {\vec q \cdot \vec k_a \over q^2}\, \langle \delta_{\vec k_1} ... \delta_{\vec k_N} \rangle^{c'} \, .
\end{eqnarray}
In the context of this Newtonian derivation, we would
like to think of the form expressed in
Eq.\ (\ref{KRPPrelation1}) and Eq.\ (\ref{KRPPrelation00}) as
more fundamental, since it is $\pi$ that experiences
a nonlinear shift in the symmetry transformation, acting
as the pion, and since the expression in in Eq.\ (\ref{KRPPrelation3}) contains two non-relativistic consistency relations, one trivial and one nontrivial.
It is also worth stressing that Eq.\ (\ref{KRPPrelation1}) does not
constrain
$\partial^2/\partial q^2$ of $[\langle \pi_{\vec q} \delta_{\vec k_1} ... \delta_{\vec k_N}
\rangle^{c'} / P_\pi (q)]$ in the soft limit, i.e.\
Eq.\ (\ref{KRPPrelation2}) can in principle contain
$O(q^2)$ corrections, and Eq.\ (\ref{KRPPrelation3})
can contain $O(q^0)$ corrections.

Let us close this derivation by observing that the only assumption
made about the hard modes is how they transform under
the symmetry (in particular, the linear part of their transformation;\
see Eqs.\ \ref{deltashiftN}, \ref{pideltashiftFourier}).
Thus, suppose we have some observable ${\cal O}$
whose linear transformation under the spatial translation is:
\begin{eqnarray}
\label{DeltaOshift}
\Delta_{\rm lin.}{\cal O}_{\vec k} =  i n^j g^j {\cal O}_{\vec k} \, .
\end{eqnarray}
For instance, if ${\cal O}$ is the galaxy overdensity, we expect $g^j
= k^j$, but $g^j$ could take other forms for other observables.
Exactly the same derivation then gives
the Newtonian translation consistency relation in a more general form:
\begin{eqnarray}
\label{KRPPrelation4}
\boxed{\lim_{\vec q \rightarrow 0} \, 
{\partial \over \partial q^j}
\left[ {\langle \pi_{\vec q} {\cal O}_{\vec k_1}  ... {\cal O}_{\vec k_N} \rangle^{c'}
\over P_\pi (q)} \right] = -
\sum_{a=1}^N {D(\eta_a) \over D'(\eta)} g_a^j \, \langle
{\cal O}_{\vec k_1} ... {\cal O}_{\vec k_N} \rangle^{c'} \, ,}
\end{eqnarray}
where we have allowed the possibility that the N hard modes correspond to
different observables, thus potentially a different $g_a^j$ for
each $a = 1,
..., N$. 
Corollaries -- analogs of Eqs.\ \ref{KRPPrelation2},
\ref{KRPPrelation3} -- follow in the same way:
\begin{eqnarray}
\label{KRPPrelation5}
\lim_{\vec q \rightarrow 0} \, 
\left[ {\langle \pi_{\vec q} {\cal O}_{\vec k_1}  ... {\cal O}_{\vec k_N} \rangle^{c'}
\over P_\pi (q)} \right] = -
\sum_{a=1}^N {D(\eta_a) \over D'(\eta)} \vec q \cdot \vec g_a \, \langle
{\cal O}_{\vec k_1} ... {\cal O}_{\vec k_N} \rangle^{c'} \, ,
\end{eqnarray}
and
\begin{eqnarray}
\label{KRPPrelation6}
\lim_{\vec q \rightarrow 0} \, 
\left[ {\langle \delta_{\vec q} {\cal O}_{\vec k_1}  ... {\cal O}_{\vec k_N} \rangle^{c'}
\over P_\pi (q)} \right] = -
\sum_{a=1}^N {D(\eta_a) \over D(\eta)} {\vec q \cdot \vec g_a \over 
q^2} \, \langle
{\cal O}_{\vec k_1} ... {\cal O}_{\vec k_N} \rangle^{c'} \, ,
\end{eqnarray}
where it should be understood that Eq.\ (\ref{KRPPrelation5}) 
and Eq.\ (\ref{KRPPrelation6}) contain
$O(q^2)$ and $O(q^0)$ corrections respectively.
Phrased as such, the consistency relation is fairly robust:\ the detailed dynamics
of the hard modes has no relevance;\ it matters not whether the corresponding
observables are astrophysically messy or highly nonlinear.
All we need to know is how they transform under a spatial translation.
To understand this robustness better, it is helpful to study concrete
examples, which is the subject of the next section.

\subsection{Robustness and Limitations of the Consistency Relation(s)}
\label{robustness}

To understand better the robustness of the consistency relation, it is
instructive to ask the question: when does it fail?
As we will see, the consistency relation stands on three legs:\ 
the existence of the time-dependent translation symmetry,
the single-field initial condition, and the adiabatic mode condition.
All three are necessary in order for the consistency relation to hold.
Here, we focus on the KRPP consistency relation as a specific example, but
the points we raise are general, pertaining to other consistency
relations (\S \ref{diffGR}) as well.

\vspace{0.2cm}

\noindent {\bf 1.} {\bf Initial condition:\ the single-field
  assumption.} A crucial step in the derivation is
Eq.\ (\ref{derivO}):\ that linear transformations of a collection of
hard modes, represented by ${\cal O}_{\vec k}$, can be
considered equivalent to placing the same hard modes in
the presence of a soft mode -- the pion $\pi_{\vec q}$. 
That this single soft mode is sufficient to account for all
the transformations of the hard modes is an assumption about
initial conditions. In the context of inflation, the assumption is often
phrased as that of a single field or a single
clock. In our Newtonian LSS context, in addition to keeping only the growing modes, essentially the assumption
is that of Gaussian initial conditions.\footnote{Note that in the inflationary context, Gaussianity is also assumed -- for the wave-function in the far past.}
More precisely, one demands that
the initial condition does not contain a coupling between soft
and hard modes beyond that captured by Eq.\ (\ref{derivO}).
We will follow the inflation terminology and call this 
the {\it  single-field assumption}.

\vspace{0.2cm}

\noindent {\bf 2.} {\bf Adiabatic mode condition.} Another crucial
ingredient in 
the derivation is the
{\it adiabatic mode condition}, that the symmetry
transformation have the correct time-dependence so that
the nonlinear shift $\Delta_{\rm nl.} \pi$ may be the long
wavelength limit of an actual physical mode. 
We stress that this is an additional 
requirement on top of demanding a symmetry, with some non-trivial implications.
To spell them out, it is
useful to have a concrete example. Since the consistency relation
is purported to be robust, in the sense that the hard modes can
be highly nonlinear and even astrophysically complex, let us
write down a system of equations that allow for these complexities:
\begin{eqnarray}
\label{messysystem}
\delta_{(a)}' + \vec \nabla \cdot [(1 + \delta_{(a)}) \vec v_{(a)}]
= R_{(a)} \quad , \quad 
\vec v_{(a)} {}' + 
\vec v_{(a)} \cdot \vec \nabla \vec v_{(a)}+ {\cal H} \vec v_{(a)} = - \vec
\nabla \Phi + \vec F_{(a)} \, .
\end{eqnarray}
Here $a$ labels the species:\ for instance, it can be dark matter,
populations of galaxies, baryons and so on.
$R_{(a)}$ represents a source term for the density evolution.
For dark matter, we expect $R_{(a)} = 0$ (barring significant
annihilation or decay). For galaxies, $R_{(a)}$ quantifies the effect
of galaxy formation and mergers. 
All particles are subjected to the same gravitational force
plus a species-dependent force $\vec F_{(a)}$. 
The gravitational potential $\Phi$ is sourced by the total mass
fluctuation:
\begin{eqnarray}
\nabla^2 \Phi = 4 \pi G a^2 \bar\rho \delta_T \, ,
\end{eqnarray} 
where $\delta_T$ represents the effective total mass density
fluctuation from all particles.

A natural generalization of the (time-dependent) translational symmetry from the previous
section would be:
\begin{eqnarray}
\label{KRPPsymmGen}
&& \eta \rightarrow \tilde\eta = \eta \,\,\,  , \,\,\,
x^i \rightarrow \tilde x {}^i = x^i + n^i \,\,\, , \,\,\,
v_{(a)} {}^i \rightarrow \tilde v_{(a)} {}^i = v_{(a)} {}^i + n^i {}'  \,\,\, , \,\,\,
\Phi \rightarrow \tilde \Phi = \Phi - ({\cal H} n^i {}' + n^i {}'') x^i
\,\,\, , \,\,\,
\nonumber \\
&& \delta_{(a)} \rightarrow \tilde \delta_{(a)} = \delta_{(a)}\,\,\, , \,\,\,
\delta_T \rightarrow \tilde \delta_T = \delta_T \,\,\, , 
\,\,\,
R_{(a)} \rightarrow \tilde R_{(a)} = R_{(a)} \,\,\, , \,\,\,
F_{(a)} {}^i \rightarrow \tilde F_{(a)} {}^i = F_{(a)} {}^i
\end{eqnarray}
This means that $R_{(a)}$ and $\vec F_{(a)}$ remain invariant under this
symmetry. For instance, they are invariant
if $R$ and $\vec F$ 
depend only on $\delta$, on the spatial gradient of
$\vec v$, or on the second gradients of
$\Phi$. $R$ and $\vec F$ could even have explicit
dependence on time $\eta$. What would violate invariance is
if $R$ or $\vec F$ depends on $\vec v$ with
no gradients -- that is, unless the dependence is of the form
$\vec v_{(a)} - \vec v_{(b)}$ (in which case the shifts in the
velocities of the two different species $(a)$ and $(b)$
cancel out). Specializing to the case of galaxies, what this
means is that the number density evolution
and dynamics of the galaxies does not care about the
absolute size of the velocity, but only about the velocity difference
(either between neighbors, or between species). The only
context in which the absolute size of velocity plays a role is 
through Hubble friction -- this is the origin of the ${\cal H}$ 
dependent term in the nonlinear shift of $\Phi$.
In other words, Hubble friction aside, {\it galaxy formation and dynamics
is frame invariant}, which seems a fairly safe assumption.
For instance, dynamical friction or ram pressure, which no doubt
exerts an influence on galaxies, should depend on velocity difference.

Thus, let us assume Eq.\ (\ref{KRPPsymmGen}) is a symmetry of our
system -- note that this is a symmetry regardless of the
time-dependence of $n^i$.
As emphasized earlier,
this is not enough to guarantee the validity of  the consistency relation.
To derive the consistency relation, $n^i$ must have the
correct time-dependence: 
$n^i {}'$ (the nonlinear shift in $v^i$)
must match the time-dependence of a physical long-wavelength
velocity perturbation.\footnote{
This typically means $n^i$ should satisfy Eq.\ (\ref{nadiabatic}), i.e.\
$n^i$ must evolve like the linear growth factor $D$. 
This holds if all particles, on large scales, evolve like dark matter
and if gravity is the only long-range force. See further discussion below.
}
This has to hold for all species, meaning
the {\it same} $n^i {}'$ matches the long-wavelength velocity
perturbation of each and everyone of the species.
In other words, {\it all species should move with the same velocity on
large scales.}
This leads to two subtleties, which are best illustrated by
assuming an explicit form for $\vec F$. Consider:
\begin{eqnarray}
\label{Fa}
\vec F_{(a)} = - c_s {}_{(a)}^2 \vec \nabla 
{\,\rm ln}(1 + \delta_{(a)}) - \beta_{(a,b)} (\vec v_{(a)}- \vec v_{(b)})
- \alpha_{(a)} \vec \nabla \varphi 
\quad , \quad \nabla^2 \varphi = 8\pi G
\sum_a \alpha_{(a)} a^2 \bar\rho_a \delta_a\, .
\end{eqnarray}
The first term on the right of the expression
for $\vec F$ represents some sort of pressure -- 
$c_s$ is the sound speed -- this would be relevant
if the subscript $(a)$ represents baryons at finite temperature.
The second term represents some sort of friction that
depends on the velocity difference between two species, with
a coefficient $\beta$.
The third term represents an additional fifth force, mediated
by the scalar $\varphi$, with a coupling $\alpha$.
The scalar $\varphi$ obeys a Poisson-like equation.
In scalar-tensor theories, the tensor part of the theory mediates
a universal gravitational force (described by the gravitational
potential
$\Phi$), but the scalar need not be universally coupled:\ hence
we allow the coupling $\alpha_{(a)}$ to depend on the species
(see e.g.\ \cite{HNS,HN}).
The form for the additional force $\vec F$ proposed in Eq.\ (\ref{Fa})
is fairly generic:\ counting derivatives, we can see that the pressure
term goes like $\partial \delta$, whereas the other two
terms go like $\partial^{-1} \delta$. 
In terms of the symmetry transformation, one can see that
\begin{eqnarray}
\varphi \rightarrow \tilde\varphi = \varphi \, 
\end{eqnarray}
is compatible with Eq.\ (\ref{KRPPsymmGen}).
Thus, we have a system (Eq.\ \ref{messysystem}) that respects the translational
symmetry spelled out in Eq.\ (\ref{KRPPsymmGen}), even if
many different kinds of forces are
present, including non-gravitational or modified gravitational ones
such as in Eq.\ (\ref{Fa}).
We wish to see how, despite the presence of the (time-dependent) translational
symmetry, there can still be a breakdown of the consistency relation,
due to obstructions in satisfying the adiabatic mode condition  --
that the velocity perturbations of all species should be equal on large
scales.

\vspace{0.2cm}

\noindent {\bf 2a.} {\bf Soft dynamics constraint.} 
In the long wavelength limit, one can ignore the pressure term
compared to the other two terms in the expression for $\vec F_{(a)}$. 
Let us first focus on the fifth-force term. This term is at the same
level in derivative as the normal gravitational force
$-\vec\nabla\Phi$, and thus both have to be taken into account
on large scales. The problem with a long-range fifth-force is the non-universal
coupling:\ if there is a different coupling $\alpha_{(a)}$ for each
kind of particles, the different species will move with different
velocities even on large scales. This means no {\it single} $n^i {}'$
can possibly generate long-wavelength velocity perturbations for all
species. In other words, unless the soft (large-scale) dynamics obeys the equivalence
principle, the consistency relation would be violated, as emphasized
by \cite{Kehagias:2013rpa,Creminelli:2013nua}. 
We stress that, in our example, the violation of equivalence
principle occurs {\it without} the violation of the translation
symmetry described by Eq.\ (\ref{KRPPsymmGen}).
The fact that the consistency relation is not obeyed is entirely
because of the failure to satisfy the adiabatic mode condition
when the equivalence principle is violated.
The friction term (second term on the right of Eq.\ \ref{Fa}), on
the other hand, is compatible with the adiabatic mode condition --
it simply vanishes if the velocities of different species are equal, 
and is therefore consistent with the large scale requirement that all species
flow with the same velocity.
To sum up, the {\it soft dynamics constraint} is:\ for the consistency relation to be valid, 
the dynamics on large scales must be consistent with
all species moving with the same velocity.

\vspace{0.2cm}
\noindent {\bf 2b.} {\bf Squeezing constraint.} Let us next turn
to the pressure term (first term on the right of Eq.\ \ref{Fa}).
Since different species have different sound speeds, this also
leads to differences in velocity flows.
This is relatively harmless though, since the pressure term
becomes subdominant on large scales. Thus, there is no
problem with the adiabatic mode condition, which is really a condition
on motions in the soft limit $q \rightarrow 0$. 
The presence of pressure does lead to a practical
limitation on the application of the consistency relation, however.
The consistency relation is a statement
about an (N+1)-point function in the squeezed limit
$q \ll k_1, ..., k_N$.
There is the practical question of how small $q$ has to be.
An important requirement is:\ $q$ must be sufficiently small
such that the velocity perturbations of different species have the same time-dependence
as that generated by a single $n^i {}'$. In the present context, it means
$q < {\cal H}/{c_s}$, i.e.\ the length scale must be above the Jeans
scale.\footnotemark 
We refer to this as the {\it squeezing constraint}:\
the soft leg of the consistency relation must be sufficiently soft
that any difference in force on the different species becomes
negligible. This is worth emphasizing, because clearly dark matter
and baryons are subject to different forces:\ while that does not by itself
lead to the breakdown of the consistency relations, one has to
be careful to make sure that the squeezed correlation function is
sufficiently squeezed.

\vspace{0.2cm}

\noindent {\bf 3.} {\bf Galaxy-biasing.} It is also instructive to
approach the subject of consistency relation violation from the
viewpoint of galaxy-biasing. What kind of galaxy-biasing 
would lead to the violation of consistency relation?
We can only address this question in the perturbative regime, but
it nonetheless provides some useful insights. Suppose the galaxy overdensity
$\delta_{(a)}$ (of type $a$) and matter density $\delta$ are related by:
\begin{eqnarray}
\delta_{(a)} {}_{\vec k}= b_{(a)} \left(\delta_{\vec k} + \int 
{d^3 k' \over (2\pi)^3} {d^3 k'' \over (2\pi)^3} (2\pi)^3 \delta_D
(\vec k - \vec k' - \vec k'') W_{(a)} (\vec k', \vec k'') 
\left[ \delta_{\vec k'} \delta_{\vec k''} - \langle \delta_{\vec k'}
  \delta_{\vec k''} \rangle\right] \right) \, ,
\end{eqnarray}
where $b_{(a)}$ is a linear bias factor (independent of momentum)
and $W_{(a)}$ is 
a kernel that describes a general quadratic bias.
To the lowest order in perturbation theory, it can be shown that
the bispectrum between three types of galaxies $a$, $b$ and $c$,
at momenta $\vec q, \vec k_1, \vec k_2$ and times $\eta, \eta_1,
\eta_2$ respectively, is 
\begin{eqnarray}
\label{Babc}
&& B_{(abc)} (\vec q, \eta; \vec k_1, \eta_1; \vec k_2, \eta_2)
= 
\nonumber \\
&& \,\, 2 b_{(a)} b_{(b)} b_{(c)} P_\delta (k_1, \eta, \eta_1)
P_\delta (k_2, \eta, \eta_2)
\Big[ \left( {5\over 7} + {1\over 2} \hat k_1
    \cdot \hat k_2 \left( {k_1 \over k_2} + {k_2 \over k_1} \right) +
    {2\over 7} (\hat k_1 \cdot \hat k_2)^2 \right) 
+ W_{(a)} (- \vec k_1 , - \vec k_2) \Big] 
\nonumber \\
&& \,\, + 2 b_{(a)} b_{(b)} b_{(c)} P_\delta (q, \eta, \eta_1)
P_\delta (k_2, \eta_1, \eta_2)
\Big[ \left( {5\over 7} + {1\over 2} \hat q
    \cdot \hat k_2 \left( {q\over k_2} + {k_2 \over q} \right) +
    {2\over 7} (\hat q \cdot \hat k_2)^2 \right) 
+ W_{(b)} (- \vec q , - \vec k_2) \Big] 
\nonumber \\
&& \,\, + 2 b_{(a)} b_{(b)} b_{(c)} P_\delta (q, \eta, \eta_2)
P_\delta (k_1, \eta_1, \eta_2)
\Big[ \left( {5\over 7} + {1\over 2} \hat q
    \cdot \hat k_1 \left( {q\over k_1} + {k_1 \over q} \right) +
    {2\over 7} (\hat q \cdot \hat k_1)^2 \right) 
+ W_{(c)} (- \vec q , - \vec k_1) \Big] 
\end{eqnarray}
where $P_\delta$ is the linear mass power spectrum -- its
two time-arguments signify the fact that the
two $\delta$'s involved can be at different times. 
We are interested in the $\vec q \rightarrow 0$ limit:
\begin{eqnarray}
\label{Babcsqueezed}
&& \lim_{\vec q \rightarrow 0}
b_{(a)} {B_{(abc)} (\vec q, \eta; \vec k_1, \eta_1; \vec k_2, \eta_2)
\over P_{(aa)} (q, \eta, \eta)} \nonumber \\
&& = E + {D(\eta_1) \over D(\eta) } 
\left[ {{\vec q \cdot \vec k_2} \over q^2} + 2W_{(b)} (-\vec q, -\vec
  k_2) + O(q^0) \right] P_{(bc)} (k_2, \eta_1, \eta_2) \nonumber \\
&& \quad + {D(\eta_2) \over D(\eta) } 
\left[ {{\vec q \cdot \vec k_1} \over q^2} + 2W_{(c)} (-\vec q, -\vec
  k_1) + O(q^0) \right] P_{(bc)} (k_1, \eta_1, \eta_2) \nonumber \\
&& = E - \left[ {D(\eta_1) \over D(\eta) } 
\left( {{\vec q \cdot \vec k_1} \over q^2} - 2W_{(b)} (-\vec q, \vec
  k_1) \right)  + {D(\eta_2) \over D(\eta) } 
\left( {{\vec q \cdot \vec k_2} \over q^2} - 2W_{(c)} (-\vec q, \vec
  k_2) \right) + O(q^0) \right]
P_{(bc)} (k_1, \eta_1, \eta_2) \, , \nonumber \\
&& {\rm where} \quad E \equiv
2 {D(\eta)^2 \over D(\eta_1) D(\eta_2)}
{b_{(a)}^2 \over b_{(b)} b_{(c)}}
{P_{(bc)} (k_1, \eta_1, \eta_2) P_{(bc)}(k_2, \eta_1, \eta_2) \over
P_{(aa)} (q, \eta, \eta)} \left[ O(q^2) + W_{(a)}(- \vec k_1, - \vec
k_2) \right] \, .
\end{eqnarray}
We have used $P_{(aa)} (q, \eta, \eta) = (D(\eta)/D(\eta_1))
b_{(a)}^2 P_\delta (q, \eta, \eta_1)$, $P_{(bc)} (k_1, \eta_1, \eta_2) = b_{(b)} b_{(c)}
P_\delta (k_1, \eta_1, \eta_2)$ and so on (appropriate only
perturbatively). 
Comparing this expression with the consistency relation
expressed in Eq.\ (\ref{KRPPrelation6}) (identifying $\vec g_a$ with
$\vec k_a$), 
we see that the two
agree if $E$, $W_{(b)}$ and $W_{(c)}$ can be ignored, and $b_{(a)} =
1$. A number of comments are in order.\footnotetext{Since the Jeans scale changes with time in general,
the formal requirement is therefore that the soft-mode be longer
than the Jeans scale at all times:\ $q < (\mathcal{H}/{c_s})_{rec.}$ where the maximum size of the sound horizon may be conservatively estimated to be the size at recombination. In practice, since most of the present
day non-Gaussianity is generated at late times, it is typically sufficient to require that the contribution from early times be subdominant.  Using second order perturbation theory, we estimate that $q < {\mathcal{H}/{c_s}}_{rec.}(D(\eta_{obs.})/D(\eta_{rec.}))$ is a sufficient parametric condition for Eq.\ \ref{KRPPrelation1} to be valid (at least away from the equal time limit).

} 

First, let us focus on the case with no galaxy biasing, so that
Eq.\ (\ref{Babcsqueezed}) simply constitutes a perturbative check
of the consistency relation for the mass overdensity
i.e.\ Eq.\ (\ref{KRPPrelation3}).
We see that the term $E$ can be ignored compared
to the terms kept only if the soft power spectrum
is not too blue:
assuming $P_{(aa)}(q) \sim q^n$ for small $q$, 
the validity of the consistency relation requires $n < 3$. 
This is a limitation on the consistency relation that is not
often emphasized. In practice though, the realistic power spectrum has 
no problem satisfying this requirement.

Let us next consider the effects of galaxy biasing.
The second point we would like to raise is that
the soft mode must be kept unbiased. 
There are two reasons for this, one trivial, the other less so.
The trivial reason is that the left hand side of the consistency
relation has to be corrected by a factor of $b_{(a)}$, the linear
bias factor for the soft mode. This is not a big problem:\ one
can obtain an estimate of the linear bias and correct
the consistency relation when comparing against observations
of the galaxy bispectrum. The more non-trivial problem is
the presence of the quadratic bias kernel $W_{(a)}$ in $E$.
Consider for instance a local biasing model of the form:
$\delta_{(a)} = b_{(a)} \delta + b_{(a), 2} \delta^2/2$ in real space,
where $b_{(a)}$ and $b_{(a), 2}$ are constants, typically referred
as the linear and quadratic bias factors.
In this case $W_{(a)} = b_{(a), 2}/(2 b_{(a)})$ has no momentum
dependence,
and so $E$ contains a contribution that goes like $q^{-n}$ for
$P_{(aa)} \sim q^n$. This means one needs $n < 1$ for $E$ to be
negligible compared to the terms we keep in the consistency relation.
On the largest scales, $n$ approaches $1$, though observations suggest
it is slightly less than $1$. On smaller scales (but still keeping
$q \ll k_1, ..., k_N$), the relevant $n$ is on the safe side.
Nonetheless, this perturbative check suggests that one should
be careful in using the a biased observable for the soft-mode.

Henceforth, let us assume the soft-mode is unbiased but the hard modes
are biased, in which case $E$ is safely negligible in the squeezed
limit as long as $n < 3$. The third point we wish to raise is that
the validity of the consistency relation requires the hard modes
be biased in a way that is not too infrared-divergent: 
$W_{(b)} (-\vec q, \vec k_1)$ and $W_{(c)} (-\vec q, \vec k_2)$ cannot
contain terms that go like $k_1/q$ or $k_2/q$ i.e.
\begin{equation}
\label{Wcondition}
\boxed{
W_{(b)} (-\vec q, \vec k), W_{(c)} (-\vec q, \vec k) \, < \, {1/q} \, 
}
\end{equation}
in the $q \rightarrow 0$ limit.
As mentioned above,
the local biasing model typically assumed in LSS studies
implies the kernels $W_{(b)}$ and $W_{(c)}$ are momentum independent,
and is thus consistent with the consistency relation.
It is worth emphasizing that the word {\it local} in local biasing
is a bit misleading: it merely states that the galaxy density at a
given point in real space is related to the mass density at the same
point. In reality, galaxies form out of the collapse of larger
regions, influenced by the tidal field of the environment: there
are therefore good reasons to believe that galaxy biasing is at some
level {\it non-local}, i.e.\ the galaxy density at a given point is affected
by the mass density at other points. This non-locality is not
non-locality in the field theory sense, in that there is nothing non-local in
the dynamics, and the so-called non-local galaxy bias arises completely
out of local processes. 
A violation of the consistency relation requires more than a non-local
galaxy bias though. It requires the non-local biasing
kernel to be infrared divergent. This does not appear to be so easily
obtained in a random-walk halo-biasing model, even if tidal effects
are taken into account \cite{Sheth:2012fc}. One way it arises is in a model in which
galaxies are born with a velocity bias, as pointed out by \cite{Chan:2012jj}:\
the quadratic kernel $W_{(c)}$ (or $W_{(b)}$), for some galaxy population with a
velocity bias of $b_v^*$ at birth (i.e.\ the galaxy velocity equals
$b_v^*$ times the dark matter velocity when the galaxy forms):
\begin{eqnarray}
\label{Wvelocityb}
W_{(c)} (\vec q, \vec k) \sim 2 b_{(c)}^{-1} (b_v^* - 1) \left( {D_*
    \over D} \right)^{3/2} 
\, \hat q \cdot \hat k \left( {q \over k} + {k
    \over q} \right) \, ,
\end{eqnarray}
where $D_*$ is the linear growth factor at the time of birth, and $D$
is growth factor at the time of interest. We display only the
term that has a dipolar dependence on the angle between
$\vec q$ and $\vec k$, and have taken the late time limit.
This has precisely the kind of infrared divergence in the $q \ll k$
limit which would invalidate the consistency relation.
It is interesting that this is also an example where we should have
expected a violation of the consistency relation based on earlier
arguments -- the existence of a scale-independent
velocity bias $b_v^*$ means dark matter and
galaxies do not flow in the same way, even on large scales.
This violates the adiabatic mode condition, and so it is not
a surprise that the consistency relation fails. Realistically, velocity bias
is present at some level of course, but is expected to approach
unity on sufficiently large scales,
unless of course the equivalence principle is violated \cite{Creminelli:2013nua}. 
As a general statement, we can say that
a non-local galaxy bias that is more
infrared-divergent than Eq.\ (\ref{Wcondition}) is what one needs to
violate the consistency relation. It is interesting to
ask whether there are other ways to physically generate such a galaxy bias besides
through equivalence principle violations.
This naturally brings us to the issue of selection.

It is worth emphasizing that the galaxy bias is also partly a
selection bias: one chooses to study galaxies of a certain luminosity,
color, morphology or some other property of interest. The question is then:\ 
can one choose the galaxy
sample in such a way as to violate Eq.\ (\ref{Wcondition})? 
What if one chooses galaxies based on their motions, for instance,
selecting galaxies that have systematically large velocities? 
It would seem by hand we have introduced a velocity bias, and thus
a violation of the consistency relation. This is actually not a
violation in the technical sense. Choosing galaxies based on their
motions can be thought of as weighing the galaxies by velocities,
i.e.\ $\delta_g \rightarrow \delta_g (1 + \pi)$. From the point of
view of violating the consistency relation, it is most relevant to
consider weighing by the large scale velocity. In that case, it is not
surprising one finds additional terms that diverge in the
squeezed limit of the correlation function -- this is because we have
included in the correlation function additional soft modes that
carry with them additional powers $1/q$.\footnote{We thank Paolo Creminelli and Guido D'Amico for discussions
  on this point.}



\vspace{0.2cm}

\noindent {\bf 4.} {\bf Robustness.} Let us close this section
by reiterating how robust the consistency relation is.
As long as the underlying assumptions -- the existence of the
time-dependent translation symmetry, single field initial condition and
adiabatic mode condition -- are satisfied, the consistency
relation is robust.
The hard or high momentum modes can be those of any LSS observable
(or even mixtures of observables), referred to as ${\cal O}_{\vec k}$
in Eq.\ (\ref{KRPPrelation6}). 
No assumption is made about the size of ${\cal O}_{\vec k}$:\
it matters not at all how nonlinear or non-perturbative these high
momentum observables are.
Indeed, we do not even need to know their detailed dynamics: 
all we need to know is how they transform under the symmetry in
Eq.\ (\ref{DeltaOshift}).\footnote{If the observable turns out to
  transform differently from Eq.\ (\ref{DeltaOshift}), one can
go back to the more fundamental Eq.\ (\ref{backWave}) to figure
out the correct consistency relation.}
Their evolution can be a lot more complicated than
that of dark matter (e.g.\ Eq.\ \ref{messysystem}).
In other words, ${\cal O}_{\vec k}$ can be astrophysically messy
observables, such as those associated with galaxies. 
The presence of pressure effects, multiple components,
multiple-streaming,\footnote{
The soft-mode, by the assumption of growing mode initial condition, is
potential flow, but the hard modes need not be, and can even involve
multiple streams. See footnote \ref{potentialflow}.
}
 star formation,
supernova explosions, etc.\ does not lead to violations of the
consistency
relations, as long as the adiabatic mode condition -- 
i.e.\ the soft dynamics constraint and the squeezing constraint -- is
satisfied.
This is why the LSS consistency
relations are interesting:\ 
they provide a reliable window into the non-perturbative,
astrophysically complex regime.

\subsection{A Simple Fluid Lagrangian for LSS}
\label{fluidLagLSS}

The time-dependent translation symmetry laid out above
was justified at the level of the equations of motion.
It would be useful to see the same at the level of the action.
In this section, we provide the action that describes the dark 
(i.e.\ pressureless) matter
dynamics under gravity.\footnote{We are indebted to Alberto Nicolis and Filippo Vernizzi
who showed us this action.}
We should stress that, for our discussion
of the consistency relation, the action is not strictly necessary;\ the
equations of motion are as good a guide to the symmetry.
Moreover, the action we will write down concerns only dark matter;\ it
does not cover realistic observables such as galaxies, while the
consistency relation applies regardless of the complex astrophysics
that might be present in such observables.
Nonetheless, the dark matter action is useful for conceptual
understanding. We provide it here for completeness, and
connect it with a more well known fluid action in Appendix
\ref{fluidAppendix}.
For simplicity, we assume potential flow;\ an extension
to allow for vorticity should be straightforward, along the lines
of \cite{Dubovsky:2005xd}.
Readers not interested in the action perspective can skip to
\S \ref{diffGR} -- the rest of the paper does not depend on this
section. 



Let us motivate the construction of the action by reducing
the standard pressureless LSS equations 
(\ref{NewtonianLSS}) into a single
equation for the velocity potential $\pi$.
The Euler equation can be integrated once to give:
\begin{eqnarray}
\label{Phipi}
\Phi = - {1\over a} \left[ (a\pi)' + {1\over 2}a (\nabla \pi)^2\right] \, ,
\end{eqnarray}
where $(\nabla\pi)^2$ stands for $\partial_i \pi \partial_i \pi$.\footnote{We use $\nabla_i$ and $\partial_i$ interchangeably, preferring the former where there is the danger of confusing
$\partial$ with the space-time derivative.}
The Poisson equation then gives us:
\begin{eqnarray}
\label{Deltapi}
\delta = - {2 M_P^2\over \bar\rho a^3} \nabla^2 \left[(a\pi)' +
  {1\over 2}a (\nabla \pi)^2\right]
\, .
\end{eqnarray}
The continuity equation can thus be turned into a single equation for $\pi$:
\begin{eqnarray}
- {2M_P^2} \nabla^2\left[ (a\pi)' +
  {1\over 2}a (\nabla \pi)^2\right]'
+ \bar\rho a^3 \nabla^2 \pi
- {2M_P^2} \nabla_i \left( 
\nabla_i \pi \nabla^2 \left[ (a\pi)' +
  {1\over 2}a (\nabla \pi)^2\right] \right) = 0 \, .
\end{eqnarray}
This is a complicated looking equation, but it is not
too difficult to guess the form of the associated action:
\begin{eqnarray}
\label{Spi}
\boxed{
S = - \int d^4 x \,  \left[ \, {1\over 2} \bar\rho a^4 (\nabla \pi)^2
+ M_P^2 \left( \nabla \left[(a\pi)' +
  {1\over 2}a (\nabla \pi)^2\right] \right)^2 \, \right] \, .}
\end{eqnarray}
The overall normalization (and sign) is arbitrary from the point of
view of reproducing the desired equation of motion, but is
chosen to conform to a more general action discussed in Appendix \ref{fluidAppendix}.
It is straightforward to check that this action is invariant under
the time-dependent translation symmetry discussed earlier, namely:
\begin{eqnarray}
\eta \rightarrow \tilde \eta = \eta \quad , \quad
x^i \rightarrow \tilde x^i = x^i + n^i \quad , \quad
\pi \rightarrow \tilde \pi = \pi + n^i {}' x^i \, .
\end{eqnarray}

The dynamics of the velocity potential
$\pi$ is completely fixed by this action.
From this point of view, the $\pi$ equation of motion
has the interpretation of the continuity equation, if 
$\delta$ is {\it defined} by Eq.\ (\ref{Deltapi});\
the gravitational potential $\Phi$ is {\it defined} by
Eq.\ (\ref{Phipi}) so as to reproduce the Poisson equation.
With this understanding, the action takes a fairly simple
form:
\begin{eqnarray}
S = \int d^4 x \, {\bar\rho a^4 \over 2}   
\left( \Phi\delta - \vec v^2 \right) \, ,
\end{eqnarray}
i.e.\ the Lagrangian is the difference between what resembles
potential energy and kinetic energy, though with an unexpected overall
sign,
which can be understood from the larger context of a fluid with
pressure
(see Appendix \ref{fluidAppendix}).

\section{Consistency Relations from Diffeomorphisms -- 
General Relativistic Symmetries}
\label{diffGR}

The time-dependent translation 
symmetry noted by KRPP (Eq.\ \ref{KRPPsymm}) appears to be a global
symmetry of the Newtonian LSS equations.
(Or, more generally, the time-dependent translation as described by
Eq. \ref{KRPPsymmGen} is a symmetry of the equations of motion
for dark matter and galaxies.)
Our goal in this section is to place it in a larger context:\ the claim
is that this symmetry is actually part of a diffeomorphism in the
context of general relativity. This perspective is useful for two
reasons:\ first, it helps us make contact with the earlier work on consistency
relations in inflation, which are based on diffeomorphism invariance; second, 
diffeomorphism invariance allows us to systematically write down further
consistency relations.
The earlier work generally uses the $\zeta$-gauge, alternatively
referred to as the unitary or comoving gauge.
On the other hand, in LSS studies, the Newtonian gauge is
the more natural one to use.
Here, we take advantage of the fact that the full list
of consistency relations are already known in the unitary or $\zeta$-gauge
\cite{Hinterbichler:2013dpa},
and transform each known symmetry in $\zeta$-gauge
into a symmetry in the Newtonian gauge.
This way, we will obtain an infinite tower of consistency
relations in the Newtonian gauge.
We emphasize that
we could equally well proceed by directly working in the Newtonian
gauge, and obtain the same results (see
\cite{Creminelli:2013mca} on the dilation and special conformal
consistency relations obtained this way).
One might wonder why writing down consistency relations in
the Newtonian gauge is useful if we already know what they are in the unitary
gauge. It has to do with the taking of the Newtonian limit, a subject
we will discuss later in this section, and in \S \ref{discuss}.


In the interest of generality, we allow the presence of multiple
components of which pressureless matter/dust is one. We assume
adiabatic initial conditions in the sense that all components
fluctuate in the same way in the long wavelength limit:\ in particular
their velocity potentials coincide in this limit.
We give in \S \ref{symmetryshifts}
the general prescription for transforming symmetries
known in the unitary gauge to symmetries in Newtonian gauge.
In \S \ref{scalarrelations} we focus on the dilation and the special
conformal symmetries, which are the symmetries that generate
only scalar modes, and we show how the KRPP Newtonian consistency relation
arises as the sub-Hubble limit of the latter. We comment
on the robustness and limitations of the consistency relations
in \S \ref{robustnessGR}, adding a relativistic twist to some
of the comments made earlier. We also discuss the taking
of the Newtonian/sub-Hubble limit. We close with 
\S \ref{vectortensorrelations} on further consistency relations
that form an infinite tower -- they generally involve the tensor modes.
We comment on why there is no useful sub-Hubble limit in these cases.

\subsection{Symmetry Transformations from Diffeomorphisms}
\label{symmetryshifts}

Here we are interested in symmetry transformations coming from residual gauge/coordinate transformations
(i.e.\ diffeomorphisms) 
\begin{eqnarray}
x' {}^\mu = x^\mu + \xi^\mu \, ,
\end{eqnarray}
that are allowed even after we have applied
the usual gauge-fixing.
In the context of inflation, a common gauge is the
unitary or $\zeta-$gauge:
\begin{eqnarray}
ds^2 = ... + a^2 e^{2\zeta} (e^\gamma)_{ij} dx^i dx^j \, , \delta \phi = 0,
\end{eqnarray}
where we have omitted the time-time and time-space components
of the metric which are obtainable from the given space-space parts
by solving the Hamiltonian and momentum constraints. Here,
$\zeta$ represents the scalar perturbation
and the transverse traceless $\gamma_{ij}$ represents the tensor
perturbation. Vector perturbations are ignored because
they are not generated by single field models (a brief discussion
of vector modes can be found in Appendix \ref{vectormodes}). 
The equal time surface is chosen so that the matter field, which we have 
called $\phi$, has no spatial fluctuation.
For our application, there can in general be multiple components, in which
case $\delta\phi$ is chosen to vanish for one of them.
To be concrete, let us choose this to be the dark matter fluid, i.e.\
we model it as a fluid described by a Lagrangian of the form
$P(X)$, where $P$ is some function of $X \equiv -
(\partial\phi)^2$. 
The velocity potential $\pi$ is related to $\phi$ by
$\delta\phi = \phi - \bar\phi = - \bar\phi' \pi$, where
$\bar\phi$ is the background, and $\bar\phi'$ is its
conformal time derivative (see Appendix \ref{fluidAppendix}).\footnote{By describing the dark matter using a single
fluid field $\phi$, we are ignoring orbit-crossing and also
vorticity. As is clear in the Newtonian discussion, neither one of these
assumptions is strictly necessary. We make them only to simplify
the general relativistic discussion.}

The full list of residual diffeomorphisms that respect the unitary gauge
is worked out in H2K. 
Since the unitary gauge is a complete gauge-fixing
for diffeomorphisms that vanish at spatial infinity, the residual
diffeomorphisms must be those that do not vanish at infinity.
They take the form:
\begin{eqnarray}
\xi^0 = 0 \quad , \quad \xi^i = \xi^i_{\rm unit.} \sim x^n \, .
\end{eqnarray}
No time-diffeomorphism is allowed since that would
violate the $\delta\phi = 0$ (or $\pi = 0$) unitary gauge condition,
and the allowed spatial diffeomorphism, which we refer to as
$\xi^i_{\rm unit.}$), goes like $x^n$, where $n=1, 2, ...$.
We will give explicit expressions for $\xi^i_{\rm unit.}$
later. They satisfy:\footnote{This holds only to the lowest
order in tensor modes. See H2K, and discussion in footnote \ref{higherordercorrections}.}
\begin{eqnarray}
\label{scalartensorsymmUnit}
{\rm scalar \, \, + \,\, tensor \, \, symmetries:} 
\quad 
\nabla^2 \xi_{\rm unit.}^i + {1\over 3} \partial_i (\partial_k
\xi^k_{\rm unit.}) = 0 \, .
\end{eqnarray}
This set of symmetries contains subsets that
only generate (nonlinearly) scalar modes,
and subsets that only generate tensor modes:
\begin{eqnarray}
\label{scalarOnlytensorOnlyUnit}
{\rm scalar \,\, symmetries}: \quad \partial_i \xi_{\rm unit.}^j
+ \partial_j \xi_{\rm unit.}^i - {2\over 3} \delta_{ij}
\partial_k \xi^k_{\rm unit.} = 0 \nonumber \\
{\rm tensor \,\, symmetries}: \quad \partial_i \xi^i_{\rm unit.} = 0
\, \, , \, \, \nabla^2 \xi^i_{\rm unit.} = 0 \, .
\end{eqnarray}
The spatial diffeomorphism
$\xi^i_{\rm unit.}$ can be considered to be time-independent.\footnote{Adiabatic mode conditions in the unitary gauge
actually make $\xi^i_{\rm unit.}$ time-dependent in general.
As shown in H2K, its time-independent part alone is sufficient
to deduce the consistency relations. We will implement
the adiabatic mode conditions separately in the Newtonian gauge
computation.}

In LSS studies, it is more common to employ the Newtonian gauge instead:
\begin{eqnarray}
ds^2 = a^2 \left[ - (1 + 2\Phi) d\eta^2 + 2 S_i dx^i d\eta
+ ( (1 - 2\Psi) \delta_{ij} +
\gamma_{ij} ) dx^i dx^j \right] \, ,
\end{eqnarray}
where we no longer impose $\pi = 0$, 
$\Phi$ and $\Psi$ are the scalar modes,
the transverse traceless $\gamma_{ij}$ denotes the tensor
modes as before, and the divergence-free
$S_i$ represents the vector modes (which is set to zero in this paper).
Here, we work perturbatively in the metric perturbations,
since the Newtonian-gauge metric perturbations are expected to be small even in the highly
nonlinear regime where the density fluctuation
$\delta$ is large, and including higher order metric perturbations
corrects the consistency relations by negligible amounts.\footnote{See footnote \ref{higherordercorrections} for a more
  detailed discussion of this point.
}
Under a small diffeomorphism $\xi^\mu$, 
the nonlinear transformations of the metric fluctuations are:\footnote{The net (linear $+$ nonlinear) transformation
of the metric is given by
$\Delta g_{\mu\nu} = - \xi^\alpha\partial_\alpha g_{\mu\nu} - g_{\alpha\mu} \partial_\nu \xi^\alpha
- g_{\alpha\nu} \partial_\mu \xi^\alpha$.}
\begin{eqnarray}
\label{metricshifts}
&& \Delta_{\rm nl.} \Phi = - \xi^0 {}' - {\cal H} \xi^0 \quad , \quad
\Delta_{\rm nl.} \Psi = {\cal H} \xi^0 + {1\over 3} \partial_k \xi^k \, ,\nonumber \\
&& \Delta_{\rm nl.} g_{0i} = a^2 (\partial_i \xi^0 - \partial_0 \xi^i) \quad ,
\quad
\Delta_{\rm nl.} g_{ij} - {1\over 3} \delta_{ij} \Delta_{\rm nl.} g_{kk}
= - a^2 \left( \partial_i \xi^j + \partial_j \xi^i - {2\over 3}
  \delta_{ij} \partial_k \xi^k \right) \, .
\end{eqnarray}

Given each symmetry in the unitary gauge, it is straightforward
to deduce the corresponding symmetry in the Newtonian gauge.
Let us break it down into a number of steps.
First, we begin with the metric in Newtonian gauge, where
$\pi = \pi_0 \ne 0$. 
We assume $\Psi = \Phi$, in the absence of anisotropic stress.\footnote{This is an adiabatic mode condition in the Newtonian gauge.
See discussions in Appendix \ref{GRadiabatic}.}
To convert to the unitary gauge, we
apply a time-diffeomorphism $\xi^0 = -\pi_0$ to make the scalar
field $\phi$ spatially homogeneous. 
Second, we apply the known unitary-gauge symmetry transformation
$\xi^i = \xi^i_{\rm unit.}$. 
Third, we wish to return to Newtonian gauge.
The first and second steps in general make
$\Psi \ne \Phi$. To restore equality, we apply an additional
time-diffeomorphism
$\xi^0 = \pi_0 + \xi^0_{\rm add.}$.
We also need to ensure $g_{0i} = 0$
(no vector modes\footnote{
The absence of vector modes is assumed in two places.
Assuming 
$\nabla^2 \xi^i_{\rm unit.} + \partial_i (\partial
\cdot \xi)/3 = 0$ means there is no vector mode
in the spatial part of the metric.
In addition, our choice of $\xi^\mu_{\rm add.}$ 
ensures there is no vector mode in
the space-time part of the metric either.}), 
and thus an additional spatial diffeomorphism
$\xi^i_{\rm add.}$ may be necessary. 
It is shown in Appendix \ref{GRadiabatic}
that the requisite additional time- and space-diffeomorphisms
are:
\begin{eqnarray}
\label{xi0xiiadd}
\xi^0_{\rm add.} = - {1\over 3c} D' \partial_i \xi^i_{\rm unit.}
\quad , \quad
\xi^i_{\rm add.} = {1\over c} D \nabla^2 \xi^i_{\rm unit.} \, .
\end{eqnarray}
Here, $D$ is the linear growth factor satisfying the following
equation:
\begin{eqnarray}
\label{DcEqt}
D'' + 2{\cal H} D' - c = 0 \, ,
\end{eqnarray}
where $c$ is a constant (independent of time and space).
In other words, the following diffeomorphism is a symmetry of Newtonian gauge:
\begin{eqnarray}
\label{xiTot}
\xi^0 = \xi^0_{\rm add.} \quad , \quad \xi^i = \xi^i_{\rm unit.} +
\xi^i_{\rm add.} \, ,
\end{eqnarray}
where $\xi^i_{\rm unit.}$ is the residual (time-independent) diffeomorphism
allowed by the unitary gauge (Eqs.\ \ref{scalartensorsymmUnit} \&
\ref{scalarOnlytensorOnlyUnit}).
Furthermore, it can be shown that
this diffeomorphism
satisfies the adiabatic mode conditions, i.e.\ the perturbations that
are nonlinearly generated match the time-dependence of
very soft (growing) physical modes. This is why the linear growth factor $D$ appears
in the diffeomorphism. The derivation is given in Appendix
\ref{GRadiabatic}. (The attentive reader might wonder why the linear growth
factor $D$ -- a quantity that shows up in the Newtonian discussion of
sub-Hubble perturbations -- appears also in a general relativistic
discussion, and how Eq.\ (\ref{DcEqt}) is related to 
the more familiar growth equation (Eq.\ \ref{nadiabatic}). 
This is discussed in Appendix \ref{piEOMderive}).
An important underlying assumption is that all fluid components
move with the same velocity in the soft limit. 
Under this assumption, it is shown in
Appendix \ref{piEOMderive} that the velocity, or velocity
potential $\pi$, evolves as:
\begin{eqnarray}
\lim_{\vec q \rightarrow 0} \pi_{\vec q} \propto D' \, .
\end{eqnarray}
In the context of a general relativistic discussion,
this statement (strictly speaking) holds in the super-Hubble limit
$q \ll {\cal H}$. What is interesting is that for the $\pi_{\vec q}$
of pressureless matter, this statement holds also for sub-Hubble (but linear) scales . It is this fact that makes
an interesting Newtonian consistency relation possible.\footnote{The fact that the soft $\pi$ is proportional to $D'$ is
nicely consistent with $\xi^0 \propto D'$, since
$\Delta_{\rm nl.} \pi = \xi^0$ (see
Eq.\ \ref{pishiftGRnl}).}

For the purpose of deducing the consistency relations, we also
need to know how other LSS observables transform under
a diffeomorphism. 
From the way a scalar should transform, one can see 
the velocity potential $\pi \equiv -\delta\phi/\bar\phi'$
should transform by 
\begin{eqnarray}
\label{pishiftGRnl}
\Delta\pi = \Delta_{\rm lin.} \pi + \Delta_{\rm nl.} \pi \quad {\rm
  where} \quad
\Delta_{\rm lin.} \pi = - \xi^0 {(\bar\phi' \pi)' \over \bar\phi'} -
\xi^i \partial_i \pi \quad , \quad
\Delta_{\rm nl.} \pi = \xi^0 \, .
\end{eqnarray}
We will mostly need only the nonlinear part of the $\pi$
transformation.
As emphasized above, the assumption of potential flow is
not strictly necessary. The nonlinear transformation of the velocity
can also be deduced by transforming the $4-$velocity $U^\mu$:\footnote{One can use $U^\mu = (1-\Phi, v^i)/a$, valid to the lowest order
  in velocity and perturbations, with the understanding that $v^i = dx^i /d\eta$.
In this paper, by relativistic effects, we are generally interested in
effects on super-Hubble scales as opposed to effects associated
with high peculiar velocities.}
\begin{eqnarray}
\Delta_{\rm nl.} v^i = \xi^i \,{}'
\end{eqnarray}

Another LSS observable of interest is the mass density fluctuation
$\delta$. Its transformation is:
\begin{eqnarray}
\label{deltashift}
\Delta \delta = \Delta_{\rm nl.} \delta + \Delta_{\rm lin.} \delta
\quad , \quad
\Delta_{\rm nl.} \delta =  - \xi^0 {\bar\rho' \over \bar\rho} \, \, , \, \,
\Delta_{\rm lin.} \delta =  - \xi^\mu \partial_\mu \delta - \xi^0
{\bar \rho' \over \bar\rho} \delta \, .
\end{eqnarray}
One could set $- \bar\rho'/\bar\rho = 3{\cal H}$ for $\bar\rho$
that redshifts like pressureless matter, but we will keep the discussion
general.
The generalization to the galaxy density fluctuation
$\delta_g$ (or the fluctuation of any component) is immediate:
\begin{eqnarray}
\label{deltagshift}
\Delta \delta_g = \Delta_{\rm nl.} \delta_g + \Delta_{\rm lin.} \delta_g
\quad , \quad
\Delta_{\rm nl.} \delta_g =  - \xi^0 {\bar\rho_g' \over \bar\rho_g} \, \, , \, \,
\Delta_{\rm lin.} \delta_g =  - \xi^\mu \partial_\mu \delta_g - \xi^0
{\bar \rho'_g \over \bar\rho_g} \delta_g\, ,
\end{eqnarray}
where $\bar\rho_g$ is the mean galaxy number density.
In both cases, the linear part of the transformations would
resemble more what one expects for a scalar if we consider
$\delta\rho = \bar\rho \delta$ instead of $\delta$:
\begin{eqnarray}
\Delta_{\rm nl.} \delta\rho = - \xi^0 \bar\rho' \,\, , \,\,
\Delta_{\rm lin.} \delta\rho = - \xi^\mu\partial_\mu \delta\rho \, .
\end{eqnarray}

\subsection{Scalar Consistency Relations}
\label{scalarrelations}

Let us first derive the consistency relations that involve only scalar modes, i.e.\ where only scalar modes are nonlinearly generated.
Recall from \S \ref{symmetryshifts} that the scalar symmetries
take the form:
\begin{eqnarray}
\xi^0 = \xi^0_{\rm add.} \quad , \quad \xi^i = \xi^i_{\rm unit.} +
\xi^i_{\rm add.} \, ,
\end{eqnarray}
with $\xi^i_{\rm unit.}$ and $\xi^\mu_{\rm add.}$ satisfying:
\begin{eqnarray}
\label{scalarsymmCond}
&& \partial_i \xi_{\rm unit.}^j
+ \partial_j \xi_{\rm unit.}^i - {2\over 3} \delta_{ij}
\partial_k \xi^k_{\rm unit.} = 0 \, , \\
\label{addDiff}
&& \xi^0_{\rm add.} = - {1\over 3c} D' \partial_i \xi^i_{\rm unit.}
\quad , \quad
\xi^i_{\rm add.} = {1\over c} D \nabla^2 \xi^i_{\rm unit.} \, ,
\end{eqnarray}
where $D$ is the linear growth factor obeying
$D'' + 2{\cal H}D' - c = 0$, with $c$ being a constant.


As discussed before, since the unitary gauge
is a complete gauge-fixing for diffeomorphisms that vanish
at spatial infinity, the residual diffeomorphism 
of interests must be one where $\xi^i_{\rm unit.}$ does not vanish
at infinity. Following H2K, we can express $\xi^i_{\rm unit.}$ as
a power series:
\begin{eqnarray}
\label{H2Kpowerseries}
\xi^i_{\rm unit.} = \sum_{n=0}^\infty {1\over (n+1)!}
M_{i\ell_0 ... \ell_n} x^{\ell_0} ... x^{\ell_n} \, ,
\end{eqnarray}
where each $M_{i\ell_0 ... \ell_n}$ represents a constant
coefficient, symmetric in its last $n+1$ indices.
As pointed out by \cite{Creminelli:2012ed,Hinterbichler:2012nm}, the
only scalar symmetries are those associated with $n=0$: $\xi^i_{\rm unit.} \sim x$
(dilation) and $n=1$: $\xi^i_{\rm unit.} \sim x^2$ (special conformal
transformation).

\subsubsection{The Dilation Consistency Relation}
\label{Dilation}

Dilation is described by $\xi^i_{\rm unit.} = \lambda x^i$ where $\lambda$
is a constant. Plugging this into Eq.\ (\ref{addDiff}) tells us
$\xi^0_{\rm add.} = - (\lambda / c)D'$ and $\xi^i_{\rm add.} = 0$. In
other words, the net residual diffeomorphism in Newtonian gauge is
\begin{eqnarray}
\label{dilationsymmSummarize}
\xi^0 = \epsilon \quad , \quad \xi^i = \lambda x^i
\quad {\rm with} \quad \epsilon \equiv - {\lambda \over c} D' \, ,
\end{eqnarray}
where $\lambda$ is a constant.
This symmetry involves a spatial dilation $+$ an accompanying time
translation, with the two related by a differential equation:
$\epsilon' + 2{\cal H}\epsilon + \lambda = 0$.
We will refer to the resulting consistency relation 
simply as the dilation consistency relation, even though the symmetry
involves more than spatial dilation.
To deduce the associated consistency relation, we employ
Eq.\ (\ref{backWave}). 
Two pieces of information are needed to use it. One is the
nonlinear shift of $\pi$ in Fourier space, obtained by taking
the Fourier transform of Eq.\ (\ref{pishiftGRnl}):
\begin{eqnarray}
\Delta_{\rm nl.} \pi^*_{\vec q} = (2\pi)^3 \delta_D (\vec q)
\, \epsilon
\, .
\end{eqnarray}
The other piece of information we need is the linear transformation
of the high momentum observable(s).
Here, let us use the density fluctuation $\delta_{\vec k}$ as the
observable at high momentum. By a Fourier transform of
Eq.\ (\ref{deltashift}),
we find
\begin{eqnarray}
\Delta_{\rm lin.} \delta_{\vec k}
= \left[ - \epsilon \left( {\bar\rho' \over \bar\rho} + \partial_\eta \right)
+ \lambda (3 + \vec k \cdot \partial_{\vec k} ) \right] \delta_{\vec k} \, .
\end{eqnarray}
Plugging these two pieces into the master equation (\ref{backWave}),
we see that
\begin{eqnarray}
\lim_{\vec q \rightarrow 0} 
{\langle \pi_{\vec q} \delta_{\vec k_1} ... \delta_{\vec k_N} \rangle^c
\over P_\pi (q)} \epsilon(\eta)
= \sum_{a=1}^N \left[ -\epsilon(\eta_a) \left( {\bar\rho' \over \bar\rho}
    \Big|_{\eta_a} + \partial_{\eta_a} \right) + \lambda
\left(3 + \vec k_a \cdot \partial_{\vec k_a} \right) \right] \langle
\delta_{\vec k_1} ... \delta_{\vec k_N} \rangle^c \, ,
\end{eqnarray}
where the time-dependence should be understood as follows:\
the soft $\vec q$ mode is evaluated at time $\eta$, while the hard
mode $\vec k_a$ is evaluated at time $\eta_a$, meaning each hard mode
can be at a different time. This is why the $\epsilon$ on the left is
at time $\eta$ -- it is associated with the nonlinear shift in $\pi$
and therefore the soft mode -- and the $\epsilon$'s on the right are evaluated at the
respective $\eta_a$, since each is associated with the linear
transformation of the corresponding hard mode.

The connected N- and (N+1)-point functions on both sides contain
the momentum conserving delta function. Its removal
requires some care since the derivatives with respect to
momentum on the right hand side act on the delta function:
\begin{eqnarray}
\sum_{a=1}^N \vec k_a \cdot \partial_{\vec k_a}
\delta_D (\vec k_1 + ... + \vec k_N)
= -3 \delta_D (\vec k_1 + ... + \vec k_N) \, ,
\end{eqnarray}
which can be established by rewriting the delta function as
$(2\pi)^{-3} \int d^3 x \, e^{i (\vec k_1 + ... + \vec k_N) \cdot \vec
  x}$, and integrating by parts.
Thus, removing the delta function on both sides, with
$\langle ... \rangle^{c'}$ representing the connected correlation function without $\delta_D$, we have\footnote{\label{kN} Note that $\langle \delta_{\vec k_1} ... \delta_{\vec k_N}
  \rangle^{c'}$ should be understood to be a function of only $N-1$
  momenta. For instance, we can think of $\vec k_N = - \vec k_1 - \vec
  k_2 ... - \vec k_N$. Thus the derivative $\partial_{\vec k_N}$,
  keeping $\vec k_1, ... , \vec k_{N-1}$ fixed, vanishes.
This point is particularly important for the higher consistency relations.
See \cite{Goldberger:2013rsa}.}
\begin{eqnarray}
\label{dilationConsistency0}
\lim_{\vec q \rightarrow 0} 
{\langle \pi_{\vec q} \delta_{\vec k_1} ... \delta_{\vec k_N} \rangle^{c'}
\over P_\pi (q)} \epsilon(\eta)
= \left( 3 \lambda (N-1) + \sum_{a=1}^N \left[ -\epsilon(\eta_a) \left( {\bar\rho' \over \bar\rho}
    \Big|_{\eta_a} + \partial_{\eta_a} \right) + \lambda
\vec k_a \cdot \partial_{\vec k_a} \right] \right) \langle
\delta_{\vec k_1} ... \delta_{\vec k_N} \rangle^{c'} \, ,
\end{eqnarray}
with the understanding that $\epsilon$ and (the constant) $\lambda$ are related by
Eq.\ (\ref{dilationsymmSummarize}), and where the N-point function depends on the time associated with each of the N modes. 
Using the relation,
we can rewrite the {\it dilation consistency relation} as
\begin{eqnarray}
\label{dilationConsistency}
\boxed{
\lim_{\vec q \rightarrow 0} 
{\langle \pi_{\vec q} \delta_{\vec k_1} ... \delta_{\vec k_N} \rangle^{c'}
\over P_\pi (q)}
= - \left({3c\over D'(\eta)} (N-1) + \sum_{a=1}^N \left[ {D'(\eta_a)\over
      D'(\eta)}
\left( {\bar\rho' \over \bar\rho}
    \Big|_{\eta_a} + \partial_{\eta_a} \right) + {c \over D'(\eta)}
\vec k_a \cdot \partial_{\vec k_a} \right] \right) \langle
\delta_{\vec k_1} ... \delta_{\vec k_N} \rangle^{c'} \, ,}
\end{eqnarray}
with the understanding that $c = D'' + 2{\cal H}D'$ is a constant.
It is trivial to generalize the consistency relation by 
changing the hard modes from
$\delta$ for the mass density
to $\delta_g$ for the galaxy density:\ simply change
the mean mass density
$\bar\rho$ on the right hand side to 
the mean galaxy number density $\bar\rho_g$ .
This consistency relation can be further rewritten in different forms.
We will postpone this discussion until after we discuss the special
conformal consistency relation.

\subsubsection{The Special Conformal Consistency Relation --
  Containing the Newtonian Translation Consistency Relation}
\label{SCTConsistency}

Next, we consider the special conformal transformation:\
$\xi^i_{\rm unit.} = 2 \vec b \cdot \vec x x^i
- b^i \vec x \cdot \vec x$, where $\vec b$ is a constant vector.
Plugging this into Eq.\ (\ref{addDiff}), we see that the requisite
accompanying diffeomorphism is:
$\xi^0_{\rm add.} = -(2/c) D' \vec b \cdot \vec x$
and $\xi^i_{\rm add.} = -(2/c) D b^i$. 
Putting everything together, we see that the symmetry is:
\begin{eqnarray}
\label{SCTsymmSummarize}
\xi^0 = n^i {}' \, x^i \quad , \quad \xi^i = n^i + 2 {\vec b} \cdot {\vec
  x} \, x^i - b^i \, \vec x \cdot \vec x \quad {\rm with} \quad
n^i \equiv - {2\over c} D b^i
\end{eqnarray}
where $\vec b$ is a constant vector.
We refer to the implied consistency relation
as the special conformal consistency relation, even
though the full symmetry transformation involves a time
diffeomorphism and a spatial translation in addition to the
special conformal transformation -- these transformations
are related via $n^i {}'' + 2{\cal H} n^i {}' + 2 b^i = 0$.
As we will see,
the Newtonian translation consistency relation is contained in here.

Once again, we employ the master equation
(Eq.\ \ref{backWave}), for which we need the nonlinear transformation
of the velocity potential $\pi$ and the linear transformation
of the hard modes -- as in the case of dilation, 
we choose the observable to be the density fluctuation $\delta$
for the hard modes. Under the current symmetry transformation,
we have:
\begin{eqnarray}
\Delta_{\rm nl.} \pi^*_{\vec q} = i \vec n {}' \cdot \partial_{\vec q} [
(2\pi)^3 \, 
\delta_D (\vec q) ]\, ,
\end{eqnarray}
\begin{eqnarray}
\Delta_{\rm lin.} \delta_{\vec k} = i \left[
\vec
n {}' \cdot \partial_{\vec k}\left( {\bar \rho' \over \bar\rho} + \partial_\eta\right) + \vec n \cdot \vec k      
- (6 \vec b \cdot \partial_{\vec k} + 2 b^j
k^i \partial_{k^j} \partial_{k^i}
- \vec b \cdot \vec k \, \nabla^2_{k})
\right] \delta_{\vec k} \, .
\end{eqnarray}
Substituting the above into Eq.\ (\ref{backWave}), we see that the
left hand side (LHS) is
\begin{eqnarray}
\label{LHSSCT}
{\rm LHS}\, = 
\lim_{\vec q \rightarrow 0} (- i \vec n {}' (\eta)) \cdot \partial_{\vec
  q} \left[ (2\pi)^3 \delta_D (\vec q + \vec k_1 + ... + \vec k_N) \right]
{\langle \pi_{\vec q} \delta_{\vec k_1} ... \delta_{\vec k_N} \rangle^{c'}
\over P_\pi (q)} \nonumber \\
- i  (2\pi)^3 \delta_D (\vec q + \vec k_1 + ... + \vec k_N) 
\vec n {}' (\eta) \cdot \partial_{\vec q} 
\left[ {\langle \pi_{\vec q} \delta_{\vec k_1} ... \delta_{\vec k_N} \rangle^{c'}
\over P_\pi (q)} \right] \, .
\end{eqnarray}
The right hand side (RHS) is
\begin{eqnarray}
\label{RHSSCT}
{\rm RHS} \, = \sum_{a=1}^N
i \left[ {\vec n} {}' (\eta_a)
  \cdot \partial_{\vec k_a}\left( {\bar\rho' \over \bar\rho} \Big|_{\eta_a}
    + \partial_{\eta_a} \right) + 
\vec n(\eta_a) \cdot \vec k_a    
- (6 \vec b \cdot \partial_{\vec k_a} + 2 b^i
k_a^j \partial_{k_a^j} \partial_{k_a^i}
- \vec b \cdot \vec k_a \, \nabla^2_{k_a})
\right] \langle \delta_{\vec k_1} ... \delta_{\vec k_N} \rangle^c \, .
\end{eqnarray}
The connected N-point function $\langle \delta_{\vec k_1}
... \delta_{\vec k_N} \rangle^c$
contains an overall momentum conserving delta function.
The momentum derivative acts non-trivially on it.
To simplify, it is useful to know:
\begin{eqnarray}
\sum_{a=1}^N \left( 2 b^j
k_a^i \partial_{k_a^j} \partial_{k_a^i}
- \vec b \cdot \vec k_a \, \nabla^2_{k_a} \right) \delta_D (\vec
k_{\rm tot.}) = - 6 \, \vec b \cdot \partial_{\vec k_{\rm tot.}}
\delta_D (\vec k_{\rm tot.}) \, ,
\end{eqnarray}
where $\vec k_{\rm tot} \equiv \vec k_1 + ... + \vec k_N$, and
which can be proved by rewriting the delta function as the spatial
integral of a plane wave. 
The term $\sum_a 6 \vec b \cdot \partial_{\vec k_a}
\delta_D (\vec k_{\rm tot.})$ can be rewritten as
$6 N \vec b \cdot \partial_{\vec k_{\rm tot.}} \delta_D (\vec k_{\rm
  tot.})$; indeed, any
$\partial_{k_a^i} \delta_D (\vec k_{\rm tot.})$ can be written as
$\partial_{k_{\rm tot.}^i} \delta_D (\vec k_{\rm tot.})$. 
Then there are terms that involve one momentum derivative on the
delta function and one momentum derivative on the N-point function:
\begin{eqnarray}
&& \sum_{a=1}^N 2 b^i k_a^j \, \partial_{k_a^j} \langle \delta_{\vec k_1}
... \delta_{\vec k_N} \rangle^{c'} \, 
\partial_{k_a^i} \delta_D (\vec k_{\rm tot.}) \, 
+ 2 b^i \partial_{k_a^j} \delta_D (\vec k_{\rm tot.})
\, 
\left( k_a^j \partial_{k_a^i} \langle \delta_{\vec k_1}
  ... \delta_{\vec k_N} \rangle^{c'} 
- k_a^i \partial_{k_a^j} \langle \delta_{\vec k_1}
  ... \delta_{\vec k_N} \rangle^{c'} \right) \nonumber \\
&& \quad = \left[ 2 \vec b \cdot \partial_{\vec k_{\rm tot.}}
\delta_D ({\vec k_{\rm tot.}}) \right] \, \sum_{a=1}^N \vec k_a \cdot \partial_{\vec k_a} \langle \delta_{\vec k_1}
  ... \delta_{\vec k_N} \rangle^{c'}  \, ,
\end{eqnarray}
where we have used rotational invariance of the N-point function to
remove the last two terms on the first line. With this understanding,
Eq.\ (\ref{RHSSCT}) can be expressed as
\begin{eqnarray}
&& {\rm RHS} = -i \Big[ \partial_{k^i_{\rm tot.}} (2\pi)^3\delta_D (\vec
  k_{\rm tot.})\Big] 
\Big[ 6 b^i (N-1) + \sum_{a=1}^N \left(-n^i {}' (\eta_a) 
\left( {\bar\rho' \over \bar\rho} \Big|_{\eta_a} + \partial_{\eta_a}
\right)
+ 2 b^i \vec k_a \cdot \partial_{\vec k_a} \right)   \Big]
\langle \delta_{\vec k_1} ... \delta_{\vec k_N} \rangle^{c'} \nonumber
\\
&& \quad \quad \quad - i (2\pi)^3 \delta_D (\vec k_{\rm tot.}) 
\sum_{a=1}^N
\Big[ -\left( {\bar\rho' \over \bar\rho} \Big|_{\eta_a}
    + \partial_{\eta_a} \right) {\vec n} {}' (\eta_a)
  \cdot \partial_{\vec k_a} -
\vec n(\eta_a) \cdot \vec k_a  
\nonumber \\
&& \quad \quad \quad \quad \quad + (6 \vec b \cdot \partial_{\vec k_a} + 2 b^i
k_a^j \partial_{k_a^j} \partial_{k_a^i}
- \vec b \cdot \vec k_a \, \nabla^2_{k_a})
\Big] \langle \delta_{\vec k_1} ... \delta_{\vec k_N} \rangle^{c'} \, ,
\end{eqnarray}
The first line of the above can be equated with
the first line of Eq.\ (\ref{LHSSCT}), since what
multiplies the derivative of the delta function
on both sides replicates the dilation consistency relation
Eq.\ (\ref{dilationConsistency0}). 
Note that $n^i$ and $b^i$ are related by Eq.\ (\ref{SCTsymmSummarize}).
Using this, eliminating the dilation consistency relation
from both sides,\footnote{This is a general pattern:\
one can obtain the correct consistency relation for
a given symmetry by using the master equation (\ref{backWave}),
{\it and} simply ignoring the delta functions on both sides.
At first sight, this might appear dangerous as there are derivatives
acting on the delta functions, but they invariably multiply
consistency relations from symmetries at the lower levels, and so can
be removed.
} and removing the delta function,
we obtain the {\it special conformal consistency relation}:
\begin{eqnarray}
\label{SCTConsistency}
\boxed{
\begin{split}
\lim_{\vec q \rightarrow 0} \partial_{q^i} 
\left[ {\langle \pi_{\vec q} \delta_{\vec k_1} ... \delta_{\vec k_N} \rangle^{c'}
\over P_\pi (q)} \right] = - \sum_{a=1}^N
\Big[ \left( {\bar\rho' \over \bar\rho} \Big|_{\eta_a}
    + \partial_{\eta_a} \right) {D' (\eta_a) \over D' (\eta)}
  \partial_{k_a^i} +
{D(\eta_a) \over D'(\eta)} k_a^i \\
+ {c \over D'(\eta)} \left( 3 \partial_{k_a^i} + 
k_a^j \partial_{k_a^j} \partial_{k_a^i}
- {1\over 2} k_a^i \, \nabla^2_{k_a} \right)
\Big] \langle \delta_{\vec k_1} ... \delta_{\vec k_N} \rangle^{c'} \, ,
\end{split}}
\end{eqnarray}
with the understanding that $c = D'' + 2{\cal H}D'$ is a constant (Eq.\ \ref{Dc}). Just as in the case of the dilation consistency
relation, this consistency relation can be easily generalized to
the hard modes being the galaxy overdensity -- changing
$\delta$ to $\delta_g$, and changing $\bar\rho$ to $\bar\rho_g$. 
Examining the terms on the right hand side,
we see that in the sub-Hubble limit, i.e.\ 
$k \gg {\cal H}$, the term that dominates on the right hand side
is $- \sum_a (D(\eta_a)/D'(\eta)) k_a^i \langle \delta_{\vec k_1} ...
\delta_{\vec k_N} \rangle^{c'}$, reproducing the translation
consistency relation (Eq.\ \ref{KRPPrelation1}) 
derived from the Newtonian equations. We will have more to say
about the non-relativistic limit in \S \ref{robustnessGR}.

At the level of the spatial diffeomorphisms, dilation and
special conformal transformations exhaust the list of purely scalar
symmetries, since it is only dilation and special conformal
transformations that
respect the second expression of Eq.\ (\ref{scalarsymmCond}) and do
not generate vector or tensor modes.
It is also worth noting that the special conformal transformation
consistency relation strictly speaking receives (small) corrections on the
right hand side, a point to which we will return (footnote
\ref{higherordercorrections}).

\subsection{Robustness and Limitations of the Consistency Relations - 
a Relativistic Perspective and the Newtonian Limit}
\label{robustnessGR}

It is useful to pause, and reflect on the fully relativistic
consistency relations derived so far. Some of our discussions here
mirror the earlier ones in the Newtonian context (\S
\ref{robustness}), but with a relativistic twist.
We also discuss the issue of taking the Newtonian, i.e.\ sub-Hubble,
limit.

\vspace{0.2cm}

\noindent {\bf 1. Newtonian limit.} The special conformal consistency relation
Eq.\ (\ref{SCTConsistency}) is the relativistic analog of the Newtonian
translation consistency relation Eq.\ (\ref{KRPPrelation1}).
The former reduces to the latter in the sense that:
\begin{eqnarray}
\label{SCTConsistencySubHubble}
\lim_{\vec q \rightarrow 0} \, 
{\partial \over \partial q_j}
\left[ {\langle \pi_{\vec q} \delta_{\vec k_1}  ... \delta_{\vec k_N} \rangle^{c'}
\over P_\pi (q)} \right] = -
\sum_{a=1}^N {D(\eta_a) \over D'(\eta)} k_a {}_j \, \langle
\delta_{\vec k_1} ... \delta_{\vec k_N} \rangle^{c'}  \times \left( 1 +
O({\cal H}^2/k^2) \right)\, ,
\end{eqnarray}
where the ${\cal H}^2/k^2$-suppressed terms
can be ignored in the sub-Hubble limit.
Note that the unsuppressed (Newtonian) terms are
of the order of ${k\over {\cal H}}\langle \delta_{\vec k_1}
... \delta_{\vec k_N} \rangle^{c'}$. 
Similarly, we can think of the dilation consistency relation
Eq.\ (\ref{dilationConsistency}) as the relativistic analog of
Eq.\ (\ref{KRPPrelation00}). The dilation consistency
relation takes the form:
\begin{eqnarray}
\label{dilationConsistencySubHubble}
\lim_{\vec q \rightarrow 0} \, {\langle \pi_{\vec q} \delta_{\vec k_1}  ... \delta_{\vec k_N} \rangle^{c'}
\over P_\pi (q)} = O\left( {k \over {\cal H}} \langle \delta_{\vec k_1}
... \delta_{\vec k_N} \rangle^{c'} \right)
\times O\left( {\cal H}^2 /k \right) = 
O\left( q {k \over {\cal H}} \langle \delta_{\vec k_1}
... \delta_{\vec k_N} \rangle^{c'} \right)
\times O({\cal H}/k)({\cal H}/q)
\, , 
\end{eqnarray}
which can be compared against $q \, \times$
Eq.\ (\ref{SCTConsistencySubHubble}).
We can see that the right hand side of
the above expression is $O({\cal H}/k) O({\cal H}/q)$ 
times $q \, \times$
Eq.\ (\ref{SCTConsistencySubHubble}).
In the sub-Hubble limit where ${\cal H}$ is small
compared to both $q$ and $k$, it is therefore consistent to think
of Eq.\ (\ref{dilationConsistencySubHubble}) as vanishing --
reducing to Eq.\ (\ref{KRPPrelation00}).\footnote{Eq.\ (\ref{KRPPrelation00}) was derived using the shift
symmetry $\pi \rightarrow \pi + b$, where $b$ is a constant.
The reader might wonder how that argument breaks down in the
relativistic context. The point is that a constant shift in $\pi$ has to be
accompanied by a time-dependent shift in $\Phi$
(see e.g.\ Eq.\ \ref{Phipi}). Such a time-dependent shift is not
a symmetry of the kinetic term for the metric once time-derivatives
are taken into account, unless coordinates change too.
It is interesting to note that $\pi \rightarrow \pi + b/\bar\phi'$ is
a symmetry (see Appendix \ref{fluidAppendix}) because of the shift symmetry in $\phi$;\ however, this symmetry does not correspond to the growing mode vacuum and therefore does not lead to a consistency relation. See Appendix \ref{vectormodes} for a further discussion.
}

\vspace{0.2cm}

\noindent {\bf 2. Combining consistency relations.} 
It is worth pointing out that, just as in the Newtonian case where
Eqs.\ (\ref{KRPPrelation00}) and (\ref{KRPPrelation1}) can be combined
into a single equation (\ref{KRPPrelation2}),
the general relativistic dilation and special conformal consistency relations can be
combined into:
\begin{eqnarray}
\label{SummedConsistencyRelation}
\boxed{
\begin{split}
\lim_{\vec q \rightarrow 0} 
{\langle \pi_{\vec q} \delta_{\vec k_1} ... \delta_{\vec k_N} \rangle^{c'}
\over P_\pi (q)}
= - \Big({3c\over D'(\eta)} (N-1) + \sum_{a=1}^N \Big[ {D'(\eta_a)\over
      D'(\eta)}
\left( {\bar\rho' \over \bar\rho}
    \Big|_{\eta_a} + \partial_{\eta_a} \right) 
\left(1 + \vec q \cdot \partial_{\vec k_a} \right) 
+ {D(\eta_a) \over D'(\eta)} \vec q \cdot \vec k_a \\
+ {c \over D'(\eta)} 
\left( \vec k_a \cdot \partial_{\vec k_a} 
+ 3 \vec q \cdot \partial_{\vec k_a} + q^i
k_a^j \partial_{k_a^j} \partial_{k_a^i}
- {1\over 2} \vec q \cdot \vec k_a \nabla^2_{k_a} \right)
\Big] \Big) \langle
\delta_{\vec k_1} ... \delta_{\vec k_N} \rangle^{c'} 
\, ,
\end{split}}
\end{eqnarray}
where the constant $c = D'' + 2{\cal H} D'$. 

\vspace{0.2cm}

\noindent {\bf 3. Alternative pions.} Recall that $\pi$, $\delta$ and $\Phi$
all shift nonlinearly under the symmetries of interest.\footnote{We could also discuss the nonlinear shift of $\delta_n$, which coincides with $\delta$ for pressureless matter.  (See Appendix \ref{piEOMderive}.)}
One might wonder whether we could have derived the
consistency relation with $\delta$ or $\Phi$ playing the role
of the pion instead. The answer is affirmative. 
Let us compare these nonlinear shifts:\
$\Delta_{\rm nl.} \pi = \xi^0$,
$\Delta_{\rm nl.} \delta = - \xi^0 \bar\rho'/\bar\rho$, 
$\Delta_{\rm nl.} \Phi = - \xi^0 {}' - {\cal H}\xi^0$. 
Recalling that $\xi^0 \propto D'$,
we see that $\Delta_{\rm nl.} \delta = -\bar\rho'/\bar\rho \times
\Delta_{\rm nl.} \pi$, and $\Delta_{\rm nl.} \Phi = - (D'' + {\cal
    H}D')/D' \times \Delta_{\rm nl.} \pi$. 
One can thus run the same arguments as before, and arrive
at essentially the same consistency relation
Eq.\ (\ref{SummedConsistencyRelation}), 
with the right hand side unaltered,
but the left hand side replaced by
\begin{eqnarray}
\label{LHSalternaterho}
{\rm LHS \,} \rightarrow
-{\bar\rho' \over \bar\rho} \Big|_\eta \times \lim_{\vec q \rightarrow 0} {\langle \delta_{\vec q} \delta_{\vec k_1}
... \delta_{\vec k_N} \rangle^{c'} \over  P_\delta (q)} \, ,
\end{eqnarray}
or
\begin{eqnarray}
\label{LHSalternatePhi}
{\rm LHS\,} \rightarrow - {D'' + {\cal H}D'\over D'} \Big|_\eta \times \lim_{\vec q \rightarrow 0} {\langle \Phi_{\vec q} \delta_{\vec k_1}
... \delta_{\vec k_N} \rangle^{c'} \over P_\Phi (q)} \, .
\end{eqnarray}
The consistency relations expressed using $\pi$, $\delta$ or $\Phi$ as
the soft pion are all equivalent -- with one important caveat, which
is related to the squeezing constraint.

\vspace{0.2cm}

\noindent {\bf 4. Squeezing constraint.} 
The reader might want to consult \S \ref{robustness} for 
a parallel discussion of the squeezing constraint in the Newtonian
context. The consistency relation, whether expressed in terms of
$\pi$ as in Eq.\ (\ref{SummedConsistencyRelation}), or expressed
in terms of $\delta$ or $\Phi$ as in Eq.\ (\ref{LHSalternaterho})
or (\ref{LHSalternatePhi}), is a
statement about the $\vec q \rightarrow 0$ limit.
In the relativistic context, this means, in addition to $q \ll
k_{1}, ... , k_{N}$, the soft mode should strictly speaking be
super-Hubble, i.e.\ $q <  {\cal H}$. On the other hand, in LSS 
we are typically interested
in sub-Hubble modes, so the question arises:\
under what conditions does the consistency relation remain valid
when all modes, including the soft one,
are sub-Hubble (while maintaining the hierarchy
$q \ll
k_{1}, ... , k_{N}$)? 

We show in Appendix \ref{piEOMderive}
a special fact about the velocity potential $\pi$ for pressureless
matter: it has the same time dependence $\propto D'$ 
(Eqs.\ \ref{piC} \& \ref{Dc}) regardless of whether
the wave-mode of interest is inside or outside the Hubble radius.
(When it is outside the Hubble radius, this statement is true
throughout the entire history of the universe; when it is inside the
Hubble radius, this statement is strictly true only after radiation domination.)
The consistency relation written in terms of the matter $\pi$
(\ref{SummedConsistencyRelation}) can therefore be safely
taken inside the Hubble radius, even for the soft mode, since
the same diffeomorphism is capable of generating the correct
$\pi$ regardless of whether it is on super- and sub-Hubble scales.
In this limit,
we recover the Newtonian consistency relation: the term
$- \sum_a [D(\eta_a)/D'(\eta)] \vec q \cdot \vec k_a
\langle \delta_{\vec k_1} ... \delta_{\vec k_N} \rangle^{c'}$
dominates on the right hand side.\footnote{There is one subtlety though:\
for a wave-mode that enters the Hubble radius during radiation
domination, its time-evolution deviates from $D'$ during part of its
history, and so strictly speaking the consistency relation does not
apply if the soft mode belongs to this category. 
An alternative way to put it is this:\
when the wave-mode is within the Hubble radius (or more precisely,
within the sound horizon) during radiation domination, 
neither the matter nor the radiation moves with a velocity that
agrees with $D'$.  A diffeomorphism that obeys the
adiabatic mode conditions (e.g.\
Eq.\ \ref{dilationsymmSummarize} for dilation, or
Eq.\ \ref{SCTsymmSummarize} for special conformal transformation) 
cannot generate the correct velocity for
either component. Even in this
case, we expect
the consistency relation to still be a good approximation 
in the late universe, to the extent that most of the late-time non-Gaussianity
is generated after radiation domination. We thank Paolo
Creminelli for discussions on this point.}
Similar statements hold for $\Phi$ as the soft pion.

The same is not true for $\delta$ (here, we focus on
the {\it matter} $\delta$ as the soft mode):\
from the continuity equation (\ref{deltaConserve}), 
it is evident that the time dependence of $\delta$
(which is the same as $\delta_n$ for pressureless matter)
depends on whether the wave-mode is inside or outside the
Hubble radius. The consistency relation
written using $\delta$ as the soft mode takes
the form of Eq.\ (\ref{LHSalternaterho}) only for
$q < {\cal H}$. If the soft $\delta$ mode is within the horizon, 
the continuity equation tells us $\delta_{\vec q} = q^2 (D/D')
\pi_{\vec q}$, and so the
left hand side of the consistency relation should read:
\begin{eqnarray}
\label{LHSalternaterho2}
{\rm LHS} \rightarrow \lim_{\vec q \rightarrow 0} q^2 {D(\eta) \over D'(\eta)} \times
{\langle \delta_{\vec q} \delta_{\vec k_1}
... \delta_{\vec k_N} \rangle^{c'} \over  P_\delta (q)} \, ,
\end{eqnarray}
while as discussed above, the right hand side reduces to
$- \sum_a [D(\eta_a)/D'(\eta)] \vec q \cdot \vec k_a
\langle \delta_{\vec k_1} ... \delta_{\vec k_N} \rangle^{c'}$.
This reproduces the Newtonian translation consistency relation
written in terms of $\delta_{\vec q}$ (Eq.\ \ref{KRPPrelation3}). 
To conclude: the consistency relation expressed in terms of a soft
$\delta_{\vec q}$ takes a different form outside versus inside
the Hubble radius i.e.\ Eq.\ (\ref{LHSalternaterho})
versus Eq.\ (\ref{LHSalternaterho2}). 
The consistency relation expressed using the matter $\pi$ or $\Phi$ as
the soft pion maintains the same form regardless.\footnote{Using the baryon $\pi$ as the soft pion is permissible too,
as long as one stays above the Jeans scale.}

\vspace{0.2cm}

\noindent {\bf 5.} {\bf The existence of an interesting Newtonian limit.}
From the discussion above, we see that
the special conformal consistency relation has a non-trivial
Newtonian limit (i.e.\ the right hand side is non-vanishing), whereas
the dilation one does not.
What is the underlying reason?
From Eq.\ (\ref{addDiff}), we see that for a given unitary-gauge
transformation $\xi^i_{\rm unit.}$, the corresponding
residual diffeomorphism in Newtonian gauge is
\begin{eqnarray}
\xi^0 \sim {\cal H}^{-1} \partial_i \xi^i_{\rm unit.} \quad , \quad
\xi^i \sim \xi^i_{\rm unit.} + {\cal H}^{-2} \nabla^2 \xi^i_{\rm
  unit.} \, .
\end{eqnarray}
The associated consistency relation, making use of the
relation $\delta_{\vec q} \sim q^2 \pi_{\vec q}
/ {\cal H}$ in the sub-Hubble
limit, can be written schematically as:
\begin{eqnarray}
\lim_{\vec q \rightarrow 0} {q^2 \over {\cal H}^2} {\langle \delta_{\vec q} \delta_{\vec k} ... 
\rangle^{c'}
\over P_\delta (q)} \left[ \partial_i \xi^i_{\rm unit.} \right]_q
\sim 
\left( \left[\partial_i \xi^i_{\rm unit.}\right]_k
+ k_i \left[\xi^i_{\rm unit.}\right]_k + k_i {\cal H}^{-2} 
\left[ \nabla^2
\xi^i_{\rm unit.} \right]_k \right) \langle \delta_{\vec k}
... \rangle^{c'} \, .
\end{eqnarray}
Here, $[ \, \, ]_k$ denotes the Fourier transform
of the quantity of interest at momentum $k$, with 
the delta function removed. 
For instance, for $\xi^i_{\rm unit.} \sim x^{n+1}$,
we have
$[\xi^i_{\rm unit.} ]_k \sim k^{-n-1}$,
$[\partial_i \xi^i_{\rm unit.}]_k \sim k^{-n}$,
$[\partial_i \xi^i_{\rm unit.}]_q \sim q^{-n}$, 
and $[\nabla^2 \xi^i_{\rm unit.}]_k \sim k^{-n+1}$. 
This gives
\begin{eqnarray}
\label{qkratio}
\lim_{\vec q \rightarrow 0} {\langle \delta_{\vec q} \delta_{\vec k} ... 
\rangle^{c'}
\over P_\delta (q)} \sim
\left({q \over k}\right)^n \left( {{\cal H}^2\over q^2} + {k^2 \over
    q^2} \right) \langle \delta_{\vec k}
... \rangle^{c'} \, .
\end{eqnarray}
In the sub-Hubble (and squeezed) 
limit where ${\cal H} \ll q \ll k$, this suggests we have
the dimensionless ratio $\langle \delta_{\vec q} \delta_{\vec k} ... 
\rangle^{c'} / (P_\delta (q) \langle \delta_{\vec k} ... \rangle^{c'})
\sim (q/k)^n (k/q)^2$. 
For $n=1$, the special conformal case, this reproduces correctly
the Newtonian translation consistency relation.
For $n=0$, the dilation case, this does not work. The reason
is that the $k^2/q^2$ term in Eq.\ (\ref{qkratio}), which is
the dominant term in the sub-Hubble limit, originates from
$\nabla^2 \xi^i_{\rm unit.}$, which 
vanishes for the dilation $\xi^i_{\rm unit.} = \lambda x^i$.
Our na\"ive power-counting argument also suggests there
could be additional $n > 1$ consistency relations that
are non-trivial in the Newtonian limit. As we will see in the
next section, the $n > 1$ consistency relations generally
involve tensors, which complicates taking the squeezed mode
to within the Hubble radius.

\vspace{0.2cm}

Let us close this section by emphasizing the robustness of
the consistency relations. As in the Newtonian derivation,
the general relativistic derivation makes no assumptions about
the dynamics of the hard modes -- all we need to know is how
they transform under diffeomorphisms. Thus, we expect
the consistency relations to hold even for nonlinear, 
or astrophysically messy, hard
modes (though the right hand side of
the consistency relations might need to be modified
depending on exactly how the modes of interest transform;
see comments after Eq.\ \ref{dilationConsistency} 
and in footnote \ref{higherordercorrections}). 
Besides the existence of symmetries, which according to the
general relativistic perspective are nothing but residual
diffeomorphisms, 
the two key assumptions are the same as in 
the Newtonian derivation:\ single field initial condition
and adiabatic mode conditions, in particular that all species
move with the same velocity in the soft limit.

\subsection{Consistency Relations Involving Tensor Modes}
\label{vectortensorrelations}




In this section, we move beyond dilation and special conformal
transformation to discuss residual diffeomorphisms that generate
tensor modes (with or without accompanying scalar modes).
We apply the same strategy as the one used for
the pure scalar symmetries:\ use the full set of symmetries
derived in the unitary gauge by H2K, and map each
to a symmetry in the Newtonian gauge by Eq.\ (\ref{xiTot}).

The unitary gauge residual diffeomorphisms can be
written as (Eq.\ \ref{H2Kpowerseries}):
\begin{eqnarray}
\xi^i_{\rm unit.} = \sum_{n=0}^\infty {1\over (n+1)!} M_{i\ell_0 ... \ell_n}
x^{\ell_0} ... x^{\ell_n} \, ,
\end{eqnarray}
where the constant coefficients $M$ satisfy:
\begin{eqnarray}
\label{Mthird}
M_{i \ell \ell \ell_2 \cdots \ell_n} = - \frac{1}{3}M_{\ell i \ell
  \ell_2 \cdots
  \ell_n} \, .
\end{eqnarray}
This condition is derived by substituting the power series into 
Eq.\ (\ref{scalartensorsymmUnit}):
$\nabla^2 \xi^i_{\rm unit.} + \partial_i (\partial_k \xi^k_{\rm
  unit.}) / 3 = 0$.
Note that $M$ by definition
is symmetric in its last $n+1$ indices. 
Since we are interested in $M$ that generates tensor modes
(in addition to possibly scalar modes), we should impose an additional
{\it adiabatic transversality condition}:
\begin{eqnarray}
\label{tensortransverse}
\hat{q}^i\left[
M_{ij \cdots}(\hat{q}) + M_{ji \cdots}(\hat{q}) - \frac{2}{3}M_{kk
  \cdots}(\hat{q})\delta_{ij}\right] = 0 \, .
\end{eqnarray}
This condition can be understood as enforcing that the tensor generated by our diffeomorphism be extensible to the $\vec{q} \to 0$ limit of a transverse physical tensor mode.  Imagine an $M$ that is nearly constant but tapers off to zero at sufficiently large $x$:\ while a constant $M$ yields (derivatives of) a delta function peaked at $\vec{q} = 0$ in Fourier space, a tapering $M$ yields a smoothed-out version thereof.  A tensor mode at a small but finite momentum should be transverse to its own momentum.  We demand that even as we take the $\vec{q} = 0$ limit (allowing the tapering of $M$ to occur at larger and larger distances), transversality continues to hold, keeping the direction $\hat{q}$ fixed.  This is the content of Eq.\ (\ref{tensortransverse}).
The choice of $\hat q$ is arbitrary;\ one could for instance choose it
to point in the $z$ direction.
In addition, if
the diffeomorphism generates {\it only} tensor modes,  
further conditions on $M$ come from
Eq.\ (\ref{scalarOnlytensorOnlyUnit}):\
$\partial_i \xi^i_{\rm unit.} = 0 \,\, \& \,\,
\nabla^2 \xi^i_{\rm unit.} = 0$, which implies
\begin{eqnarray}
\label{Mtracefree}
M_{\ell\ell \ell_1\ell_2 \cdots\ell_n} = 0 \quad \& \quad
M_{i \ell\ell\ell_2 \cdots\ell_n} = 0 \, ,
\end{eqnarray}
i.e.\ $M$ is traceless over any pairs of indices, which
also trivially satisfies Eq.\ (\ref{Mthird}).

As discussed
in \S \ref{symmetryshifts}, 
for each unitary gauge residual diffeomorphism, there
is a corresponding one in Newtonian gauge:
\begin{eqnarray}
\xi^\mu = \xi^\mu_{\rm unit.} + \xi^\mu_{\rm add.} \quad {\rm where}
\quad \xi^0_{\rm add.} = - {1\over 3c} D' \partial_i \xi^i_{\rm unit.}
\quad , \quad
\xi^i_{\rm add.} = {1\over c} D \nabla^2 \xi^i_{\rm unit.} \, ,
\end{eqnarray}
with $D$ being the linear growth factor and where $c$ is a constant
satisfying
$D'' + 2{\cal H}D' - c = 0$. Thus, at the level $n$, the
Newtonian gauge diffeomorphism is:
\begin{eqnarray}
\label{diffMgeneral}
\boxed{
\xi^0 = - {D'\over 3c n!}  M_{\ell\ell\ell_1 ... \ell_n}
x^{\ell_1} ... x^{\ell_n} \quad , \quad
\xi^i = {1\over (n+1)!} M_{i\ell_0\ell_1 ... \ell_n} x^{\ell_0}
... x^{\ell_n}
+ {D\over c (n-1)!} M_{i\ell\ell\ell_2 ... \ell_n} x^{\ell_2}
... x^{\ell_n} \, .}
\end{eqnarray}
This expression holds for all $n$, with the exception of $n=0$,
in which case the last term on the right for $\xi^i$ is absent.
One thing which is immediately clear is that
for purely tensor symmetries --- diffeomorphisms that
generate only tensor modes --- we have $\xi^0_{\rm add.}
= \xi^i_{\rm add.} = 0$, and so they are identical
in the unitary gauge and the Newtonian gauge, as expected.
It is worth emphasizing that Eqs.\ (\ref{diffMgeneral})
and (\ref{Mthird}) are general:\ they apply to symmetries
that generate only scalar modes, 
 or only tensor modes, or both.
For symmetries that generate tensor modes,
adiabatic transversality expressed in Eq.\ (\ref{tensortransverse})
is an additional requirement, and
Eq.\ (\ref{Mtracefree}) applies if {\it only} tensor modes are
generated.

To derive the corresponding consistency relations, we need
a master equation analogous to Eq.\ (\ref{backWave}) but generalized to
allow for the possibility of tensor modes:
\begin{eqnarray}
\label{backWavetensor}
\int {d^3 q\over (2\pi)^3} \, \left[ {\langle \pi_{\vec q} {\cal O}_{\vec k_1}  ... {\cal O}_{\vec k_N} \rangle^c
\over P_\pi (q)} \Delta_{\rm nl.} \pi_{\vec q}^*
+ \sum_{s} {\langle \gamma^s_{\vec q} {\cal O}_{\vec k_1}  ... {\cal O}_{\vec k_N} \rangle^c
\over P_\gamma (q)} \Delta_{\rm nl.} \gamma^s_{\vec q} {}^* \right]
= \Delta_{\rm lin.} \langle  {\cal O}_{\vec k_1}  ... {\cal O}_{\vec k_N} \rangle^c
\, .
\end{eqnarray}
Here, the label $s$ denotes one of the two possible tensor polarization states;\
a given tensor perturbation $\gamma_{ij} (\vec q)$ can be decomposed
as $\gamma_{ij} (\vec q) = \sum_s \epsilon^s_{ij} (\hat q)
\gamma^s_{\vec q}$, where the symmetric traceless polarization tensor
$\epsilon^s_{ij} (\hat q)$ obeys $\hat q^i \epsilon^s_{ij} (\hat q) =
0$
and $\epsilon^s_{ij} (\hat q) \epsilon^{s'}_{ij} (\hat q)^* = 2
\delta^{ss'}$. The tensor power spectrum is defined by
$\langle \gamma^s_{\vec q} \gamma^{s'}_{\vec q'} \rangle = (2\pi)^3
\delta_D(\vec q + \vec q') \delta^{ss'} P_\gamma (q)$.
Note that
\begin{eqnarray}
\Delta_{\rm nl.} \gamma_{ij} = - \left( \partial_i \xi^j + \partial_j
  \xi^i
- {2\over 3} \delta_{ij} \partial_k \xi^k \right) \, .
\end{eqnarray}
Eq.\ (\ref{backWavetensor}) can alternatively be written as:
\begin{eqnarray}
\label{backWavetensor2}
\boxed{
\int {d^3 q\over (2\pi)^3} \, \left[ {\langle \pi_{\vec q} {\cal O}_{\vec k_1}  ... {\cal O}_{\vec k_N} \rangle^c
\over P_\pi (q)} \Delta_{\rm nl.} \pi_{\vec q}^*
+ {1\over 2} 
{\langle \gamma_{ij} (\vec q) {\cal O}_{\vec k_1}  ... {\cal O}_{\vec k_N} \rangle^c
\over P_\gamma (q)} \Delta_{\rm nl.} \gamma_{ij} (\vec q) {}^* \right]
= \Delta_{\rm lin.} \langle  {\cal O}_{\vec k_1}  ... {\cal O}_{\vec k_N} \rangle^c
\, ,}
\end{eqnarray}
using the fact that $\Delta_{\rm nl.} \gamma^s_{\vec q} {}^* = 
\Delta_{\rm nl.} \gamma_{ij} (\vec q)^* \epsilon^s_{ij} (\hat q) / 2$. 
Let us step through a few low $n$ examples to get 
a feel for the kind of consistency relations that arise from
these diffeomorphisms. The discussion follows that of H2K,
with suitable deformations to the Newtonian gauge.

For $n=0$, $M_{i\ell_0}$ can be written as a trace (dilation), 
an antisymmetric part (which does not generate a nonlinear shift
in the metric\footnote{Such a transformation would correspond to a time-independent rotation.  See Appendix \ref{vectormodes} though for the corresponding decaying mode, which although visible corresponds to a nonstandard choice of initial conditions.}) and a symmetric traceless part (anisotropic rescaling, which
generates tensor perturbations). 
Note that $n=0$ is a special case, in the sense that
the last term of Eq.\ (\ref{diffMgeneral}) does not exist
(because $\nabla^2 \xi^i_{\rm unit.} = 0$).
Focusing on a symmetric traceless $M_{i\ell_0}$, there are 5
independent components. 
Imposing the adiabatic transversality condition
$\hat q^i M_{i\ell_0} (\hat q) = 0$
reduces the number of independent tensor modes to 2.
Thus, at the level of $n=0$, we have one pure scalar
and 2 pure tensor symmetries.
It is straightforward to infer 
the corresponding symmetries in Newtonian gauge:\ 
dilation gets deformed
as discussed in \S \ref{Dilation}; the purely tensor symmetries
take exactly the same form in the two gauges. 
The $n=0$ anisotropic rescaling tensor consistency relation reads:
\begin{eqnarray}
\lim_{\vec q \rightarrow 0} 
{\langle \gamma^s_{\vec q} \, \delta_{\vec k_1} ... 
\delta_{\vec k_N} \rangle^{c'} \over P_{\gamma} (q)}
= -{1\over 2} \epsilon^s_{ij} (\hat q)^* \sum_{a=1}^N k_a^i \partial_{k_a^j}
\langle \delta_{\vec k_1} ... \delta_{\vec k_N} \rangle^{c'} \, ,
\end{eqnarray}
a relation first pointed out by Maldacena \cite{Maldacena:2002vr}.
To derive this, we use $\Delta_{\rm nl.} \gamma_{i\ell_0} (\vec q)^* = - 2
M_{i\ell_0} (2\pi)^3 \delta_D (\vec q)$, which gives 
$\Delta_{\rm nl.} \gamma^s_{\vec q} {}^* = - 2(2\pi)^3 \delta_D (\vec
q)$ if we choose $M_{i\ell_0} = \epsilon^s_{i\ell_0} (\hat q)^*$; the linear
transformation of $\delta$ under the diffeomorphism
$\xi^i = M_{i\ell_0} x^{\ell_0}$ is
$\Delta_{\rm lin.} \delta_{\vec k} = M_{i\ell_0} k^i \partial_{k^{\ell_0}}
\delta_{\vec k}$. 

For $n=1$, there are 3 purely scalar symmetries (the special
conformation transformations) and 4 purely tensor ones.
The special conformal transformations correspond to
\begin{eqnarray}
M_{i\ell_0\ell_1} = 2(b^{\ell_1} \delta_{i\ell_0} + b^{\ell_0}
\delta_{i\ell_1} - b^i \delta_{\ell_0\ell_1}) \, .
\end{eqnarray}
This is manifestly symmetric between $\ell_0$ and $\ell_1$, and
satisfies Eq.\ (\ref{Mthird}). We see that plugging this into
Eq.\ (\ref{diffMgeneral})
reproduces Eq.\ (\ref{SCTsymmSummarize}), and thus
the special conformal consistency relation of Eq.\ \ref{SCTConsistency} follows.
Each of the 4 tensor symmetries come from an $M_{i\ell_0\ell_1}$ that
is symmetric between $\ell_0$ and $\ell_1$, fully traceless over any
pair of indices, and transverse in the sense of
Eq.\ (\ref{tensortransverse}). The corresponding $n=1$ tensor consistency
relation reads:
\begin{eqnarray}
\lim_{\vec q \rightarrow 0} M_{i\ell_0\ell_1} {\partial \over \partial q^{\ell_1}}
\left[ {\langle \gamma_{i\ell_0} (\vec q) \delta_{\vec k_1}
    ... \delta_{\vec k_N} \rangle^{c'} \over P_\gamma (q)} \right]
= - M_{i\ell_0\ell_1} \sum_{a=1}^N \left[ {1\over 2} k_a^i {\partial^2 \over
\partial k_a^{\ell_0} \partial k_a^{\ell_1}} \right] \langle \delta_{\vec k_1}
... \delta_{\vec k_N} \rangle^{c'} \, ,
\end{eqnarray}
where the dependence on $M_{i\ell_0\ell_1}$ can be
removed by applying suitable projectors (see H2K).

For each $n \ge 2$, there are 4 purely tensor symmetries
and 2 mixed symmetries where both $\pi$ and
$\gamma$ transform nonlinearly.
In general, any $n \ge 0$ consistency relation reads:
\begin{eqnarray}
\label{consistencyGeneral}
\boxed{
\begin{split}
\lim_{\vec{q} \to 0}& M_{i\ell_0\ell_1 ... \ell_n}
{\partial^n \over \partial q^{\ell_1} ...
\partial q^{\ell_n}}
\Big[ {\langle \gamma_{i\ell_0} (\vec q) \delta_{\vec k_1}
... \delta_{\vec k_N} \rangle^{c'} \over P_\gamma (q)}
+ \delta_{i\ell_0} {D'(\eta) \over 3c}
{\langle \pi_{\vec q} \delta_{\vec k_1} ... \delta_{\vec k_N} \rangle^{c'}
\over P_\pi (q)} \Big] \\
= & - M_{i\ell_0 \ell_1 ... \ell_n} 
\sum_{a=1}^N \Big[ \delta_{i\ell_0} 
{\partial^n \over \partial k_a^{\ell_1} ... \partial k_a^{\ell_n}}
- \delta_{i\ell_0} {\delta_{n0} \over N}
+ {k_a^i \over n+1} {\partial^{n+1} \over 
\partial k_a^{\ell_0} ... \partial k_a^{\ell_n}} 
\\ &+ \delta_{i\ell_0} {D(\eta_a)' \over 3c} 
\left( {\bar\rho' \over \bar\rho} \Big|_{\eta_a} + \partial_{\eta_a}
\right) {\partial^n \over \partial k_a^{\ell_1} ... \partial
  k_a^{\ell_n}}
\\
& - {n\over c} \delta_{\ell_0\ell_1} \left( {D(\eta_a)}
-{D(\eta)} (1-\delta_{n1})\right)
k_a^i {\partial^{n-1} \over \partial k_a^{\ell_2} ...\partial
  k_a^{\ell_n}}
\Big] 
 \langle \delta_{\vec k_1} ... \delta_{\vec k_N} \rangle^{c'} \,
.
\end{split}}
\end{eqnarray}
This is our most general result:\ each Newtonian
gauge residual diffeomorphism described by Eq. (\ref{diffMgeneral})
gives rise to a consistency relation given by
Eq.\ (\ref{consistencyGeneral}).\footnote{
\label{higherordercorrections}
With the exception of the dilation consistency relation ($n=0$
with $M_{i\ell_0} \propto \delta_{i\ell_0}$),
these consistency relations in general
receive corrections on the right hand side which
either involve replacing one of the hard modes by 
a hard (scalar or tensor) metric perturbation, or
involve higher powers of the metric perturbations.
These corrections arise because the associated diffeomorphisms
generally need to be corrected order by order in metric
perturbations (H2K, \cite{Creminelli:2013mca}). What we focus on in this paper are the lowest
order terms in the diffeomorphisms (i.e.\ metric-independent
contributions). 
Even in the nonlinear regime where density perturbations are large,
the metric perturbations are in general small. Thus,
the corrections to the consistency relations
are negligible in applications where
the hard modes are density (as opposed to metric) perturbations
on sub-Hubble scales.
}
The consistency relations can also be written in a form in which
the matrix $M$ is projected out (see H2K).
The familiar dilation and special conformal consistency
relations are contained here and the $M$'s in those
cases take a form that projects out the tensor
term on the left hand side.
Note that the soft modes are assumed to be at time $\eta$,
while the hard modes are at time $\eta_a$ for each momentum $\vec k_a$.
Purely tensor consistency relations
follow from those $M$'s that are fully traceless,
hence setting the scalar contributions on the left
hand side to be zero (and zeroing out terms proportional
to the Kroenecker delta on the right hand side as well).
For $n \geq 2$, there are choices of $M$ ($2$ for each $n \ge 2$) that
have a structure that gives both non-vanishing tensor
and scalar contributions on the left hand side.
It is worth pointing out that on the right hand side,
the first set of terms (second line) are time-independent;\ they
originate from the unitary gauge diffeomorphisms.
\footnote{The term $- \delta_{i\ell_0} \delta_{n0}/N$
arises from the removal of delta functions. See H2K for
discussion.}
The second set of terms (third line) originate from
$\xi^0_{\rm add.}$, the additional time diffeomorphism
that is necessary to keep us in Newtonian gauge.
Likewise, the last set of terms (fourth line) come from $\xi^i_{\rm add.}$, and we have used the pure tensor consistency relation at level $(n-2)$ to move the terms proportional to $D(\eta)$ to the right hand side.

Let us study the taking of the Newtonian, i.e.\ sub-Hubble, limit.
As explained in \S \ref{robustnessGR}, it is helpful to rewrite
the consistency relations using $\delta_{\vec q} \sim q^2 \pi_{\vec
  q}/{\cal H}$ (the precise relation
is $\delta_{\vec q} = q^2 \pi_{\vec q} D / D'$ for
the $\delta$ and $\pi$ of pressureless matter in the sub-Hubble limit).
Recalling that $c = D'' + 2{\cal H}D' \sim {\cal H}^2$,
we see that Eq.\ (\ref{consistencyGeneral})
na\"ively has a sub-Hubble limit of the schematic form:
\begin{eqnarray}
\lim_{\vec q \rightarrow 0} 
\Big\{ {{\cal H}^2 \over q^2} {\langle \gamma_{\vec q} \delta_{\vec k} ... \rangle^{c'}
\over P_\gamma (q)} + 
{\langle \delta_{\vec q} \delta_{\vec k_1} ... \rangle^{c'}
\over P_\delta(q)}\Big\}
\sim 
\left( {q \over k} \right)^n
\left( {{\cal H}^2 \over q^2} + n{k^2 \over q^2} \right)
\langle \delta_{\vec k} ... \rangle^{c'} \, ,
\end{eqnarray}
where we have equated $\partial_q \sim 1/q$ and $\partial_k \sim 1/k$.
Of the terms on the right hand side, the term suppressed in
the sub-Hubble limit by (${\cal H}^2/q^2$) arises from
the second and third lines of Eq. (\ref{consistencyGeneral}),
and the unsuppressed term comes from the last line of
Eq.\ (\ref{consistencyGeneral}). 
It is also worth noting that the unsuppressed term is
in general non-vanishing even if all the hard modes are at the same time, as long as the soft mode is at a different time; the $n=1$ case (that gives rise to the KRPP consistency relation)
is an exception rather than the rule.

At
first sight, this suggests there is a non-trivial Newtonian
limit for each $n > 0$, with the $n=1$ case (KRPP) being one
example. This is not the case because of the presence of
tensor modes. In all $n \ge 2$ cases where the diffeomorphism
generates a soft scalar, the same diffeomorphism generates
a soft tensor as well.
The tensor equation of motion 
$\gamma_{ij}'' + 2{\cal H}\gamma_{ij}' - \nabla^2 \gamma_{ij} = 0$
(Eq.\ \ref{tensoradiabatic}) tells us that (1) ignoring 
$\nabla^2$, $\gamma_{ij} = {\,\rm const.}$ is the growing mode
solution (or more properly, the dominant mode solution; the other mode
decays);\ (2) allowing for a small $\nabla^2$, 
the growing mode tensor solution gets corrected by a term proportional
to $D$ (see Appendix 
\ref{GRadiabatic});\ (3) when $\nabla^2$ is important,
the tensor mode oscillates with an amplitude that decays as
$1/a$. Cases (1) and (2) pertain to super-Hubble modes
while case (3) has to do with sub-Hubble ones.
The purely tensor consistency relations follow
from diffeomorphisms that are time-independent which generate
tensor modes of type (1). The mixed scalar-tensor consistency
relations follow from diffeomorphisms that generate tensor
modes of type (2) (see footnote \ref{tensortimedep}). 
In neither case are we allowed to take the soft tensor mode to within
the Hubble radius. This is in contrast with the purely scalar
consistency relations (such as dilation and special conformal
transformation),
where the time-dependence of the soft $\pi_{\vec q}$ remains
the same whether it is outside or inside the Hubble radius.
One might be tempted to say:\ within the Hubble radius, the
tensor mode decays anyway, so why not just drop
the tensor term from the consistency relations? 
This is not allowed because the tensor mode enters in
both the numerator and denominator of $\langle \gamma_{\vec q}
... \rangle^{c'}/ P_\gamma (q)$. In general this ratio is independent
of the amplitude of the tensor mode;\ the consistency relations
express precisely this fact.

\section{Discussion}
\label{discuss}

Let us give a summary of our main results.

\vspace{0.2cm}

\noindent {\bf 1.} Consistency relations, between
a squeezed (N+1)-point function and an N-point function,
are a generic consequence of nonlinearly realized symmetries, which shift the fields of interest by amounts that
do not depend on the fields.  They can be thought of as
the LSS analogs of soft-pion theorems in particle physics,
with the squeezed (soft) mode playing the role of a soft pion.
The master formula for deriving consistency relations for
any such nonlinearly realized symmetries is given in
Eq.\ (\ref{backWave}).\footnote{That is for the case of having a single pion;
see Eq.\ (\ref{backWavetensor2}) for an example that involves two
independent soft pions (the scalar mode and the tensor mode(s)).}

\vspace{0.2cm}

\noindent {\bf 2.} In this paper, we focus on nonlinearly realized
symmetries that involve a change of coordinates;\ they are
diffeomorphisms from the point of view of general relativity.
These are residual diffeomorphisms that are allowed even after the usual 
gauge-fixing;\ they all share the property that they do not vanish
at spatial infinity. From earlier work in the unitary gauge
(or $\zeta-$gauge, where
the equal time surface is chosen to be one where
the matter perturbation vanishes, and where for us `matter' means
the dark or pressureless matter), it is known that there is an infinite number of
such symmetries, of the schematic form $x^i \rightarrow x^i + M^i x^{n+1}$,
where $n=0, 1, 2, ...$ and the detailed
index structures of $M^i$ and $x^{n+1}$ 
are suppressed (H2K). Associated with theses change of coordinates
are both linear and nonlinear shifts in the fields, from which we derive consistency relations. For each $n$,
the consistency relation (in unitary gauge) takes the schematic form:
\begin{eqnarray}
\lim_{\vec q \rightarrow 0} \partial_q^n
\left[ 
{\langle \zeta_{\vec q} {\cal O}_{\vec k_1}
... {\cal O}_{\vec k_N} \rangle^{c'} \over P_\zeta (q)}
+ \Theta (n > 1)
{\langle \gamma_{\vec q} {\cal O}_{\vec k_1}
... {\cal O}_{\vec k_N} \rangle^{c'} \over P_\gamma (q)}
\right]
\sim \partial_k^n \langle {\cal O}_{\vec k_1}
... {\cal O}_{\vec k_N} \rangle^{c'}  \, ,
\end{eqnarray}
where $\zeta$ and
$\gamma$ are the (soft) curvature and tensor perturbations,
$P_\zeta$ and $P_\gamma$ are their respective power spectra, and
${\cal O}$ represents some observables 
of interest at hard momenta ${\vec k_1}, ... {\vec k_N}$. 
The correlation functions are connected and have the overall
momentum-conserving delta functions removed. For $n \le 1$,
the tensor term on the left hand side is absent, hence
the step function $\Theta (n > 1)$.

For LSS applications, especially when one is interested in sub-Hubble
scales, it is useful to have the analogous relations
in Newtonian gauge where the matter fluctuations are non-vanishing.
We give the prescription for mapping each unitary gauge
symmetry to its Newtonian gauge counterpart. 
This is given in Eq.\ (\ref{diffMgeneral}). The diffeomorphism
becomes $x^0 \rightarrow x^0 + {\cal H}^{-1} M x^n$ and
$x^i \rightarrow x^i + M^i x^{n+1} + {\cal H}^{-2} M^i x^{n-1}$,
where some of the suppressed
indices of $M$ need to be internally contracted;\ the factors
of ${\cal H}^{-1}$ and ${\cal H}^{-2}$ are meant to indicate
coefficients that are time-dependent, and the powers of
Hubble reflect the order of magnitude of these coefficients.
From these diffeomorphisms, the Newtonian gauge consistency
relations read schematically:
\begin{eqnarray}
\lim_{\vec q \rightarrow 0} \partial_q^n
\left[ 
{\cal H}^{-1} {\langle \pi_{\vec q} {\cal O}_{\vec k_1}
... {\cal O}_{\vec k_N} \rangle^{c'} \over P_\pi (q)}
+ \Theta (n>1)
{\langle \gamma_{\vec q} {\cal O}_{\vec k_1}
... {\cal O}_{\vec k_N} \rangle^{c'} \over P_\gamma (q)}
\right]
\sim 
\left[ \partial_k^n + n{\cal H}^{-2} \partial_k^{n-2} \right]
\langle {\cal O}_{\vec k_1}
... {\cal O}_{\vec k_N} \rangle^{c'}  \, ,
\end{eqnarray}
where $\pi_{\vec q}$ is the soft velocity potential.
The precise form is given in Eq.\ (\ref{consistencyGeneral}). 
Because $\pi_{\vec q}$ is dimensionful (velocity $v^i = \partial_i \pi$), it
is useful to rewrite this using the dimensionless
$\tilde \delta_{\vec q} \equiv q^2 \pi_{\vec q}D/D' \sim q^2 \pi_{\vec
  q}/{\cal H}$:\footnote{ 
We are careful to state this as a definition for $\tilde
\delta_{\vec q}$ -- it is equal to what we normally call $\delta_{\vec
  q}$ the matter density fluctuation only if ${\vec q}$ is
sub-Hubble.}
\begin{eqnarray}
\lim_{\vec q \rightarrow 0} \partial_q^n
\left[ 
q^2 {\cal H}^{-2} {\langle \tilde \delta_{\vec q} {\cal O}_{\vec k_1}
... {\cal O}_{\vec k_N} \rangle^{c'} \over P_{\tilde\delta} (q)}
+ \Theta (n > 1)
{\langle \gamma_{\vec q} {\cal O}_{\vec k_1}
... {\cal O}_{\vec k_N} \rangle^{c'} \over P_\gamma (q)}
\right]
\sim 
\left[ \partial_k^n + n{\cal H}^{-2} \partial_k^{n-2} \right]
\langle {\cal O}_{\vec k_1}
... {\cal O}_{\vec k_N} \rangle^{c'}  \, .
\end{eqnarray}
Two noteworthy points:\ first, the right hand
side should always be understood to contain
corrections that vanish in the $\vec q \rightarrow 0$
limit i.e. we expect $O(q)$ corrections to the right;
second, it appears that a non-trivial 
sub-Hubble limit exists (by sending 
${\cal H} \rightarrow \infty$) for all $n > 0$, but
there is a subtlety.

\vspace{0.2cm}

\noindent {\bf 3.} The subtlety has to do with the adiabatic mode
condition. The consistency relations make three assumptions:\
the existence of nonlinearly realized symmetries, the single field initial
condition, and the adiabatic mode condition. The last says that
the soft mode generated nonlinearly by our symmetry transformation
must satisfy the equation of motion at a low but finite momentum,
i.e.\ the symmetry generated soft mode must have the correct
time dependence to match that of a long wavelength physical mode.
A corollary is that since each of our symmetries is
a diffeomorphism, the {\it same} diffeomorphism must generate
all the soft modes in the problem. This means that in a 
universe with multiple particle species, all the particles must
move with the same velocity on large scales, implying that the equivalence
principle is obeyed. (It is worth stressing that the equivalence principle needs
to hold only on large scales; baryons or galaxies can
move differently from dark matter on small scales.)
This also means the sub-Hubble or Newtonian limit must be taken
with care. Recall that the general relativistic consistency
relations are strictly speaking $\vec q \rightarrow 0$ statements,
and thus valid for super-Hubble scales $q < {\cal H}$.
For the relations to continue to hold even as one takes $q > {\cal H}$
(keeping $k \gg q$ of course), the soft mode must maintain the
same time dependence across the Hubble radius.
The soft $\pi_{\vec q}$ has this property, and can be safely taken
to be sub-Hubble.
The same is not true for the soft tensor mode $\gamma_{\vec q}$.
Thus, consistency relations which involve the tensor mode
strictly hold only if the soft mode is super-Hubble.
It is only in the special cases of $n=0, 1$
(dilation and special conformal transformation)
that the tensor term is absent from the left hand side, and
the corresponding purely scalar consistency relations hold
even within the Hubble radius (i.e.\ for ${\cal H} \ll q \ll k$). 
Of these two cases, $n=0$ does not give a non-trivial right hand side;
only $n=1$ does in the sub-Hubble limit -- this gives precisely
the Newtonian consistency relation obtained by
KRPP (Eq.\ \ref{KRPPrelation1}). In fact, it is worth emphasizing that the KRPP relation as originally expressed (Eq.\ \ref{KRPPrelation3}) should be thought of as containing two separate non-relativistic consistency relations:\ the lack of a $1/q^2$ pole is the sub-Hubble limit of the dilation relation and the leading $1/q$ term comes from the special conformal transformation, which (suitably generalized to Newtonian gauge) reduces
to a time-dependent spatial translation in this limit.

\vspace{0.2cm}

\noindent {\bf 4.} The fact that the symmetries of interest are
diffeomorphisms, albeit ones that do not vanish at spatial infinity,
suggest the consistency relations are fairly robust:\ no detailed
dynamical assumptions need to be made about the hard modes, and
the only information we need is how they transform.
They could be highly non-perturbative and even astrophysically
complex,
such as galaxy observables on small scales (though some care
is needed if the hard modes are metric as opposed to density
fluctuations, see footnote \ref{higherordercorrections}).
We demonstrate this in the context of the Newtonian
consistency relation by writing down explicitly a model of galaxy
dynamics that allows for galaxy formation, mergers, dynamical friction
and so on (\S \ref{robustness}). From the symmetry standpoint,
the key assumption is that the dynamics and formation process of galaxies
be frame invariant, i.e.\ aside from Hubble friction, the only possible
dependence on velocity arises from gradients thereof or on velocity
difference between different species.
From the standpoint of the adiabatic mode condition, the key assumption
is that all objects fall in the same way on large scales.
From the practical standpoint, in terms of checking the consistency
relation at a small momentum $q$ (as opposed to a literally vanishing
$q$), it is important to ensure $q$ is sufficiently squeezed such that
the soft modes evolve in a way consistent with the adiabatic mode
condition. Checking the consistency relation observationally will
allow us to test some fundamental assumptions in cosmology --
in particular the single-field initial condition and the equivalence principle
\cite{Kehagias:2013rpa,Creminelli:2013nua}.

\vspace{0.2cm}

\noindent {\bf 5.} We carry out a perturbative check on the robustness
of the Newtonian consistency relation, by including a galaxy bias at the quadratic level
(\S \ref{robustness}). For the consistency relation to hold, we find
that any non-local quadratic bias cannot be too infrared divergent:
supposing the galaxy density $\delta_g$ is related
to the mass density $\delta$ by
$\delta_g {}_{\vec k} \sim b [ \delta_{\vec k} + \int d^3 k'
(2\pi)^{-3} \, W (\vec k', \vec k-\vec k')
(\delta_{\vec k'} \delta_{\vec k - \vec k'} - \langle \delta_{\vec k'}
\delta_{\vec k - \vec k'} 
\rangle) ]$, the quadratic kernel $W(\vec k', \vec k - \vec k')$
must grow slower than $1/k'$ in the small $k'$ limit
(Eq.\ \ref{Wcondition}). Thus, a local quadratic bias where
$W$ is independent of momenta, which is a form of bias commonly invoked,
respects the consistency relation.
Interestingly, the one case we know of where a $W \sim 1/k'$ behavior
occurs is one where galaxies are born with a velocity bias -- which is
consistent with our understanding that having a large scale
velocity bias violates the adiabatic mode condition
\cite{Peloso:2013spa}.
The consistency relation allows the hard modes to be astrophysically
complex observables such as the galaxy density $\delta_g$. 
But the soft mode strictly speaking should still be that
of the dark (pressureless) matter. 
On the other hand, from an observational standpoint, galaxy density
is much easier to measure. We show that the soft mode can be
a galaxy observable provided that (1) the consistency relation
is corrected by a multiplicative linear bias factor;\ (2) 
the galaxy power spectrum on large scales has
a spectral slope of $n < 1$ 
(where $n \equiv d{\,\rm ln\,} P(q) /d{\,\rm ln\,} q$;\ see \S \ref{robustness}). 

\vspace{0.2cm}

There are a number of outstanding questions for future investigations.
The symmetries we are using are gauge symmetries:\ why is it that
we manage to derive physically relevant statements out of gauge
redundancies? We know consistency relations are not empty statements,
because there are models that violate them, for
instance models of inflation that involve multiple fields
\cite{Creminelli:2004yq}. The answer presumably has to do with two
aspects of the consistency relations:\ (1) they make certain physical
assumptions about the initial condition, in particular
single field initial condition;\ (2) while a strictly zero-momentum
mode is unobservable, a soft mode with a small but non-zero momentum
is observable, and consistency relations are ultimately statements
about the dominant terms in a correlation function with a soft mode.
It would be useful to clarify the curious role of gauge
symmetries in physical statements.
Further clarity in 
the derivation of  the consistency relations
is desirable for another reason:\ we see from our perturbative
check in \S \ref{robustness} that, even without galaxy bias,
the validity of the consistency relations requires the soft power
spectrum to be not too blue ($n < 3$). Why this should be so is
not easy to see from the background wave argument.
Can one see this from other arguments in the literature, such
as the operator formalism (H2K), operator product expansion
\cite{Assassi:2012zq}, or the effective action approach
\cite{Goldberger:2013rsa}?

Lastly, in this paper, we focus exclusively on nonlinearly realized
symmetries that originate from diffeomorphisms.
Could the LSS dynamics have other nonlinearly realized symmetries?
Recently, it was pointed out by \cite{Valageas:2013zda}
and \cite{Kehagias:2013paa} that, indeed, further nonlinearly realized
symmetries exist, albeit ones that involve the transformation
of parameters as well. Can there be more?

\section*{Acknowledgements}

We have benefited a great deal from discussions with Lasha Berezhiani,
Paolo Creminelli, Justin Khoury, Alberto Nicolis,
Jorge Nore\~na, Marko Simonovi\'c, and especially Filippo Vernizzi who collaborated with us
on parts of the project. Our paper has a substantial overlap with the paper \cite{Creminelli:2013mca} by 
Paolo Creminelli, Jorge Nore\~na, Marko Simonovi\'c \& Filippo
Vernizzi. 
Our viewpoint is largely shaped by
joint work with Justin Khoury \& Kurt Hinterbichler, and Walter Goldberger \& Alberto Nicolis.
We further thank Guido D'Amico, Kurt Hinterbichler, 
Marco Peloso, Massimo Pietroni, Rafael Porto, Roman Scoccimarro, David
Spergel and Matias Zaldarriaga for useful discussions.
We thank David Spergel for organizing a non-Gaussianity workshop at Princeton
University where there were many interesting discussions.
We thank the Kavli Institute for Theoretical Physics, 
as well as the Lorentz Center at Universiteit Leiden, for hospitality
while this work was in progress.  This work is
supported in part by the United States Department of Energy under DOE grant DE-FG02-92-ER40699, and by NASA under NASA ATP grant NNX10AN14G.  

\appendix

\section{A Lagrangian for Fluid with Pressure}
\label{fluidAppendix}

In this Appendix, we connect the LSS Lagrangian discussed in \S
\ref{fluidLagLSS} with a more commonly used fluid Lagrangian.
We will continue to work within the zero vorticity regime. 
For generalizations to include a non-vanishing vorticity, see \cite{Dubovsky:2005xd}.
Let us consider the following action:
\begin{eqnarray}
\label{actionstart}
S = \int d^4 x \sqrt{-g} \left[ {1\over 2} M_P^2 R + {\cal P}(X) +
  ... \right] \, ,
\end{eqnarray}
where the first term is the Einstein-Hilbert action, and
${\cal P}(X)$ is the fluid action, where ${\cal P}$ is some function of $X \equiv \sqrt{-
  g^{\mu\nu} \partial_\mu \phi \partial_\nu \phi}$ with $\phi$
describing 
the single degree of freedom of an irrotational fluid.\footnote{
The vanishing of vorticity can be expressed covariantly as
$\epsilon^{\mu \nu \rho \sigma} u_\nu \partial_{\rho}u_{\sigma} = 0$.}
The $...$ stands for other possible matter or energy content in the
universe, i.e.\ the background expansion need not be determined
solely by the ${\cal P}(X)$ fluid in question.
This is a completely relativistic action, and has been used by many
authors \cite{Boubekeur:2008kn,Creminelli:2008wc}. Our goal here is to take the non-relativistic limit,
and connect the result with the action in \S \ref{fluidLagLSS} (Eq.\ \ref{Spi}).

We assume a metric of the form:
\begin{eqnarray}
ds^2 = a^2 \left[ - (1 + 2\Phi) d\eta^2 + (1 - 2\Psi) d\vec x^2
\right] \, .
\end{eqnarray}
The fluid energy-momentum $T_{\mu\nu}$ can be obtained from the fluid
action by $\sqrt{-g} T_{\mu\nu} = - 2 \delta S_{\rm fluid}/\delta g^{\mu\nu}$:
\begin{eqnarray}
T_{\mu\nu} = 2 {\cal P}_{,X} \partial_\mu \phi \partial_\nu\phi +
g_{\mu\nu} {\cal P}  \, 
\end{eqnarray}
which is the energy-momentum of a perfect fluid, with the 4-velocity $U^\mu$,
energy density $\rho$ and pressure $P$ given by:
\begin{eqnarray}
U^\mu = {- \partial^\mu \phi \over \sqrt{X}} \quad , \quad
\rho = 2 X {\cal P}_{,X} - {\cal P}
\quad , \quad P = {\cal P} \, .
\end{eqnarray}
A fluid with an equation of state $P = w \rho$ can be modeled by
a ${\cal P}(X)$ of the form:
\begin{eqnarray}
{\cal P}(X) \propto X^{1+w \over 2w} \, .
\end{eqnarray}
We are interested in the case of a small $w$.
Let us split $\phi$ into a background $\bar\phi (\eta)$ and
perturbation:
\begin{eqnarray}
\label{deltaphi}
\phi = \bar\phi + \delta\phi \quad , \quad \pi \equiv -
\delta\phi/\bar\phi' \, ,
\end{eqnarray}
where we have defined $\pi$ in terms of the field fluctuation
$\delta\phi$. This definition is consistent with the interpretation
of $\pi$ as the velocity potential, as can be seen by working out
$U^\mu$ in terms of $\phi$ and equating
$U^\mu = a^{-1} (1, \vec v)$ i.e. $v_i = \nabla_i \pi$ to the lowest
order in perturbations.
The background $\bar\phi$ obeys:
\begin{eqnarray}
\partial_\eta (a^4 \bar\rho/\bar\phi') = 0 \, ,
\end{eqnarray}
which implies $\bar\phi' \propto a^{1-3w}$, using the fact that
$\bar\rho \propto a^{-3(1+w)}$. 
We denote by $\bar X$ the value of $X$ evaluated at $\phi = \bar\phi$.
Using the fact that $P = w\rho$, we find
$1 + \delta = (1 + [\delta X/\bar X])^{(1+w)/2w}$ which implies
\begin{eqnarray}
{2w \over 1 + w} {\,\rm ln} (1 + \delta) = {\,\rm ln}
\left( 1 + {\delta X \over \bar X} \right)
\end{eqnarray}
Assuming both $w$ and $\delta X/\bar X$ are small, but without
assuming
$\delta$ is small, we can approximate this by
\begin{eqnarray}
\label{wdeltaX}
2w {\,\rm ln} (1 + \delta) \sim  {\delta X \over \bar X} \, .
\end{eqnarray}
Let us write out $\delta X /\bar X$ explicitly in terms of the
metric and $\phi$ fluctuations:
\begin{eqnarray}
\label{dX/X}
{\delta X \over \bar X} = - 2 \left[ \Phi + {1\over a} (a \pi)' 
+ {1\over 2} (\nabla \pi)^2 \right] \, ,
\end{eqnarray}
where we have approximated $\bar\phi' \propto a$ (for small $w$), 
assumed $\Phi \sim v^2 \lesssim \pi' \sim {\cal H} \pi \ll 1$, and ignored
terms to higher order
(we regard $\Phi^2$, $\Phi'$ and $w\Phi$ as both higher order). Eqs.\ (\ref{wdeltaX}) and (\ref{dX/X}) combined
give:
\begin{eqnarray}
\label{Euler2}
-w {\,\rm ln} (1+\delta) = \Phi + {1\over a} (a \pi)' 
+ {1\over 2} (\nabla \pi)^2 \, ,
\end{eqnarray}
Applying the spatial gradient on this equation reproduces the Euler
equation in the presence of pressure (Eq.\ \ref{Fa}),
upon identifying $w$ with $c_s^2$, the sound speed squared.

We are interested in rewriting the action in Eq.\ (\ref{actionstart})
in terms of the fluctuations. In other words, we are not so much interested
in the background as in the dynamics of the fluctuations.
Thus, we ignore the background term in $\sqrt{-g} [M_P^2 R/2 + {\cal P}(X)]$.
We also remove (tadpole) terms that are linear in fluctuations --
they only serve to multiply the background equation of motion. Thus,
we have (in the sub-Hubble, non-relativistic limit):
\begin{eqnarray}
&& S = S_{\rm EH} + S_{\rm fluid} \nonumber \\
&& S_{\rm EH} = \int d^4 x \, a^2 M_P^2 (\nabla_i \Psi \nabla_i \Psi - 2 \nabla_i \Psi
\nabla_i \Phi) \nonumber \\
&& S_{\rm fluid} = \int d^4 x \, w a^4 \bar\rho 
\left[ {1 + w \over 2w} {\delta X \over \bar X} + \left( \delta - {1 +
    w \over 2w} {\delta X \over \bar X} \right) \right] \, ,
\end{eqnarray}
where $S_{\rm EH}$ comes from the Einstein-Hilbert action, and
$S_{\rm fluid}$ comes from the fluid part of the action.
For the latter, we have used the fact that
${\cal P}(X) = w \rho = w \bar\rho (1 + \delta)$, and removed
the background piece $w\bar\rho$. We add and subtract 
$(1 + w)\delta X /(2w \bar X)$ to facilitate the removal of tadpole
terms from expanding out
$(1 + \delta) = (1 + \delta X/\bar X)^{1+w/2w}$. 
We are to understand the last line as follows:\
the first $(1 + w)\delta X /(2w \bar X)$ should be understood
to have the linear fluctuations removed, while the second
$(1 + w)\delta X /(2w \bar X)$ has all terms in it.\footnote{The determinant $\sqrt{-g}$ contains terms of order $\Phi$ and $\Phi^2$.
Terms of order $\Phi$ multiplying the background are removed as
tadpoles. Surviving terms can be seen to multiply at least one factor
of $w$ or of $v^2 \sim \Phi$ (the latter with no compensating $1/w$), and so are
small compared to what we keep (which are or order
$v^2$ or $v^2 (v^2/w)$).}
The fluid part of $S$ is therefore
\begin{eqnarray}
\label{fluidF}
S_{\rm fluid} = \int d^4 x \, a^4 \bar\rho
\left[ - {1\over 2} (\nabla\pi)^2 
+ F \right]   \, ,
\end{eqnarray}
where $F$ is
\begin{eqnarray}
\label{Fdef}
F \equiv w \left( \delta - {1 \over 2w} {\delta X \over \bar X}
\right) \, ,
\end{eqnarray}
with $\delta$ and $\delta X$ understood to be expressible in terms of
$\Phi$ and $\pi$ using Eqs.\ (\ref{wdeltaX}) and
(\ref{dX/X}). We have already verified
that the Euler equation with pressure holds (Eq.\ \ref{Euler2}) --
from the point of view of the action $S_{\rm fluid}$, this merely
serves as a definition for $\delta$. 
Let us verify we obtain the Poisson and continuity equations by
varying the action.
First, we see that $\Psi$ can be integrated out by setting $\Psi =
\Phi$. In other words, let us work with the action:
\begin{eqnarray}
\label{SfluidF}
\boxed{
S = \int d^4 x \, \left\{
- a^2 M_P^2 (\nabla \Phi)^2
+ a^4 \bar\rho
\left[ - {1\over 2} (\nabla\pi)^2 
+ F \right]  \right\} \, .
}
\end{eqnarray}
The variation $\Delta F$ when we vary $\Phi$, using 
Eqs.\ (\ref{wdeltaX}) and
(\ref{dX/X}), is
\begin{eqnarray}
\Delta F = {1\over 2} {\Delta \delta X \over \bar X} \delta = - \delta
\Delta \Phi \, ,
\end{eqnarray}
giving us
\begin{eqnarray}
\Delta S = \int d^4 x \, [2 a^2 M_P^2 \nabla^2\Phi - a^4 \bar\rho
\delta] \Delta \Phi \, ,
\end{eqnarray}
and therefore the Poisson equation. This assumes that the only fluctuations
sourcing $\Phi$ is from the ${\cal P}(X)$ fluid, which of course
can be relaxed. The $\pi$ equation of motion, on the
other hand, follows from
\begin{eqnarray}
\Delta F =  {1\over 2} {\Delta \delta X \over \bar X} \delta 
= - \left[ {1\over a} (a\Delta\pi)' + \nabla_i \pi \nabla
  \Delta \pi \right] \delta \, ,
\end{eqnarray}
which together with the variation $\Delta\left(-\frac{1}{2}(\nabla \pi)^2\right)$ gives us the continuity equation $\delta' + \nabla_i [(1 + \delta)\nabla_i \pi] = 0$. 

The action in Eq.\ (\ref{fluidF}) is a bit hard to use, because $F$
involves a fairly nonlinear function of the fields.
There are two possible simplifications.

We are interested in the $w \rightarrow 0$ limit.
Keeping $\delta$ finite, Eq.\ (\ref{Euler2}) tells us
\begin{eqnarray}
\Phi =  - {1\over a} \left[ (a \pi)' 
+ {1\over 2} a (\nabla \pi)^2 \right] \, 
\end{eqnarray}
which is just the pressureless Euler equation again.
Sending $F \rightarrow 0$, and substituting the above into
Eq.\ (\ref{SfluidF}), we obtain:
\begin{eqnarray}
S = - \int d^4 x \,  \left[ \, {1\over 2} \bar\rho a^4 (\nabla \pi)^2
+ M_P^2 \left( \nabla \left[(a\pi)' +
  {1\over 2}a (\nabla \pi)^2\right] \right)^2 \, \right] 
\end{eqnarray}
reproducing Eq.\ (\ref{Spi}) that we wrote down in
\S \ref{fluidLagLSS}. This justifies the normalization and sign
that was adopted there.

The other possible simplification is to
expand out $(1 + \delta) = (1 + \delta X/\bar X)^{1+w/2w}$
to second order in $\delta X/\bar X$. We have resisted
doing so earlier, because doing so effectively assumes
$\delta X/\bar X$ is parametrically smaller than $w$
(which is itself small). This
is equivalent to assuming small $\delta$, something we
might not want to impose. It is nonetheless instructive to see
what results:

\begin{eqnarray}
\label{fluidS}
\boxed{
S = \int d^4 x \, \left\{- a^2 M_P^2 (\nabla \Phi )^2 \, + \,
 {1\over 2} a^4 \bar\rho 
\left[ -(\nabla \pi)^2
+ c_s^{-2} \left( {1\over a} (a\pi)' + \Phi + {1\over 2} (\nabla \pi)^2
\right)^2 \right] \right\}
\, ,}
\end{eqnarray}
where we have set $w = c_s^2$.
In the context of this action, we treat $\delta$ as
{\it defined} by:
\begin{eqnarray}
\label{Euler3}
- c_s^2 \delta = \Phi + {1\over a} (a \pi)' 
+ {1\over 2} (\nabla \pi)^2 \, .
\end{eqnarray}
Notice how this differs from Eq.\ (\ref{Euler2}) in replacing
${\rm ln\,}(1+\delta)$ by $\delta$ on the left hand side.
The reason for this definition is so that the $\Phi$ equation
of motion gives the Poisson equation as usual.
The $\pi$ equation of motion can be seen to give the continuity
equation. In other words, the full set of equations in this system
are:
\begin{eqnarray}
\label{NewtonianLSScs}
\delta' + \nabla_i [(1 + \delta) v_i] = 0 \quad , \quad 
v_i' + v_j \nabla_j v_i + {\cal H} v_i = - \nabla_i \Phi -
c_s^2 \nabla_i \delta \quad , \quad 
\nabla^2 \Phi = 4\pi G \bar\rho a^2 \delta \, .
\end{eqnarray}
This is the set of equations one expects
for a fluid with pressure, except the pressure term in the Euler
equation is slightly modified from the non-perturbative one 
displayed in Eq.\ (\ref{Fa}). 
Aside from this modification,
this system of equations has the correct nonlinear structure.
In particular, on length scales above the Jeans scale i.e.\
$k < k_J$ where $k_J^2 \equiv a^2\bar\rho/(2 M_P^2
c_s^2)$, one can ignore the pressure term, and the system
reduces exactly to the standard pressureless LSS equations 
(Eq.\ \ref{NewtonianLSS}). 
A useful feature of the action in Eq.\ (\ref{fluidS}) is that
it shows clearly $\pi$ has a kinetic term of the correct sign.

To summarize, the fluid action Eq.\ (\ref{SfluidF}) gives
the exact nonlinear equations for the perturbations of 
a fluid with pressure in the Newtonian limit.
It simplifies in the zero-pressure limit to the action in
Eq.\ (\ref{Spi}), which
gives the exact nonlinear equations for a pressureless fluid.
It can be approximated by the action in Eq.\ (\ref{fluidS})
which gives a linearized pressure term for the Euler equation,
but otherwise retains the full nonlinear structure of the exact theory.


\section{Derivation of the General Relativistic
Adiabatic Mode Conditions in Newtonian Gauge}
\label{GRadiabatic}

In this Appendix, we derive the adiabatic mode conditions appropriate
for the Newtonian gauge, and 
derive the additional diffeomorphism laid out in Eq.\ (\ref{xi0xiiadd}).

For the purpose of deriving consistency relations, it is 
important that the modes generated nonlinearly by the symmetries be 
the low momentum limit of actual physical modes, i.e.\ they must obey adiabatic
mode conditions (see \S \ref{robustness}). 
For the low momentum modes (and for them only), it is sufficient
to consider the linearized Einstein equations, and study the
time-dependence they imply for the perturbations.
The linearized Einstein equations in Newtonian gauge are:
\makeatletter
\def\@eqnnum{{\normalsize \normalcolor (\theequation)}} 
  \makeatother
{\footnotesize
\begin{eqnarray}
\label{G00}
&& -{1\over 2} \delta G^0 {}_0 = -4\pi G \delta T^0 {}_0
\rightarrow \nabla^2 \Psi - 3{\cal H} (\Psi' + {\cal H} \Phi) = 4\pi G a^2 \sum (\rho - \bar\rho)  \, ,
\\
\label{G0i}
&& {1\over 2} \delta G^0 {}_i = 4\pi G \delta T^0 {}_i \rightarrow
- \partial_i \left( \Psi' + {\cal H} \Phi \right) + {1\over 4}
\nabla^2 S_i = 4\pi G a^2 \sum
(\bar\rho + \bar P) (v_i + S_i) \, , \\
\label{Gkk}
&& {1\over 6} {\delta G^k {}_k} = {4\pi \over 3} G \delta T^k {}_k
\rightarrow \Psi'' + {\cal H} (\Phi' + 2 \Psi') + (2 {\cal H}'
+ {\cal H}^2) \Phi - {1\over 3} \nabla^2 (\Psi - \Phi) = 4 \pi G a^2
{\sum} (P - \bar P) , \\
\label{Gij}
&& \delta G^i {}_j - {1\over 3} \delta^i {}_j \delta G^k {}_k 
= 8\pi G \left( \delta T^i {}_j - {1\over 3} \delta^i {}_j \delta T^k
  {}_k \right) \rightarrow \nonumber \\
&& \left( \partial_i \partial_j - {\delta_{ij} \over 3}\nabla^2
\right) (\Psi - \Phi) - (\partial_0 + 2 {\cal H}) \partial_{(i} S_{j)}
+ (\partial_0^2  + 2{\cal H} \partial_0 - \nabla^2)
{\gamma_{ij} \over 2}
= 8\pi G \left( \delta T^i {}_j - {1\over 3} \delta_{ij} \delta T^k
  {}_k \right) .
\end{eqnarray}}
We have allowed the possibility that there might be multiple fluid
components
present (for instance dark matter, baryons, radiation, etc.), 
hence the summation on the right hand side, though we suppress
the label for each component.
Also useful are the linearized conservation equations, assuming each
fluid is individually conserved. The continuity equation 
for each fluid component is
\begin{eqnarray}
\label{deltaConserve}
\delta_n' + \partial_i v_i = 3 \Psi' \, ,
\end{eqnarray}
where $\delta_n$ is related to the density fluctuation $\delta
\equiv (\rho - \bar\rho)/\bar\rho$ by $(1 + w) \delta_n = \delta$,
with $w = P/\rho$ being the equation of state parameter of the fluid component of interest.
This definition of $\delta_n$ is motivated by
the fact that $\bar\rho \propto a^{-3(1+w)}$, and so
it is $\bar\rho^{[1/(1+w)]}$ that redshifts like $a^{-3}$, i.e.\
one can think of $n \equiv \rho^{[1/(1+w)]}$ as the ``number''
density, and of $\delta_n$ as its fractional (small) fluctuation
(for instance, for $w=1/3$, $n$ would be the number density of photons).
In deriving Eq.\ (\ref{deltaConserve}), it is useful to know
${\cal H}^2 - {\cal H}' = 4 \pi G a^2 \sum (\bar \rho + \bar P)$. 
Note also that, in an analogous manner to 
Eq.\ (\ref{deltashift}):
\begin{eqnarray}
\label{deltanShift}
\Delta_{\rm nl.} \delta_n = 3{\cal H} \xi^0 \, .
\end{eqnarray}
The relativistic Euler equation for each component is:
\begin{eqnarray}
\label{EulerEinstein}
(v_i + S_i)' + (1 - 3w) {\cal H} (v_i + S_i) = - \partial_i \Phi -
w \partial_i \delta_n \, ,
\end{eqnarray}
where we have assumed the fourth Einstein equation has a vanishing
source.\footnote{This does not strictly hold if for instance the fluid is
  made out of a collection of relativistic particles, but it is a
  reasonably good approximation in LSS. In essence,
we assume $\Phi = \Psi$, and $(a^2 S_i)' = 0$.
}
Decomposing this last equation into scalar, vector and tensor parts, we have
\begin{eqnarray}
\label{PhiPsi}
\left( \partial_i \partial_j - {\delta_{ij} \over 3}\nabla^2
\right) (\Psi - \Phi) = 0 \, , \\
\label{Sij}
(\partial_0 + 2 {\cal H}) \partial_{(i} S_{j)} = 0 \, , \\
\label{tensor}
(\partial_0^2  + 2{\cal H} \partial_0 - \nabla^2)
{\gamma_{ij} } = 0 \, .
\end{eqnarray}

Following Weinberg \cite{Weinberg:2003sw}, we demand that the (nonlinear part of the) 
symmetry-generated perturbations, as described in
\S \ref{symmetryshifts}), solve the
Einstein equations in a non-trivial way, that is, in a way that
works even if we deform those perturbations slightly away
from the zero momentum $q = 0$ limit.
(See Eq.\ \ref{nadiabatic} for the Newtonian analog of this statement.)
For scalar fluctuations, we therefore insist:
\begin{eqnarray}
\label{scalarAdb}
{\rm scalar \, \, adiabatic \, \, mode \, \, condition\,\,: \,\,} \Psi = \Phi \quad , \quad - (\Psi' + {\cal H}\Phi) = 4\pi G a^2 \sum (\bar
\rho + \bar P) \pi \, ,
\end{eqnarray}
which comes from removing spatial gradients from Eq.\ (\ref{PhiPsi})
and the scalar part of Eq.\ (\ref{G0i}). 
Similarly, the vector adiabatic mode conditions 
are (if vector modes are
present):
\begin{eqnarray}
\label{vectorAdb}
{\rm vector \, \, adiabatic \, \, mode \, \, condition\,\,: \,\,} 
(\partial_0 + 2{\cal H}) S_i = 0 \quad , \quad
(v^\perp_i + S_i)' + (1-3w) {\cal H} (v_i^\perp + S_i) = 0 \, ,
\end{eqnarray}
which comes from Eq.\ (\ref{Sij}) and the transverse
component of Eq.\ (\ref{EulerEinstein}) ($v_i^\perp$ is the transverse
part of the velocity $v_i$, i.e.\ the vorticity component).
Note how the vector modes have only a single solution, which decays. 
Since single field
inflation cannot generate vector modes, we will not consider them
further here.
The tensor adiabatic mode condition is simply the tensor
equation of motion (\ref{tensor}):
\begin{eqnarray}
\label{tensoradiabatic}
{\rm tensor \,\, adiabatic \,\, mode \,\, condition\,\, :\,\,} 
\gamma_{ij}'' + 2{\cal H} \gamma_{ij}' - \nabla^2 \gamma_{ij} = 0 \, .
\end{eqnarray}
Note that we do not wish to simply set the gradient to zero, because
we are interested in diffeomorphisms generating a $\gamma_{ij}$
that is the soft limit of a finite momentum physical mode.

Applying the above conditions to the (nonlinear part of the)
symmetry-generated perturbations (Eqs.\ \ref{metricshifts}
and \ref{pishiftGRnl}), 
we obtain:
\begin{eqnarray}
\label{xiconditionGR}
\xi^0 {}' + 2{\cal H} \xi^0 + {1\over 3} \partial_i \xi^i = 0 \quad ,
\quad \partial_i \xi^i {}' = 0 \quad , \quad
\gamma_{ij} = -\left(\partial_i \xi^j + \partial_j \xi^i - {2\over 3} \delta_{ij}
\partial_k \xi^k\right)\, .
\end{eqnarray}
The first equality enforces $\Phi = \Psi$.
The second equality enforces
the second part of the scalar adiabatic mode condition, with
the understanding that in the soft limit, all fluid components share the
same velocity perturbation $\pi$.
The third equality equates the tensor mode
with the traceless part of the spatial metric generated by
the diffeomorphism -- this holds only if a certain gauge
condition is satisfied such that the scalar contribution to the
spatial metric resides entirely in its trace (see below). 
As far as the adiabatic mode condition is concerned,
the important point is that $\gamma_{ij}$ defined this
way satisfies the tensor equation of motion (\ref{tensoradiabatic}).
These three expressions constitute the adiabatic mode conditions
on residual diffeomorphisms in Newtonian gauge.
For a diffeomorphism to respect the Newtonian gauge,
it must satisfy
\begin{eqnarray}
\label{residualdiffNewt}
\partial_i \xi^0 = \partial_0 \xi^i \quad , \quad
\nabla^2 \xi^i + {1\over 3} \partial_i
(\partial_k \xi^k) = 0
\end{eqnarray}
such that $\Delta_{\rm nl} g_{0i} = 0$ and
the traceless part of $\Delta_{\rm nl} g_{ij}$ is transverse
(see Eq.\ \ref{metricshifts}). 

As discussed in \S \ref{symmetryshifts}, 
one way to organize the set of Newtonian-gauge diffeomorphisms that
satisfy Eqs.\  (\ref{xiconditionGR}) and (\ref{residualdiffNewt}) is to
relate each such diffeomorphism to a corresponding known
residual diffeomorphism in the unitary gauge $\xi_{\rm unit.}$
(Eq.\ \ref{xiTot}):
\begin{eqnarray}
\xi^0 = \xi^0_{\rm add.} \quad , \quad \xi^i = \xi^i_{\rm unit.} +
\xi^i_{\rm add.} \, ,
\end{eqnarray}
where the time-independent $\xi^i_{\rm unit.}$ is supplemented
by a time- and space- diffeomorphism $\xi^0_{\rm add.}, \, \xi^i_{\rm
  add.}$. The time-independent unitary-gauge diffeomorphism $\xi^i_{\rm
  unit.}$ satisfies
Eq.\ (\ref{scalartensorsymmUnit}). Comparing this with
Eq.\ (\ref{residualdiffNewt}), we see that $\xi^i_{\rm add.}$
itself must satisfy the same:\ $\nabla^2 \xi^i_{\rm add.} + \partial_i (\partial_k \xi_{\rm
  add.}^k)/3 = 0$. For this reason, 
we might as well absorb any time-independent part of
$\xi^i_{\rm add.}$ into the definition of $\xi^i_{\rm unit.}$. 
From the second condition in Eq.\ (\ref{xiconditionGR}),
we see that $\partial_i \xi^i_{\rm add}$ must be independent of time.
Suppose it is equal to some function $f(x)$. One can express
$\xi^i_{\rm add.}$ as a gradient and a curl (plus possibly some
function that depends only on time). The divergence of
the gradient is what matches up with $f(x)$, i.e.\ the gradient
part is time-independent, and so by definition, it should have been
absorbed into $\xi_{\rm unit.}$ already. Thus, we can set $f(x) = 0$
and we can assume $\partial_i \xi^i_{\rm add.} = 0$ without loss
of generality. The first condition of Eq.\ (\ref{xiconditionGR}) thus
tells us
\begin{eqnarray}
\label{xi0addEqt}
\xi^0_{\rm add.} {}' + 2{\cal H}\xi^0_{\rm add.} + {1\over 3} \partial_i
\xi^i_{\rm unit.} = 0
\end{eqnarray}
Recall from Eq.\ (\ref{pishiftGRnl}) that $\Delta_{\rm nl.} \pi = \xi^0
= \xi^0_{\rm add.}$.  From Appendix \ref{piEOMderive}, 
we see that $\pi$ in the soft limit has the time
dependence of $D'$ where $D(\eta)$ is the linear growth factor satisfying:
\begin{eqnarray}
D'' + 2{\cal H} D' - c = 0
\end{eqnarray}
where $c$ is a constant. Comparing this against
Eq.\ (\ref{xi0addEqt}) and keeping only the growing solution, we see that
\begin{eqnarray}
\label{xi0add}
\xi^0_{\rm add.} = - {1\over 3c} D' \partial_i \xi^i_{\rm unit.} \, ,
\end{eqnarray}
confirming the time-diffeomorphism of Eq.\ (\ref{xi0xiiadd}). 
We can then solve for $\xi^i_{\rm add.}$ from the first
expression of Eq.\ (\ref{residualdiffNewt}) which tells us
$\partial_i \xi^0_{\rm add.} = \partial_0 \xi^i_{\rm add.}$, i.e.\
\begin{eqnarray}
\label{xiIadd}
\xi^i_{\rm add.} = - {1\over 3c} D \partial_i (\partial_k \xi^k_{\rm
  unit.})
= {1\over c} D \nabla^2 \xi^i_{\rm unit.} \, ,
\end{eqnarray}
where the second equality follows from
Eq.\ (\ref{scalartensorsymmUnit}). This confirms the
space-diffeomorphism of Eq.\ (\ref{xi0xiiadd}).
As a self-consistency check, one can see that
Eq.\ (\ref{scalartensorsymmUnit}) also implies that
$\partial_i \xi^i_{\rm add.} = 0$.
Lastly, it can also be checked that the tensor mode
created by this diffeomorphism 
(the third expression of Eq.\ \ref{xiconditionGR})
obeys the tensor equation of motion.
To see this, it is useful to note that 
because $\xi^i_{\rm unit.}$ satisfies
$\nabla^2 \xi^i_{\rm unit.} +
\partial_i (\partial_k \xi^k_{\rm unit.})/3 = 0$,
we also know $\nabla^2 \partial_k \xi_{\rm unit.}^k = 0$,
$\nabla^2 \partial_i \xi^j_{\rm unit.}
= - \partial_i \partial_j \partial_k \xi^k_{\rm unit.}/3$,
and $\nabla^2 \nabla^2 \partial_i \xi^j_{\rm unit.} = 0$.\footnote{\label{tensortimedep} In other words, the combined action of
$\xi^i_{\rm unit.} + \xi^i_{\rm add.}$ generates
a tensor mode of the form $\gamma_{ij}
= [1 + (D/c)\nabla^2] \gamma_{ij} {}_{\rm unit.}$
where $\gamma_{ij} {}_{\rm unit.}$ is the tensor mode
generated by the time-independent
unitary diffeomorphism alone. That the constant
tensor (growing) mode gets corrected at finite momentum by
a term proportional to momentum squared should not
be surprising. The time dependence can also be checked
explicitly by solving the tensor equation of motion in the
small but finite momentum limit.
}
It is worth noting that for pure tensor symmetries, where
$\partial_i \xi^i_{\rm unit} = 0$, both $\xi^0_{\rm add.}$ and
$\xi^i_{\rm add.}$ vanish, and so the pure tensor symmetries
coincide in the Newtonian gauge and unitary gauge, as they should.

\section{Derivation of the General Relativistic Velocity Equation}
\label{piEOMderive}

Our goal in this Appendix is to derive the following equation for
the velocity potential $\pi$:
\makeatletter
\def\@eqnnum{{\normalsize \normalcolor (\theequation)}} 
  \makeatother
{\footnotesize
\begin{eqnarray}
\label{piEOM}
(\pi' + 2{\cal H}\pi - C)' - 3w{\cal H} (\pi' + 2{\cal H}\pi - C)
= w (g' + {\cal H}g) 
- (1+3w) \left[ ({\cal H}^2 - {\cal H}') \pi - 4\pi G a^2 \sum 
(\bar\rho + \bar P) \pi \right] \, ,
\end{eqnarray}}
where
$g \equiv \int d\eta \nabla^2 \pi$, 
and $C$ is a constant in time but not space (determined by
initial conditions). Here,
$\pi$ refers to the velocity potential of some particular fluid component
of interest with an equation of state parameter $w$ -- except 
in the very last term where $\sum (\bar\rho + \bar P)\pi$ refers to a sum
over all fluid components.
We will use this equation to deduce useful statements about the
time-dependence of $\pi$ in the soft limit.

The continuity equation (\ref{deltaConserve}) can be
integrated once to obtain:
\begin{eqnarray}
\delta_n = 3(\Psi + C) - g \quad , \quad  g \equiv \int d\eta \nabla^2
\pi \, ,
\end{eqnarray}
where $C$ denotes some integration constant -- independent of time,
but dependent on space in general.
This can be substituted into (the scalar part of) the relativistic
Euler equation 
(\ref{EulerEinstein})
and integrated once to give
\begin{eqnarray}
\label{piPsi1}
\pi' + {\cal H} (1-3w) \pi = - (1 + 3w)\Psi - 3w C + wg \, .
\end{eqnarray}
On the other hand, (the scalar part of) the $\delta G^0 {}_i$ equation (\ref{G0i}) can be integrated once
to obtain 
\begin{eqnarray}
\label{piPsi2}
-(\Psi' + {\cal H} \Psi) = 4\pi G a^2 \sum (\bar\rho + \bar P) \pi \,
,
\end{eqnarray}
where we have assumed $\Psi = \Phi$.
One can solve for $\Psi$ from Eq.\ (\ref{piPsi1}), substitute the result
into Eq.\ (\ref{piPsi2}), and subtract from both
sides $({\cal H}^2 - {\cal H}') \pi$. This gives
Eq.\ (\ref{piEOM}). 
Note that in this derivation, 
we have not thrown away any gradient terms, i.e.\ we have not
made any super-Hubble approximation.


Equation (\ref{piEOM}) simplifies if $\pi$ happens to be the same for all
fluid components, in which case what appears within the square brackets
$[\, \,]$ sums to zero, by virtue of 
${\cal H}^2 - {\cal H}' = 4\pi G a^2 \sum (\bar\rho + \bar P)$.
This happens, for instance, if we work on
super-Hubble scales and assume adiabatic initial conditions.
One can check that this is a self-consistent solution
on super-Hubble scales, and assuming all fluid components move
with the same $\pi$, the entire right hand side of 
Eq.\ (\ref{piEOM}) vanishes, implying:
\begin{eqnarray}
\pi' + 2{\cal H} \pi - C \propto a^{3w} \, .
\end{eqnarray}
This suggests different fluid components (with different
$w$'s) evolve
differently, {\it unless} 
the proportionality constant
is in fact zero, i.e.
\begin{eqnarray}
\label{piC}
\pi' + 2{\cal H} \pi - C = 0 \, .
\end{eqnarray}
With this choice of the initial condition, it is thus consistent to have the same $\pi$
for all fluid components on super-Hubble scales.
{\it Interestingly, for pressureless matter ($w=0$), Eq.\ (\ref{piC}) holds 
even on
sub-Hubble (but linear) scales, after radiation domination}.
This can be seen by setting $w=0$ in Eq.\ (\ref{piEOM}),
and noting that during matter or cosmological constant
domination, the terms within the square brackets $[\,\,]$ still sum to zero.
This means that for a wave-mode (of pressureless
matter) that enters the Hubble
radius after radiation domination, 
Eq.\ (\ref{piC}) holds for its entire history.
For a wave-mode that enters the Hubble radius 
before matter domination, however, Eq.\ (\ref{piC}) does not
hold in the intermediate period when the mode is within
the Hubble radius during the radiation dominated phase.\footnote{
It is worth pointing out that Eq.\ (\ref{piC}), when
substituted into Eq.\ (\ref{piPsi1})
gives $\Psi = - (\pi' + {\cal H} \pi)$ -- this holds
as long as
the $wg$ term can be ignored, which can be justified
either for super-Hubble scales, or for $w=0$. 
}


As we see in \S \ref{scalarrelations},  the fact that Eq.\ (\ref{piC}) holds
for pressureless matter both inside and outside the Hubble radius
(as long as the wave-mode of interest crosses the Hubble radius
after radiation domination)
enables us to have interesting
consistency relations in the Newtonian limit.
It is also worth noting that since $\partial_i \pi$
describes the dark matter velocity on all (linear) scales, including
sub-Hubble ones, where we know the velocity scales with
time as $D'$ ($D$ being the linear growth factor), we expect
\begin{eqnarray}
\label{Dc}
D'' + 2{\cal H} D' - c = 0 \, ,
\end{eqnarray}
where $c$ is some constant whose normalization is arbitrary -- its
normalization is tied to the normalization of the growth factor $D$.
That this relation holds for the Newtonian growth factor
in a matter dominated
universe is easy to check:\ $D \propto a$. That this is true for more
general cases is less familiar. Let us check this for a universe with
a cosmological constant.

For a flat universe with pressureless matter and a cosmological constant,
the linear growth factor can be written in closed
form \cite{dodelson}:
\begin{eqnarray}
\label{Danalytic}
D = {5 \Omega_m^0 \over 2} {H(a) \over H_0} \int_0^a {d\tilde a \over
(\tilde a H(\tilde a)/H_0)^3} \, ,
\end{eqnarray}
where $\Omega_m^0$ is the matter density today, $H_0$ is the Hubble
constant today, and the normalization is chosen such that $D$ equals
the scale factor $a$ in the early universe. Note that $H = a'/a^2$
whereas
${\cal H} = a'/a$. 

Let us rewrite what we want to show, Eq.\ (\ref{Dc}), as follows:
\begin{eqnarray}
\label{Hc}
H {d\over da} \left( a^4 H {dD\over da} \right) = c \, .
\end{eqnarray}
First, note that
\begin{eqnarray}
{d \over da} \left( {H \over H_0} \right) = - {3\Omega_m^0\over 2 a^4}
{H_0 \over H} \, .
\end{eqnarray}
We can therefore work out:
\begin{eqnarray}
{dD \over da} = -{3\Omega_m^0\over 2 a^4}
{H_0 \over H} {5 \Omega_m^0 \over 2} \int_0^a {d\tilde a \over
(\tilde a H(\tilde a)/H_0)^3} + {5\Omega_m^0 \over 2} {H_0^2 \over a^3
H^2} \, .
\end{eqnarray}
Therefore,
\begin{eqnarray}
{d\over da} \left( a^4 H {dD\over da} \right)
= {5\Omega_m^0 \over 2} {H_0^2 \over H} \, ,
\end{eqnarray}
implying the desired result Eq.\ (\ref{Hc}).

\section{Decaying Modes and Vector Symmetries}
\label{vectormodes}

At several places in our discussion we have omitted the second solution to the scalar and tensor equations of motion, which decays at late times.  It is worth discussing the decaying solution in a little more detail, both for completeness and because the reasons for ignoring it are slightly subtle.  

The decaying solution arises at the linearized level because both the scalar and tensor equations of motion for the linearized modes (Eq.\ \ref{linearizeDelta} or Eq.\ \ref{Dc}, and \ref{tensoradiabatic}) are second order.  In particular, we note that Eq.\ \ref{Dc} for the linear growth factor $D(\eta)$ can allow for a decaying piece $D' \propto 1/a^2$ which is independent of $c$ as well as the strictly growing piece (which does depend on having $c \neq 0$, and which is what is usually meant by the linear growth factor).  The decaying piece corresponds to varying the lower limit of the integral in Eq.\ \ref{Danalytic}, which is arbitrary.  It is straightforward to check that this solution gives the correct decaying solution $\delta \propto \mathcal{H}/a$ for Eq.\ \ref{linearizeDelta} in the Newtonian limit, using the helpful lemma
\be
\mathcal{H}' - \mathcal{H}^2 = -4 \pi G_N a^2 \Sigma(\bar{\rho} +
\bar{P}) = -4\pi G_N \bar{\rho}_{\rm matter} \propto \frac{1}{a^2}
\ee
in a $\Lambda$CDM universe.  From Eq.\ \ref{xi0addEqt}, the most
general diffeomorphism $\xi^0_{\rm add.}, \xi^i_{\rm unit.} +
\xi^i_{\rm add.}$ involving scalar and tensor modes allowed by the adiabatic mode conditions includes the decaying piece 
\be
\xi^0_{\rm add., decay} = \frac{d(x)}{a^2}\, , \, \xi^i_{\rm add., decay} = \int^{\eta}\frac{\partial^i d(x)}{a^2} 
\ee
where $d(x)$ is harmonic because of Eqs.\ (\ref{xiconditionGR}) and (\ref{residualdiffNewt}). It is straightforward to check that this satisfies the tensor equation \ref{tensoradiabatic} in the super-Hubble limit as well.

Note that the decaying mode is independent of the time-independent (growing mode) unitary transformation, and so the diffeomorphism corresponding to a decaying mode is a separate symmetry.  We can Taylor expand as in \S \ref{vectortensorrelations} to write the most general decaying symmetries as
\be
\xi^0_{\rm decay} = \frac{1}{n!}\frac{1}{a^2} M_{\ell \ell \ell_1
  \cdots \ell_n}x^{\ell_1} \cdots x^{\ell_n}, \, \xi^i_{\rm decay} = \frac{1}{(n-1)!}\int^{\eta} \frac{1}{a^2} M^{i}_{\ell \ell \ell_2 \cdots \ell_n} x^{\ell_2} \cdots x^{\ell_n}.
\ee
for $(n \geq 1)$, where the $M$'s are constant and obey the usual transversality and adiabatic transversality conditions.  

Can we derive consistency relations for the decaying modes using these symmetries, using Eq.\ \ref{backWave} or its generalization Eq.\ \ref{backWavetensor}?  We argue that the answer is no, though it is not enough to say that these simply decay away.  Rather, keeping the decaying modes would correspond to a nonstandard choice of the initial vacuum state in the far past:\ if the decaying mode is not set to zero the energy associated with these modes (the scalar part of the action scales like $\rho a^4 \sim a^2(\nabla \Phi)^2 \sim \mathcal{H}^2/a^2$) becomes divergent at early times.  Had we chosen to ignore this problem and work within the putative vacuum containing only decaying modes, we could have, in which case the lack of the time-dependent piece would make our consistency relations look slightly different from Eq.\ \ref{consistencyGeneral}.  Note that it is not true that the consistency relations should vanish in this case because the modes decay at late times:\ this is because the consistency relation is a ratio between the (N+1)-pt function and the power spectrum, both of which decay, but the ratio on the right hand side does not.

For the sake of completeness, we discuss the case where the symmetries may involve vector modes -- unlike the scalars and tensors, these have only a decaying solution.  The condition
\be
\nabla^2 \xi^i + \frac{1}{3} \partial^i (\partial_j \xi^j) = 0.
\ee
on the spatial part of the diffeomorphism will still be obeyed, but the condition $\partial_0 \xi^i = \partial_i \xi^0$ will be violated and replaced by the weaker condition
\be
\nabla^2 \xi^0 = \partial^0 \partial_i \xi^i = 0.
\ee
where in the second equality we have made use of the second equation in Eq.\ \ref{xiconditionGR}.  Using the vector adiabatic mode conditions (Eq.\ \ref{vectorAdb}) we have
\be
(\partial_i \xi^0 - \partial_0 \xi^i)_{\rm vec.} =
\frac{\bar{\xi}^i}{a^2}, \, (\partial_i \xi^0)_{\rm vec.} \propto \frac{1}{a}
\ee
where $\bar{\xi}^i$ is transverse and time-independent.  The second of these conditions is clearly incompatible with the first condition in Eq.\ \ref{xiconditionGR}:
\be
{\xi^0}' + 2\mathcal{H}\xi^{0} + \frac{1}{3}\partial_i \xi^i = 0,
\ee
unless $(\partial_i \xi^0)_{\rm vec.}$ vanishes, and so the vector part of the symmetry will be $\bar{\xi}^i \int\frac{d\eta}{a^2} \subset \xi^i$.  We can Taylor expand
\be
\bar{\xi}^i = \sum\frac{1}{(n+1)!}\bar{M}^i {}_{\ell_0 \ell_1 \cdots \ell_n} x^{\ell_0} \cdots x^{\ell_n}
\ee
The $\bar{M}$'s are completely traceless, and the obey the usual tensor transversality conditions Eqs.\ \ref{Mthird}, \ref{tensortransverse} as well, so these are vector-tensor symmetries.  Note that they obey the tensor equation of motion Eq.\ \ref{tensoradiabatic} on superhorizon scales, though they correspond to the decaying mode solution.  For $n \geq 1$ there will be 4 such symmetries at each level.  For $n = 0$, there are additional symmetries where with $\bar{M}_{i\ell_0}$ is antisymmetric in the indices;\ these correspond to time-dependent rotations.\footnote{Note that for $n \geq 1$, the tensor structure that is antisymmetric in the first two indices and symmetric in the last $n + 1$ will vanish identically.}  They will obey the further adiabatic transversality condition
\be
\label{vectortransverse}
\hat{q}^i (\bar{M}_{i \ell_0}(\hat{q}) - \bar{M}_{\ell_0 i}) = 0
\ee
which will reduce the number of allowed polarizations from 3 to 2.  Since a localized rotation necessarily involves shearing, we need this condition to enforce transversality in addition to the antisymmetric tensor structure.

To summarize, at $n = 0$ there are two purely vector symmetries, and for $n \geq 1$ there are four vector + tensor symmetries.  Since vector modes always decay, for our choice of vacuum there are no consistency relations that involve vector modes.


\bibliographystyle{JHEP}

\begin{thebibliography}{10}


  

  
  
  
  
  
  
  
  
  
  
  
  
  
  



\bibitem{Kehagias:2013yd} 
  A.~Kehagias and A.~Riotto,
  ``Symmetries and Consistency Relations in the Large Scale Structure of the Universe,''
  Nucl.\ Phys.\ B {\bf 873}, 514 (2013)
  [arXiv:1302.0130 [astro-ph.CO]].
  
\bibitem{Peloso:2013zw} 
  M.~Peloso and M.~Pietroni,
  ``Galilean invariance and the consistency relation for the nonlinear squeezed bispectrum of large scale structure,''
  JCAP {\bf 1305}, 031 (2013)
  [arXiv:1302.0223 [astro-ph.CO]].

\bibitem{Maldacena:2002vr} 
  J.~M.~Maldacena,
  ``Non-Gaussian features of primordial fluctuations in single field inflationary models,''
  JHEP {\bf 0305}, 013 (2003)
  [astro-ph/0210603].

\bibitem{Creminelli:2004yq} 
  P.~Creminelli and M.~Zaldarriaga,
  ``Single field consistency relation for the 3-point function,''
  JCAP {\bf 0410}, 006 (2004)
  [astro-ph/0407059].
  
\bibitem{Creminelli:2012ed} 
  P.~Creminelli, J.~Norena and M.~Simonovi\'c,
  ``Conformal consistency relations for single-field inflation,''
  JCAP {\bf 1207}, 052 (2012)
  [arXiv:1203.4595 [hep-th]].
  
\bibitem{Hinterbichler:2012nm} 
  K.~Hinterbichler, L.~Hui and J.~Khoury,
  ``Conformal Symmetries of Adiabatic Modes in Cosmology,''
  JCAP {\bf 1208}, 017 (2012)
  [arXiv:1203.6351 [hep-th]].

\bibitem{Hinterbichler:2013dpa} 
(H2K) K.~Hinterbichler, L.~Hui and J.~Khoury,
  ``An Infinite Set of Ward Identities for Adiabatic Modes in Cosmology,''
  arXiv:1304.5527 [hep-th].
  
\bibitem{Assassi:2012zq} 
  V.~Assassi, D.~Baumann and D.~Green,
  ``On Soft Limits of Inflationary Correlation Functions,''
  JCAP {\bf 1211}, 047 (2012)
  [arXiv:1204.4207 [hep-th]].

\bibitem{Assassi:2012et} 
  V.~Assassi, D.~Baumann and D.~Green,
  ``Symmetries and Loops in Inflation,''
  JHEP {\bf 1302}, 151 (2013)
  [arXiv:1210.7792 [hep-th]].

\bibitem{Kehagias:2012pd} 
  A.~Kehagias and A.~Riotto,
  Nucl.\ Phys.\ B {\bf 864}, 492 (2012)
  [arXiv:1205.1523 [hep-th]].

\bibitem{Goldberger:2013rsa} 
  W.~D.~Goldberger, L.~Hui and A.~Nicolis,
  ``One-particle-irreducible consistency relations for cosmological perturbations,''
  Phys.\ Rev.\ D {\bf 87}, 103520 (2013)
  [arXiv:1303.1193 [hep-th]].
  
\bibitem{Pimentel:2013gza} 
  G.~L.~Pimentel,
  ``Inflationary Consistency Conditions from a Wavefunctional Perspective,''
  arXiv:1309.1793 [hep-th].

\bibitem{Berezhiani:2013ewa} 
  L.~Berezhiani and J.~Khoury,
  ``Slavnov-Taylor Identities for Primordial Perturbations,''
  arXiv:1309.4461 [hep-th].

\bibitem{Boubekeur:2008kn} 
  L.~Boubekeur, P.~Creminelli, J.~Norena and F.~Vernizzi,
  ``Action approach to cosmological perturbations: the 2nd order metric in matter dominance,''
  JCAP {\bf 0808}, 028 (2008)
  [arXiv:0806.1016 [astro-ph]].
  
\bibitem{Creminelli:2008wc} 
  P.~Creminelli, G.~D'Amico, J.~Norena and F.~Vernizzi,
  ``The Effective Theory of Quintessence: the w<-1 Side Unveiled,''
  JCAP {\bf 0902}, 018 (2009)
  [arXiv:0811.0827 [astro-ph]].
  
\bibitem{Creminelli:2013mca} 
  P.~Creminelli, J.~NoreÒa, M.~Simonovi\'c and F.~Vernizzi,
  ``Single-Field Consistency Relations of Large Scale Structure,''
  arXiv:1309.3557 [astro-ph.CO].

\bibitem{Peloso:2013spa} 
  M.~Peloso and M.~Pietroni,
  ``Ward identities and consistency relations for the large scale structure with multiple species,''
  arXiv:1310.7915 [astro-ph.CO].

\bibitem{Kehagias:2013rpa} 
  A.~Kehagias, J.~NoreÒa, H.~Perrier and A.~Riotto,
  ``Consequences of Symmetries and Consistency Relations in the Large-Scale Structure of the Universe for Non-local bias and Modified Gravity,''
  arXiv:1311.0786 [astro-ph.CO].
  
\bibitem{Kehagias:2013paa} 
  A.~Kehagias, H.~Perrier and A.~Riotto,
  ``Equal-time Consistency Relations in the Large-Scale Structure of the Universe,''
  arXiv:1311.5524 [astro-ph.CO].

\bibitem{Creminelli:2013poa} 
  P.~Creminelli, JÈrÙm.~Gleyzes, M.~Simonovi\'c and F.~Vernizzi,
  ``Single-Field Consistency Relations of Large Scale Structure. Part II: Resummation and Redshift Space,''
  arXiv:1311.0290 [astro-ph.CO].

\bibitem{Creminelli:2013nua} 
  P.~Creminelli, JÈrÙm.~Gleyzes, L.~Hui, M.~Simonovi\'c and F.~Vernizzi,
  ``Single-Field Consistency Relations of Large Scale Structure. Part III: Test of the Equivalence Principle,''
  arXiv:1312.6074 [astro-ph.CO].

\bibitem{Valageas:2013zda} 
  P.~Valageas,
  ``Angular averaged consistency relations of large-scale structures,''
  arXiv:1311.4286 [astro-ph.CO].

\bibitem{Kehagias:2013paa} 
  A.~Kehagias, H.~Perrier and A.~Riotto,
  ``Equal-time Consistency Relations in the Large-Scale Structure of the Universe,''
  arXiv:1311.5524 [astro-ph.CO].

\bibitem{Valageas:2013cma} 
  P.~Valageas,
  ``Consistency relations of large-scale structures,''
  arXiv:1311.1236 [astro-ph.CO].

\bibitem{Weinberg:2003sw} 
  S.~Weinberg,
  ``Adiabatic modes in cosmology,''
  Phys.\ Rev.\ D {\bf 67}, 123504 (2003)
  [astro-ph/0302326].

\bibitem{HNS}
  L.~Hui, A.~Nicolis and C.~Stubbs,
  ``Equivalence principle implications of modified gravity models,''
  Phys.\ Rev.\  D {\bf 80} 104002 (2009)
  [arXiv:0905.2966 [astro-ph.CO]].

\bibitem{HN}
 L.~Hui and A.~Nicolis,
  ``An Equivalence principle for scalar forces,''
  Phys.\ Rev.\ Lett.\ \ {\bf 105}, 231101  (2010)
  [arXiv:1009.2520 [hep-th]].

\bibitem{Sheth:2012fc} 
  R.~K.~Sheth, K.~C.~Chan and R.~Scoccimarro,
  ``Non-local Lagrangian bias,''
  Phys.\ Rev.\ D {\bf 87}, 083002 (2013)
  [arXiv:1207.7117 [astro-ph.CO]].

\bibitem{Chan:2012jj} 
  K.~C.~Chan, R.~Scoccimarro and R.~K.~Sheth,
  ``Gravity and Large-Scale Non-local Bias,''
  Phys.\ Rev.\ D {\bf 85}, 083509 (2012)
  [arXiv:1201.3614 [astro-ph.CO]].

\bibitem{Dubovsky:2005xd} 
  S.~Dubovsky, T.~Gregoire, A.~Nicolis and R.~Rattazzi,
  ``Null energy condition and superluminal propagation,''
  JHEP {\bf 0603}, 025 (2006)
  [hep-th/0512260].

\bibitem{Crocce:2005xy} 
  M.~Crocce and R.~Scoccimarro,
  ``Renormalized cosmological perturbation theory,''
  Phys.\ Rev.\ D {\bf 73}, 063519 (2006)
  [astro-ph/0509418].

\bibitem{dodelson}
 S.~Dodelson,
  ``Modern cosmology,''
  Amsterdam, Netherlands:\ Academic Pr.\ (2003), 440 pp.

\end{thebibliography}
\renewcommand{\refname}{Bibliography}
\addcontentsline{toc}{section}{Bibliography}
\providecommand{\href}[2]{#2}\begingroup\raggedright

\end{document}